%% file: article.tex
 \documentclass[a4paper,twocolumn,english,superscriptaddress,accepted=2025-07-02,amsmath,amssymb,amsfonts]{quantumarticle}
\pdfoutput=1

\usepackage[utf8]{inputenc}
\usepackage[english]{babel}
\usepackage[T1]{fontenc}
\usepackage[numbers,sort&compress]{natbib}
\usepackage{csquotes}

\newcommand{\externalizetikz}{0}   

\usepackage{adjustbox} 
\usepackage{braket}
\usepackage{qcircuit}
\usepackage{subdepth}

\usepackage[usenames,dvipsnames]{xcolor}
\usepackage{mathtools,mymath,cancel}
\usepackage[algo2e,ruled]{algorithm2e}
\SetKwComment{Comment}{/* }{ */}

\usepackage{comment}
\usepackage{fontawesome}
\DeclareUnicodeCharacter{2212}{-}
\DeclareUnicodeCharacter{0301}{\'{}}

\definecolor{myblue}{rgb}{0,0.4470,0.7410}
\definecolor{myred}{rgb}{0.8500,0.3250,0.0980}
\definecolor{myorange}{rgb}{0.9290,0.6940,0.1250}
\definecolor{mypurple}{rgb}{0.4940,0.1840,0.5560}
\definecolor{mygreen}{rgb}{0.4660,0.6740,0.1880}
\definecolor{mylightblue}{rgb}{0.3010,0.7450,0.9330}
\definecolor{mydarkred}{rgb}{0.6350,0.0780,0.1840}

\ifnum\externalizetikz=2
  \usepackage{tikzexternal}
  \tikzsetexternalprefix{fig/}
\else
  \usepackage{tikz}
  \usepackage{pgfplots}
  \usetikzlibrary{calc}
  \usetikzlibrary{intersections}
  \usetikzlibrary{patterns.meta}
  \usetikzlibrary{positioning}
  \usetikzlibrary{plotmarks}
  \tikzset{>=latex}
  \usetikzlibrary{shapes,decorations,matrix}
  \usetikzlibrary{backgrounds,fit,decorations.pathreplacing,calc}
  \pgfplotsset{
    compat=newest,
    table/header=false,
    tick label style={font=\footnotesize},
    label style={font=\small},
    legend style={font=\footnotesize},
    legend cell align=left,
    colormap={parula}{
      rgb255=(53,42,135)
      rgb255=(15,92,221)
      rgb255=(18,125,216)
      rgb255=(7,156,207)
      rgb255=(21,177,180)
      rgb255=(89,189,140)
      rgb255=(165,190,107)
      rgb255=(225,185,82)
      rgb255=(252,206,46)
      rgb255=(249,251,14)
    }
  }
  \pgfplotsset{
    myColOne/.style={myblue},
    myColTwo/.style={myred},
    myColThr/.style={myorange},
    myColFou/.style={mypurple},
    myColFiv/.style={mygreen},
    myColSix/.style={mylightblue},
    myColSev/.style={mydarkred}
  }
  \ifnum\externalizetikz=1
    \usepgfplotslibrary{external}
  \fi
\fi

\newcommand{\datfile}[1]{dat/#1.dat}

\newcommand\lseq[1]{\stackrel{{\rm LS}}{=}}
\newcommand*{\vertbar}{\rule[-0.2 ex]{0.5pt}{2.0ex}}
\newcommand{\ritz}{\lambdat}
\newcommand{\res}{r}
\newcommand{\resritz}{r}
\newcommand{\Dt}{\Delta t}
\newcommand{\Ko}{\mathcal{K}_{\Delta t}}

\input figures_commands.tex

\usepackage[colorlinks,urlcolor=blue,linkcolor=black,citecolor=blue,breaklinks=true]{hyperref}
\usepackage[capitalize,nameinlink]{cleveref}

\begin{document}
\title{Estimating Eigenenergies from Quantum Dynamics: A Unified Noise-Resilient Measurement-Driven Approach}

\author{Yizhi~Shen}
\affiliation{Applied Mathematics and Computational Research Division, Lawrence Berkeley National Laboratory, Berkeley, CA 94720, USA}
\email{yizhis@lbl.gov}
\author{Daan~Camps}
\affiliation{National Energy Research Scientific Computing Center, Lawrence Berkeley National Laboratory, Berkeley, CA 94720, USA}
\author{Aaron~Szasz}
\affiliation{Applied Mathematics and Computational Research Division, Lawrence Berkeley National Laboratory, Berkeley, CA 94720, USA}
\author{Siva~Darbha}
\affiliation{Applied Mathematics and Computational Research Division, Lawrence Berkeley National Laboratory, Berkeley, CA 94720, USA}
\affiliation{National Energy Research Scientific Computing Center, Lawrence Berkeley National Laboratory, Berkeley, CA 94720, USA}
\author{Katherine~Klymko}
\affiliation{National Energy Research Scientific Computing Center, Lawrence Berkeley National Laboratory, Berkeley, CA 94720, USA}
\author{David~B.~Williams--Young}
\affiliation{Applied Mathematics and Computational Research Division, Lawrence Berkeley National Laboratory, Berkeley, CA 94720, USA}
\author{Norm~M.~Tubman}
\affiliation{NASA Ames Research Center, Moffett Field, CA 94035, USA}
\author{Roel~{Van~Beeumen}}
\affiliation{Applied Mathematics and Computational Research Division, Lawrence Berkeley National Laboratory, Berkeley, CA 94720, USA}
\maketitle

\begin{abstract}
Ground state energy estimation in physical, chemical, and materials sciences is one of the most promising applications of quantum computing. In this work, we introduce a new hybrid approach that finds the eigenenergies by collecting real-time measurements and post-processing them using the machinery of dynamic mode decomposition (DMD). From the perspective of quantum dynamics, we establish that our approach can be formally understood as a stable variational method on the function space of observables available from a quantum many-body system. We also provide strong theoretical and numerical evidence that our method converges rapidly even in the presence of a large degree of perturbative noise, and show that the method bears an isomorphism to robust matrix factorization methods developed independently across various scientific communities. Our numerical benchmarks on spin and molecular systems demonstrate an accelerated convergence and a favorable resource reduction over state-of-the-art algorithms. The DMD-centric strategy can systematically mitigate noise and stands out as a leading hybrid quantum-classical eigensolver.
\end{abstract}



\section{\label{sec:intro}Introduction}

In domains ranging from fundamental physics to practical chemistry, the characterization of many-body quantum systems requires the calculation of low-energy states, especially ground states. By identifying ground states of lattice models describing crystalline materials, we can look for novel phases of matter such as quantum spin liquids~\cite{Balents2010, Zhou2017} and topological insulators~\cite{Moore2010,Kane2010}, or even understand the mechanisms driving high-temperature superconductivity~\cite{Bednorz1986, Kamihara2008, Keimer2015}. By finding the low-lying eigenstates of molecules, we can uncover reaction pathways and design optimized processes in various sectors from fertilizer production~\cite{nitrogen_fixation_2017} to drug discovery~\cite{Ajagekar2023}.

For strongly interacting many-body systems, quantum computers provide arguably the best tool for finding ground states. Naive classical approaches based on exact diagonalization of the many-body Hamiltonian scale exponentially in the number of quantum spins and/or electronic orbitals, while more sophisticated approaches such as tensor network simulations~\cite{SCHOLLWOCK2011, ORUS2014117} are inefficient when describing highly-entangled or high-dimensional systems. Simulation using a quantum computer can overcome these challenges, scaling polynomially in the number of spins or orbitals and faithfully representing states with any degree of entanglement. 

With a fully fault-tolerant, error-corrected quantum computer, accurate ground state energies can be determined via the well-established 
quantum phase estimation (QPE)~\cite{qpe,QPE_prl_1997,qpe_prl_1999} algorithm. QPE is theoretically optimal: it saturates the Heisenberg limit given a single copy of the input reference state, taking only $\mathcal{O}(\epsilon^{-1})$ computational time to ascertain precision $\epsilon$. However, QPE requires the use of $\mathcal{O}(\log(\epsilon^{-1}))$ ancilla qubits, making this algorithmic primitive infeasible without full fault tolerance. 

In contrast, hybrid quantum-classical algorithms may bring a quantum advantage even on near-term devices.  Hybrid strategies deploy a quantum computer for the tasks where it excels, for example evolving a wavefunction with unitary operators, and off-load other parts of a simulation to a classical computer. Notably, the variational quantum eigensolver (VQE)~\cite{McClean_2016,vqe_prl_2019,cerezo2021variational,Nicholas_2021} has emerged as a popular approach for current quantum hardware. 
However, the VQE method often relies on the heuristic selection of circuit ansatz, and can be prone to serious optimization challenges including vanishing gradients~\cite{mcclean2018barren,Wang2021,vqe_prl_2021}. As such, the potential computational advantages of VQE are not clearly understood. 

A promising alternative family of algorithms only requires access to measurements of state overlaps using a single ancilla qubit. 
These algorithms are known as early fault-tolerant because they allocate fewer qubits than full phase estimation but necessitate multiple copies of the input reference state. State-of-the-art examples within this category include variational quantum phase estimation (VQPE)~\cite{klymko2022real,shen2022real}, quantum complex exponential least-squares (QCELS)~\cite{ding2022even,ding2023simultaneous}, and quantum Lanczos~\cite{kirby2023exact}, among others. Although these algorithms have promising potential to achieve Heisenberg scaling before full fault tolerance, they also come with their own limitations. For example, some algorithms require a substantial number of shots to reduce the statistical errors. Additionally, due to the prohibitive cost of state orthogonalization, recent attention has focused on algorithms with non-orthogonal bases \cite{huggins2020non,novqe_2023}, resulting in ill-conditioned generalized eigenvalue problems. This difficulty is exacerbated by the presence of noise sources, even when accounting solely for the statistical noise from sampling on any practical hardware. Other algorithms build on the premise of a good input state that has significant overlap with the low energy eigenstates, which imposes serious extra resource overhead for initial state preparation for some problems. 

In this paper, we introduce a new hybrid algorithm for finding ground state energies and other ground state properties based on the machinery of dynamic mode decomposition (DMD) \cite{Mezi2004ComparisonOS,Mezi2005SpectralPO,Rowley2009SpectralAO,schmid_2010,Mezi2013AnalysisOF}.  
DMD processes snapshots of a dynamical system to identify the principal oscillating modes and constructs an efficient representation of the full dynamical trajectory. To apply DMD to a quantum system, where we do not have access to the full state vector, we use a vector of time-delayed measured observables as a proxy. In particular, the oscillating modes of a quantum system are its energy eigenstates: if we use expectation values of the time evolution operator as the observables, we expect DMD to return the ground state energy of the system.  We name this method observable dynamic mode decomposition (ODMD).
By fully exploiting real-time data from quantum hardware, our approach introduces a unified eigensolver framework that preserves the salient features of different classes of approaches, including variational convergence, compactness, and noise resilience.

Similar to state-of-the-art approaches such as VQPE and QCELS, ODMD only requires matrix elements of the time evolution operator, computed via state overlaps, and at most one ancilla qubit to find the ground state energy.  However, ODMD can overcome some limitations of those approaches and other hybrid algorithms. On the one hand, in contrast to subspace projection techniques such as VQPE, the classical portion of our algorithm involves a least-squares (LS) problem rather than a generalized eigenvalue problem, improving conditioning and stability against perturbative noise. More importantly, ODMD can be formulated as a functional subspace method applied to the space of observables in our setting. This connection can be established via Koopman operator analysis~\cite{Koopman1931HamiltonianSA,Koopman1932,BruntonKutz2015,BruntonKutz2016,Arabi2017}, which allows us to further generalize ODMD to a multi-observable framework for far more efficient excited state computation.

On the other hand, while ODMD shares the frequency retrieval perspective underlying QCELS and its variants, ODMD distinguishes from these recent signal processing algorithms in the following crucial aspects. First, ODMD does not hinge on the requirement that the reference state must have a dominant overlap with specific eigenstates of interest. For other signal processing algorithms targeting ground state energy, this dominance condition translates into a restricted ground state overlap greater than $\frac{1}{2}$~\cite{ding2022even,ding2023simultaneous,Ding2024quantummultiple}, which imposes an extra quantum cost for state preparation. Second, the classical cost for ODMD scales polynomially with the number of target eigenenergies due to its matrix-based post-processing. This is exponentially more favorable than that of the nonlinear approaches~\cite{ding2022even,ding2023simultaneous}. We note that although the sparse Fourier approach of~\cite{Ding2024quantummultiple} also exponentially and moreover optimally reduces such cost, it fails to provide accurate energy estimates when the reference state evolves inexactly, for example in the presence of a common depolarizing noise channel. This is because damping or any deviation from a purely sinusoidal time signal can severely limit the resolution of frequencies achievable through standard Fourier recovery. Third, ODMD reveals information about the eigenstates, whereas alternative signal processing approaches recover only the eigenenergies.

Other benefits of ODMD include the ability to use only the real parts of the overlap matrix elements, therefore reducing the number of shots by half. Furthermore, our method progressively updates the ground state estimate (or the target eigenstate estimates in general) with the addition of each new matrix element, so resources can be utilized efficiently to achieve the desired level of precision.

Our manuscript is organized as follows: In \cref{sec:background}, we give an overview of existing real-time evolution-based approaches to finding ground state energy, and we discuss their relative strengths and weaknesses. We introduce in \cref{sec:odmd} a new measurement-driven framework, which we name the observable dynamic mode decomposition (ODMD), starting from the machinery of classic dynamic mode decomposition (DMD). A rigorous examination of the ODMD convergence is found in \cref{sec:convergence}. 
For readers interested in the theoretical underpinnings of ODMD, in \cref{sec:connections} we show how the method can be reinterpreted in terms of subspace approaches in function space and in terms of matrix factorization methods from the applied mathematics literature; readers who are primarily interested in ODMD as a practical tool could skip this technical section on a first read. In \cref{sec:numerics}, we numerically demonstrate the efficacy of our approach for ground and excited state problems using representative spin and molecular many-body systems. \cref{sec:diss} provides a discussion and concluding remarks.


\section{Background and landscape of quantum methods}
\label{sec:background}

There has been a growing interest in leveraging real-time methods for finding ground state energies~\cite{parrish2019quantum, stair2020multireference, klymko2022real, cortes2022quantum, StairCortes2023}. 
The primary appeal of these methods comes from the unitarity of the time evolution operator $e^{-i H t}$. Unitarity makes real-time based approaches naturally compatible with quantum hardware, which accordingly translates to shorter-depth circuits compared to other, naturally non-unitary methods such as the power method~\cite{powermethod}, Lanczos method~\cite{kirby2023exact}, or imaginary-time implementations~\cite{motta2020determining}.

\subsection{Real-time evolution-based algorithms for ground state estimation}
\label{subsec:real_time}

Recent studies have proposed subspace methods~\cite{parrish2019quantum, stair2020multireference, klymko2022real, cortes2022quantum, shen2022real}, which project the many-body Schr\"{o}dinger equation onto a subspace of the Hilbert space constructed through real-time evolution.
These approaches yield generalized eigenvalue equations built from non-orthogonal real-time basis states. 
The subspace Hamiltonian and overlap matrix elements,
\begin{align}
    \textbf{H}_{ij} &= \braket{\boldsymbol{\phi}_i|H|\boldsymbol{\phi}_j} = \braket{\boldsymbol{\phi}_0|e^{i H t_i} H e^{-iH t_j}|\boldsymbol{\phi}_0},\label{eq:VQPE_H}\\
    \textbf{S}_{ij} &= \braket{\boldsymbol{\phi}_i|\boldsymbol{\phi}_j} = \braket{\boldsymbol{\phi}_0|e^{iH(t_i-t_j)} |\boldsymbol{\phi}_0}, \label{eq:VQPE_S}
\end{align}
are obtained via measurements performed on quantum hardware and the resulting projected eigenvalue problem, $\textbf{H} \boldsymbol{\Psi} = \Tilde{E} \textbf{S} \boldsymbol{\Psi}$, is solved classically for approximate energies $\Tilde{E}$. For example, quantum filter diagonalization (QFD) and the VQPE~\cite{klymko2022real} method have this form with a uniform timestep grid $t_j^{ } = j\Delta t$.
The multireference selected quantum
Krylov (MRSQK) algorithm~\cite{StairCortes2023} is a variant where time evolution is applied to a set of initial states in order to reduce the condition number of the resulting  generalized eigenvalue equation. As suggested by recent work~\cite{stair2020multireference, klymko2022real, shen2022real}, a relatively small number of real-time states $\{\ket{\boldsymbol{\phi}_i}\}_i$ suffice to achieve accurate estimates for the low-lying energies in many physically relevant cases, making these methods relevant for near-term calculations on quantum hardware.

A principal challenge in subspace methods is the classical solution of the generalized eigenvalue equation. 
Given that the matrix elements are measured on quantum hardware, they are inevitably subject to noise, which can become quite significant in practice. 
Due to the non-orthogonality of the real-time basis states, the noise leads to a highly ill-conditioned problem. 
One common approach to address ill-conditioned eigenvalue equations is to regularize or precondition the subspace ${\rm span} \{\ket{\boldsymbol{\phi}_i}\}_i$. For example, this can be done by pruning small singular values of the overlap matrix $\mathbf{S}$ beneath a noise-consistent threshold. However, such techniques are only viable up to a specific limit of the subspace dimension. Beyond this limit, noise accumulation in the matrix elements adversely affects the accuracy of the eigenvalue estimate as more real-time states are introduced~\cite{epperly2022theory}.

In addition to explicit subspace diagonalization, an emerging class of approaches tackles the eigenvalue problem via a sparse Fourier perspective \cite{lee2023quantum}. For example, QCELS \cite{ding2022even} derives its inspiration from real-time signal processing techniques by singling out the dominant eigenphases through the optimization,
\begin{align}
    \tilde{E} = \argmin_{\theta \in \mathbb{R} } \left\{  \min_{r \in \mathbb{C}} \sum_{i} \big\lvert \braket{\boldsymbol{\phi}_0 | e^{-i H t_i}|\boldsymbol{\phi}_0 } - re^{-i\theta t_i} \big\rvert^2 \right\},
\end{align}
where a classical minimization routine is called to search for the best amplitude and phase parameters $(r,\theta)$ fitting the quantum data. The fitted energy $\Tilde{E}$ converges rapidly to the ground state energy if the chosen initial state $\ket{\boldsymbol{\phi}_0}$ is well-aligned with the exact many-body ground state. 

Indeed, the effectiveness of many real-time approaches thus far has been mostly constrained to situations where an optimal starting or reference state $\ket{\boldsymbol{\phi}_0}$, which has a considerable overlap with the desired eigenstates, is achievable. In such scenarios, short time evolution suffices for convergence, and a basic truncation strategy can effectively remove perturbative noise and mitigate issues of ill-conditioning. However, creating starting states with substantial eigenstate overlap is difficult in general, especially for increasingly large and strongly correlated many-body systems.

\subsection{Computing overlaps}

The real-time methods described in \cref{subsec:real_time} all involve calculating expectation values of the form $\braket{\boldsymbol{\phi} | e^{-i H t} | \boldsymbol{\phi} }$, where $\ket{\boldsymbol{\phi}}$ is some input state of interest and the expectation specifies the overlap between the input and time-evolved states. In particular, the methods do not require direct extraction of time-evolved quantum states in the high-dimensional Hilbert space, which is intractable for large systems owing to an exponential cost. In contrast, since the time evolution operator is often expressible on quantum hardware, the expectation value can be efficiently evaluated. 

\input figure_hadamard.tex

The time-evolved overlaps have the more general form $\braket{\boldsymbol{\phi}|U|\boldsymbol{\phi}}$, where $U$ is a unitary operator. 
Such expectation values can be accessed using projective quantum measurements. In particular, they can be measured in the Pauli basis through the Hadamard test \cite{Cleve:1997dh} or mirror-type fidelity estimation \cite{mirror2019,cortes2022quantum}. The Hadamard test, illustrated in \cref{fig:hadamard_test}, evaluates the expectation using a single ancilla and controlled time evolution, $\ket{0}\bra{0} \otimes I + \ket{1}\bra{1} \otimes e^{-i H t}$. We measure the ancilla in the Pauli basis and record the measurement result as either $+1$ or $-1$. We perform repeated measurements and average over the $\{ \pm 1 \}$ results to converge the expectation. When the ancilla measurements are performed in the Pauli $X$ and $Y$ bases, we acquire the real and imaginary parts of the complex expectation, respectively. The mirror circuits circumvent the need for controlled evolution, which can sometimes be expensive to implement on current hardware. Instead, they employ controlled state preparation followed by unrestricted time evolution.

The real-time framework we present in \cref{sec:odmd} relies on acquiring \textit{system observables}, which we define to be scalar-valued (either real or complex) functions over the system state space. With a judicious selection of observables, we can effectively compress high-dimensional state information into lower-dimensional observable vectors, and therefore encode some of the spectral features of the underlying many-body system. The main system observables that we evaluate in our framework are the expectations $\braket{\boldsymbol{\phi} | e^{-i H t} | \boldsymbol{\phi} }$.


\section{ODMD method}
\label{sec:odmd}

As we design methods tailored to accommodate noisy quantum data, we now look beyond directly solving ill-conditioned generalized eigenvalue problems. Instead, we introduce a new optimization-free algorithm that solves a least-squares problem to achieve a controllable stability against noise. We show that the real-time observables $\braket{\boldsymbol{\phi} | e^{-i H t} | \boldsymbol{\phi} }$ give natural probes of spectral and dynamical information of a many-body quantum system. Building upon the classical machinery of DMD, we develop a noise-resilient measurement-driven framework for estimating eigenenergies from quantum dynamics.

\subsection{DMD to ODMD}
The standard dynamic mode decomposition (DMD), originally developed within the field of numerical fluid dynamics, is a measurement-driven approximation for the temporal progression of a classical dynamical system \cite{Mezi2004ComparisonOS,Mezi2005SpectralPO,Rowley2009SpectralAO,schmid_2010,Mezi2013AnalysisOF}. Specifically, DMD samples snapshots of the system at some regular time intervals $\Delta t$ and exploits them to construct an efficient representation of the full dynamical trajectory. For simplicity, we consider a system whose $N$-dimensional state manifold is $\mathbb{C}^N$. 
The optimal linear approximation for the discretized time step $k \mapsto k+1$ is expressed as the least-squares (LS) relation,
\begin{equation}
    \boldsymbol{\phi}_{k+1} \lseq{} A_k  \boldsymbol{\phi}_k , 
\end{equation}
where $\boldsymbol{\phi}_k \in \mathbb{C}^{N}$ specifies the system state at time $k\Dt$ and $A_k \in \mathbb{C}^{N \times N}$ gives the system matrix, \textit{i.e.}, the linear operator that minimizes the residual $\Vert \boldsymbol{\phi}_{k+1} - A_k \boldsymbol{\phi}_k \Vert_2$ to yield the LS relation above. Similarly, the optimal linear approximation for a sequence of successive snapshots $k = 0, 1, \ldots, K+1$ can be determined by the solution,
\begin{align}
     \begin{bmatrix}
        \vertbar & \vertbar & & \vertbar\\
        \boldsymbol{\phi}_{1} & \boldsymbol{\phi}_{2} & \cdots & \boldsymbol{\phi}_{K+1} \\
        \vertbar & \vertbar & & \vertbar\\
    \end{bmatrix} \lseq{} A \begin{bmatrix}
        \vertbar & \vertbar & & \vertbar\\
        \boldsymbol{\phi}_{0} & \boldsymbol{\phi}_{1} & \cdots & \boldsymbol{\phi}_{K} \\
        \vertbar & \vertbar & & \vertbar\\
    \end{bmatrix},
    \label{eq:dmd_lls}
\end{align}
where the system matrix $A$ minimizes the sum of squared residuals over the sequence. The optimal linear flow described by this matrix equation naturally generates approximate dynamics $\boldsymbol{\phi}_{\rm DMD}(t) = A^{\frac{t}{\Delta t}} \boldsymbol{\phi}_0$ governed by the eigenmodes of $A$. DMD-based approaches can be remarkably effective despite their formal simplicity, since they are rooted in the general Koopman operator theory developed to describe the behavior of general (non)linear dynamical systems \cite{Koopman1931HamiltonianSA,Koopman1932,BruntonKutz2015,BruntonKutz2016,Arabi2017}.

\begin{figure}[t!]
    \centering
    \resizebox{0.485\textwidth}{!}{\input{figure_quantum_vs_classical.tex}}
    \caption{Takens' embedding of scalar observables in the quantum context. The left panel depicts the time evolution of a reference quantum state $\ket{\boldsymbol{\phi}_0}$ in the many-body Hilbert space. Time-evolved states $\ket{\boldsymbol{\phi}_k} = \ket{\boldsymbol{\phi}(t_k)}$ are shown at a sequence of time steps $t_k = k\Dt$. The right panel tracks the corresponding observables $o(t)$ at the same sequence of time steps, which can be evaluated efficiently on a quantum computer. Takens' embedding establishes a connection between the state evolution and time-delayed observables.}
    \label{fig:dmd}
\end{figure}

The standard DMD approach described above for classical dynamics cannot be immediately translated to quantum dynamics. The DMD approximation of the system evolution requires complete knowledge of the state, as specified by an $N$-dimensional complex vector at each time step.  However, we do not have direct access to the full many-body quantum state. Instead, we can only access the state of a quantum system via measurement sampling of observables~\footnote{While ``observable'' is typically used in quantum mechanics to refer specifically to Hermitian operators (with real expectation value), here we use a broader definition, encompassing also complex scalar quantities that can be computed from measurements on a quantum computer.}. To address this challenge, we employ a technique motivated by Takens' embedding theorem \cite{takens1971,takens1981,Gutierrez:2021wik} to obtain an effective state vector consisting of an operator measured at a sequence of successive times.  We reformulate the linear model underpinning DMD in terms of these observable-vectors to approximate the system dynamics. (An alternative approach involves forming a state vector from multiple distinct operators~\cite{Niladri_DMD}.)

Takens' embedding theorem \cite{takens1971,takens1981} establishes a connection between the manifold of states, which an observer cannot directly access, and time-delayed measurements of an observable.
In particular, the theorem asserts, under generous conditions, that a state on an $N$-dimensional (sub)manifold can be completely determined using a sequence of at most $d_{\star} \leq 2N + 1$ time-delayed observables. The correspondence reads
\begin{align}
    \boldsymbol{\phi}(t) \leftrightarrow
     \boldsymbol{o}_{t,d_\star} = \begin{bmatrix}
   o(t) \\
    o(t+\Delta\tau) \\
     \vdots \\
    o(t+(d_*-1)\Delta\tau)
    \end{bmatrix},
    \label{eq:Takens}
\end{align}
where $\Delta\tau > 0$ is the time delay, $\boldsymbol{\phi}(t)$ is the system state, $o(t) = o[\boldsymbol{\phi}(t)]$ is the measured observable, and $\boldsymbol{o}_{t}$ is the $d_{\star}$-dimensional ``observable trajectory'' containing the dynamical information. 
The RHS of \cref{eq:Takens} is known as a $d_{\star}$-dimensional delayed embedding of the observable. 
Takens' theorem relates the evolution of microscopic degrees of freedom to the evolution history of macroscopic observables, providing a concrete probe into the dynamical properties of the system without direct access to the full states. 
Here we adopt the term Takens' embedding technique to refer to the method of applying time delays on the system observables, motivated by the rigorous results of Takens' embedding theorem. This adaptation to the quantum realm is illustrated in \cref{fig:dmd}.

\input figure_overview.tex

In anticipation of efficiently leveraging near-term quantum resources in the hybrid quantum-classical approach that we present in \cref{subsec:gs_from_odmd}, we choose the time delay in Takens' embedding technique to equal the DMD time interval, \textit{i.e.}, $\Delta\tau = \Delta t$.
Given this choice, we then measure the system along time steps $\{t_k = k\Delta t\}_{k=0}^{K+1}$ and acquire the sequence of observable trajectories $\{ \boldsymbol{o}_{t_k,d}\}_{k=0}^{K+1}$, each of some length $d \leq d_{\star}$,
\begin{align}
     \boldsymbol{o}_{t_k,d} &= \begin{bmatrix}
         o(t_k) \\
         o(t_{k+1}) \\
         \vdots \\
         o(t_{k+d-1})\\
    \end{bmatrix}, & 0 &\leq k \leq K+1.
\end{align}
By construction, the first $(d-1)$ entries of $\bfo_{t_k,d}$ are identical to the last $(d-1)$ entries of $\bfo_{t_{k-1},d}$. Consequently, the matrix assembled by arranging successive trajectories $ \boldsymbol{o}_{t_k}$ as columns
\begin{equation}
    \mathbf{X}_{k_1:k_2} = \begin{bmatrix}  \boldsymbol{o}_{t_{k_1},d} &  \boldsymbol{o}_{t_{k_1+1},d} & \cdots &  \boldsymbol{o}_{t_{k_2},d} \end{bmatrix},
\end{equation}
has a Hankel form, \textit{i.e.}, the matrix elements on each anti-diagonal are equal (see \cref{fig:overview}). In the embedding space, we can then identify the closest linear flow,
\begin{equation}
    \mathbf{X}_{1:K+1} \lseq{} A \mathbf{X}_{0:K} \implies A = \mathbf{X}_{1:K+1} (\mathbf{X}_{0:K})^{+},
    \label{eq:LS_ODMD}
\end{equation}
where $(\hspace{0.06cm}\cdot \hspace{0.06cm})^+$ denotes the Moore–Penrose pseudo-inverse. 
The system matrix $A$ assumes a companion structure with just $d$ free parameters (see \cref{eq:prony_recur} in \cref{app:sec:prony}). The approximation to the system dynamics is then stored in $d$ parameters inferred from measurements of a total of $K+d+1$ delayed observables. We hence name our least-squares embedding in the observable space the \textit{observable dynamic mode decomposition} (ODMD).

\subsection{Estimating the ground state energy}
\label{subsec:gs_from_odmd}

We now specialize the ODMD method in pursuit of a robust eigenenergy estimator for quantum systems.  With Hamiltonian $H$, the system dynamics are given by the time evolution operator $e^{-iHt}$ (with the convention of $\hbar=1$).  Let $E_0 \leq E_1 \leq \cdots \leq E_{N-1}$ denote the eigenenergies of the Hamiltonian. We consider as our observable the complex-valued overlap,
\begin{equation}
   s(k\Dt) = \bra{\boldsymbol{\phi}_0} e^{-iH k \Dt}\ket{\boldsymbol{\phi}_0}, \quad k=0,1,\ldots,
\label{eq:overlap}
\end{equation}
whose real and imaginary parts can be separately measured, \textit{e.g.}, using the Hadamard test \cite{Cleve:1997dh}. For an integer pair $(d,K)$ specifying our time-delayed embedding, this requires $K+d+1$ observables in total, each corresponding to a distinct quantum evolution circuit.

Upon arranging these measured overlaps into a pair of time-shifted Hankel matrices $\mathbf{X}, \mathbf{X}' \inC[d][(K+1)]$,
\begin{align}
    \mathbf{X}  &= \mathbf{X}_{0:K}, &
    \mathbf{X}' &= \mathbf{X}_{1:K+1},
\label{eq:hankel}
\end{align}
the eigenenergies of the Hamiltonian can then be simply estimated by solving the standard eigenvalue problem,
\begin{align}
    A \boldsymbol{\Psi}_{\ell} = \tilde{\lambda}_{\ell} \boldsymbol{\Psi}_{\ell},
    \label{eq:dmd_eigval}
\end{align}
where the matrix $A$ is defined in \cref{eq:LS_ODMD}. Intuitively, we can read off our eigenenergy estimates $\Tilde{E}_{\ell}$ from the ordering of the phases ${\rm arg}(\lambda_{\ell})$. The eigenvalue $\Tilde{\lambda}_{0} \approx e^{-i E_0 \Delta t}$ attaining the maximal phase, ${\rm arg}(\tilde{\lambda}_0) = \max_{\ell} { {\rm arg}(\tilde{\lambda}_{\ell})} = -  \min_{\ell} \Tilde{E}_{\ell} \Dt$, encodes the DMD approximation $\Tilde{E}_0$ to the true ground state energy $E_0$. For concreteness, we construct our $d \times (K+1)$ data matrices using $d = \lfloor \alpha (K+1) \rfloor$ for $\alpha=\frac{1}{2}$ fixed throughout the main text. Schematically, an overview of the ODMD approach for the ground state problem is provided in \cref{fig:overview}.

To protect against both statistical noise and quantum hardware noise, we exploit the least-squares formulation given in \cref{eq:LS_ODMD}, benefiting from its rigorous and robust regularization capabilities. Since the recovery of the system matrix $A$ requires computing the pseudo-inverse of the data matrix $\mathbf{X}$, its conditioning critically influences the stability of the solution. Here we use the singular value decomposition of $\mathbf{X}$,
\begin{align}
    \mathbf{X} = \sum_{\ell = 0}^{d-1} \sigma_\ell \boldsymbol{u}_\ell \boldsymbol{v}_\ell^{\dagger},
\label{eq:svd_x}
\end{align}
where $\sigma_{\ell}>0$ are the singular values, and $(\boldsymbol{u}_{\ell},\boldsymbol{v}_{\ell})$ specify the associated left and right singular vectors, respectively. To ensure noise-resilience, we filter out perturbative error using a thresholding procedure: we regularize by truncating singular values smaller than the cutoff, $\tilde{\delta} \sigma_{\rm max}(\mathbf{X})$, with a relative threshold $\tilde{\delta} > 0$ of our choice,
\begin{align}
    \mathbf{X} \mapsto \mathbf{X}_{\Tilde{\delta}} = \sum_{\ell: \sigma_{\ell} > \Tilde{\delta}\sigma_{\rm max}} \sigma_{\ell} \boldsymbol{u}_{\ell} \boldsymbol{v}_{\ell}^{\dagger},\label{eq:reg}
\end{align}
where $\sigma_{\rm max} = \max_{\ell} \sigma_{\ell}$ denotes the largest singular value of the data matrix $\mathbf{X}$. Incorporating the LS regularization, our measurement-driven estimation of the ground state energy is outlined in \cref{alg:hodmd}. Our algorithm takes as input the time step $\Delta t$ and noise threshold $\Tilde{\delta}$.

\begin{algorithm2e}[hbtp]
\caption{ODMD ground state energy estim.\hspace{-2em}~\label{alg:hodmd}}
\vspace{5pt}
\KwIn{Time step $\Delta t$, noise threshold $\tilde{\delta}$.}
\KwOut{Estimated ground state energy $\tilde{E}_0$.}
\vspace{5pt}
\nl $k \leftarrow 0$ \\
\While{${\rm not~converged}$}{
  \nl $s_k \leftarrow \Re s(k \Delta t)$ \Comment*{\footnotesize{quantum measurement}}
  \nl $\mathbf{X}, \mathbf{X}' \leftarrow \mathrm{Hankel}\left( s_0, s_1, \ldots, s_k \right)$ \Comment*{\footnotesize{Eq.~(\ref{eq:hankel})}}
   \nl $\mathbf{X}_{\tilde{\delta}} \leftarrow \mathbf{X}$ \Comment*{\footnotesize{least-squares regularization}}
  \nl $A \leftarrow \mathbf{X}' \mathbf{X}_{\tilde{\delta}}^{+}$ \Comment*{\footnotesize{update system matrix}}
  \nl $\displaystyle \tilde{E}_0 \Delta t \leftarrow  - \max_{1 \leq \ell \leq d_k} \Im \log(\tilde{\lambda}_{\ell})$ \Comment*{\footnotesize{update energy}}
  \nl $k \leftarrow k+1$
}
\end{algorithm2e}

The choice of the time step $\Dt$ is crucial for the optimal performance of the ODMD algorithm. Small time steps can hinder convergence due to the collection of excessively correlated information. On the other hand, there is a fundamental limitation on the maximum size of $\Delta t$, \emph{i.e.}, when we extract $\tilde{\lambda}_0\approx e^{-i E_0\Delta t}$ from ODMD, we can only solve for the leading eigenangle $E_0^{ }\Delta t$ mod $2\pi$. Thus to extract $E_0$ unambiguously, the angles $\{E_{n}^{ }\Delta t\}_{n=0}^{N-1}$ can wrap around the circle only once, which requires $(E_{N-1}^{ }-E_0^{ })\Delta t < 2\pi$, and we must also identify the location of the dividing line between upper and lower parts of the eigenangles. We show two illustrative examples in \cref{fig:no_aliasing}.  

\begin{figure}[t!]
    \centering
    \resizebox{0.485\textwidth}{!}{\input{figure_aliasing.tex}}
    \caption{Guideline to select $\Delta t$ and unambiguously extract $\tilde{E}_0$ from the ODMD eigenvalue $\tilde{\lambda}_0$, depending on known bounds on the eigenenergy spectrum, $E_\text{min}$ and $E_\text{max}$.  The region shaded in gray shows where the choice of $\Delta t$ ensures there are no eigenphases $\lambda_{n}$ and separates the top and bottom of the eigenangles $E_n \Dt$.  \textbf{Left:} When the spectral bounds are symmetric around 0, \textit{i.e.}, $\norm{H}_2^{ } \leq E_\text{max} = |E_\text{min}|$, we take $\Delta t \lesssim \pi/E_{\rm max}$ and the dividing boundary between the upper and lower parts of the eigenangles sits at $\pi$. \textbf{Right:} When the whole energy spectrum is positive with asymmetric spectral bounds $0 = E_{\rm min} < \norm{H}_2^{ } \leq E_{\rm max}$, taking $\Delta t \lesssim 2\pi/E_\text{max}$ places the dividing eigenangle boundary at $0^+$.}  
    \label{fig:no_aliasing}
\end{figure}

Note that the exact largest and smallest eigenenergies do not need to be known in advance.  Rather, if we know that $E_\text{min}^{ } \leq \langle \boldsymbol{\phi}| H| \boldsymbol{\phi} \rangle \leq E_\text{max}^{ }$ for any normalized state $\ket{\boldsymbol{\phi}}$, we can guarantee that the phase condition is satisfied by choosing $\Delta t$ so that $(E_\text{max}^{ } - E_\text{min}^{ })\Delta t < 2\pi$. Moreover, $\Delta t$ must also satisfy a further compatibility condition: as derived in \cref{app:sec:prony}, 
\begin{align}
    \Delta t < \frac{2\pi}{(E_{N-1} - E_0) + (E_1 - E_0)}\label{eq:timestep_choice}
\end{align}
in order to ensure the convergence of the ODMD estimate to the ground state energy.  Notably, the restriction on $\Delta t$ for unambiguous determination of $E_0^{ }$ from $\lambda_0^{ }$ actually follows directly from \eqref{eq:timestep_choice}. 

In practice, a good choice is to first bound the spectral range of the Hamiltonian and then linearly shift the range to be in $[-C\pi,C\pi]$, for some constant $C < 1$ close to but less than 1; the lower right phase sketch in \cref{fig:overview} shows a restriction with $C=3/4$. The time step $\Delta t$ can then be set to 1, giving rapid convergence of the algorithm while satisfying the condition of \cref{eq:timestep_choice}.  Alternatively, if the Hamiltonian has spectral norm $\norm{H}_2$, we can simply use $\Delta t < \pi/\norm{H}_2$.  Either way, the spectrum of $H \Dt$ will be guaranteed to lie in the window $(-\pi,\pi)$ and there is no ambiguity in estimating the eigenenergy $E_0$.

Finally, we leave an important remark that measuring only the real or imaginary part of the overlap, $\Re s(k\Dt)$ or $\Im s(k \Dt)$, leads to a real-valued system matrix $A$. Accordingly, the eigenvalues of $A$ contain pairs of complex conjugates. These ODMD eigenvalues, however, preserve the ground state information due to Euler's identity, for example, $\cos{(E_0 \Delta t)} = \frac{1}{2} (e^{-iE_0 \Dt} + e^{iE_0 \Dt})$. The eigenenergies can be explicitly approximated as the phases of the ODMD eigenvalues just as in the complex case. Because of the phase symmetrization $E_\ell \mapsto \pm E_\ell$, what appears to be the ground state energy could instead come from the symmetrization of the maximum energy eigenvalue, if $|E_{N-1}| > |E_0|$.  Thus to ensure that the maximal phase ODMD eigenvalue really corresponds to the ground state energy $E_0$, we need to shift the energy range of $H$ so that $|E_{N-1}| \leq |E_0|$.  This is guaranteed if the applied spectral shift satisfies, $E_0 <0$ and $|E_0| = \norm{H}_2$. Even accounting for this constant shift, the use of $\Re s(k\Dt)$ can give a reduction in the total measurement cost of up to 50\%.

\subsection{Accessing the ground state}
\label{subsec:eigenvector}

In addition to accessing the eigenenergies, the ODMD algorithm can also provide an estimate to the corresponding eigenstates. As a first step, let us recall that the companion system matrix $A$ is non-Hermitian, and it admits a spectral decomposition $A = \sum_{\ell} \Tilde{\lambda}_{\ell} \boldsymbol{\Psi}_{\ell} \boldsymbol{\Phi}_{\ell}$ with right and left eigenvectors, $\boldsymbol{\Psi}_{\ell}$ and $\boldsymbol{\Phi}_{\ell}$, respectively. The left eigenvector $\boldsymbol{\Phi}_{0} = [z_0, z_1, \ldots, z_{d-1}]$ corresponding to the eigenvalue $\Tilde{\lambda}_0$ satisfies
\begin{align}
    \boldsymbol{\Phi}_0 A = \Tilde{\lambda}_0 \boldsymbol{\Phi}_0 \implies \boldsymbol{\Phi}_0 \mathbf{X}' = \Tilde{\lambda}_0 \boldsymbol{\Phi}_0 \mathbf{X}.
    \label{eq:dmd_lefteigvec}
\end{align}
Focusing on observable snapshots from, for example, the first columns of our data matrices, \cref{eq:dmd_lefteigvec} implies
\begin{align}
   \boldsymbol{\Phi}_0 \begin{bmatrix}
   s(\Dt) \\
   s(2\Dt) \\
     \vdots \\
   s(d\Dt)
    \end{bmatrix} &= \Tilde{\lambda}_0  \boldsymbol{\Phi}_0 \begin{bmatrix}
   s(0) \\
   s(\Dt) \\
     \vdots \\
   s((d-1)\Dt)
    \end{bmatrix},
\end{align}
which can be expressed in terms of the eigenvector coordinates $z_{\ell}$ and real-time observables,
\begin{align}
    \begin{split}
        \bra{ \boldsymbol{\phi}_0} e^{-iH\Dt} & \left(\sum_{\ell=0}^{d-1} z_{\ell} e^{- i H \ell \Dt} \ket{\boldsymbol{\phi}_0} \right) \\
        &= \Tilde{\lambda}_0 \bra{ \boldsymbol{\phi}_0} \left(\sum_{\ell=0}^{d-1} z_{\ell} e^{- i H \ell \Dt}  \ket{\boldsymbol{\phi}_0} \right).
    \end{split}\label{eq:gs}
\end{align}
Since $\Tilde{\lambda}_0 \approx e^{-i E_0 \Dt}$, the dynamic mode above closely follows the ground state oscillation driven at the desired frequency, $\braket{\boldsymbol{\phi}_0| e^{-iH\Dt} |\psi_0} = e^{-iE_0 \Dt} \braket{ \boldsymbol{\phi}_0| \psi_0}$, where $\ket{\psi_0}$ is the actual ground state. By \cref{eq:gs}, we thus approximate the ground state as the weighted sum,
\begin{align}
    \ket{ \psi_0 } \approx \ket{\tilde{\psi}_0} = \frac{ \sum_{\ell=0}^{d-1} z_{\ell} e^{- i H \ell \Dt} \ket{\boldsymbol{\phi}_0}}{\norm{ \sum_{\ell=0}^{d-1} z_{\ell} e^{-i H \ell \Dt} \ket{\boldsymbol{\phi}_0}}_2},
    \label{eq:odmd_eigvec}
\end{align}
where the set of coefficients $z_\ell$ assign the relative weights to the real-time states $e^{-i H \ell \Dt} \ket{\boldsymbol{\phi}_0}$. For simplicity, we assume that the coefficients $z_{\ell}$ produce a normalized state so that the denominator of \cref{eq:odmd_eigvec} is unity. The ground state properties can hence be extracted in terms of the real-time states,
\begin{align}
    \braket{\psi_0|O|\psi_0} \approx \sum_{k, \ell=0}^{d-1} z_{k}^{\ast} z_{\ell} \braket{\boldsymbol{\phi}_0 | e^{i H k \Dt} O e^{-i H \ell \Dt} | \boldsymbol{\phi}_0},\label{eq:GS_properties}
\end{align}
for an arbitrary operator $O$. As the mode approximation typically admits small $d$ at convergence, we remark that ODMD enables us to extract a highly compact representation of the eigenstates through real-time evolution.

\subsection{Practical considerations}
\label{subsec:practical_performance}

For practical implementation, we discuss the feasibility of executing the ODMD algorithm on near-term devices, particularly when the time evolution of the initial state is inexact. Here we assess two sources of inexactness, the algorithmic error due to approximation(s) of the operator exponentials for time evolution and the device error from noisy quantum hardware.

For algorithmic error, we consider the Trotter-Suzuki product formula~\cite{Trotter} as one of the standard approaches for approximating the time evolution operator. Since our algorithm is based on time-delayed embedding, it suffices to accurately approximate the unit-timestep propagator $e^{-iH\Delta t}$. For a first-order Trotter decomposition, we have
\begin{align}
    e^{-i H \Delta t} \approx \left[ \prod_{\nu} e^{-i H_\nu \Delta t/M } \right]^{M},
\end{align}
where $H = \sum_{\nu} H_\nu$ gives a decomposition of the problem Hamiltonian into Pauli terms $H_{\nu} = c_\nu P_\nu$ (with $P_{\nu}$ being a Pauli string and $c_{\nu}$ a scaling coefficient satisfying $\lvert c_{\nu} \rvert = \lVert H_{\nu} \rVert_2$). In particular, the Trotter error $\epsilon_{\rm TS} = \mathcal{O} (\frac{\Delta t^2 \Gamma}{M})$ is controlled by a constant $\Gamma$ dependent on commutators of the terms $H_\nu$. Accordingly, the inexact evolution can be performed with circuit depth $\mathcal{O}( \frac{N_{H} \sqrt{M \Gamma} t_{\rm max}} {\sqrt{\epsilon_{\rm TS}}} )$, where $N_{H}$ is the number of Pauli terms in the Hamiltonian and $t_{\rm max}$ is the maximal evolution time.

Recall that the timestep $\Delta t$ in the ODMD algorithm is selected as $\Delta t = \mathcal{O}(\lVert H \rVert_2^{-1})$. For a typical many-body Hamiltonian, it suffices to set $M = \frac{1}{{\rm polylog}(N) \epsilon_{\rm TS}}$ in the Trotterized evolution to ensure an approximation error of $\mathcal{O}(\epsilon_{\rm TS})$, introducing a spectral perturbation of $\mathcal{O}(\epsilon_{\rm TS})$. With time steps $t_k = k \Dt$ taken at regular intervals, the total evolution time scales as $t_{\rm tot} = \mathcal{O}(t_{\rm max}^2)$ and the total gate count grows accordingly. In addition, the total shot count is proportional to
\begin{align}
    k_{\rm max} \times {\rm Measurement~cost~per~circuit} = \frac{t_{\rm max}}{\Dt \epsilon^2 },
\end{align}
where $\epsilon$ quantifies the sampling error due to finite shots per overlap measurement using the Hadamard test. The $\mathcal{O}(\epsilon^{-2})$ dependence reflects the standard measurement scaling, where mitigating statistical uncertainty requires quadratically more shots. Provided that the eigenenergy perturbation due to the Trotterized $\Dt$-evolution is sufficiently small, the ODMD estimates remain robust. As an example, choosing $t_{\rm max} = \mathcal{O}(\epsilon_{\rm TS}^{-\eta})$ with $\frac{2}{3} \leq \eta \leq 1$ and $\epsilon = \mathcal{O}(1)$ guarantees a final energy error of $\mathcal{O}(\epsilon_{\rm TS})$~\cite{esprit_2024}. Notably, the stable convergence of ODMD under inexact time evolution has been confirmed directly in recent numerical and experimental simulations of frustrated spin lattices~\cite{szasz2024groundstateenergymagnetization}.

For device error, we first analyze the elementary case of a depolarizing channel, perhaps the simplest and most ubiquitous error models. The global depolarizing channel acts as,
\begin{align}
    \Lambda_{t}[\rho;\zeta] = e^{ - \zeta \lvert t \rvert } \mathcal{U}_t \rho \mathcal{U}_t^{\dagger} +  \frac{1 - e^{ - \zeta \lvert t \rvert }}{2N} I_{2N},
\end{align}
where $\rho$ is any input state consisting of the ancilla and register qubits, $\mathcal{U}_t = \ket{0}\bra{0} \otimes I + \ket{1}\bra{1} \otimes e^{-iHt}$ denotes the controlled time evolution queried in a Hadamard test, and the positive parameter $\zeta > 0$ characterizes the noise rate. In the ODMD setup, we have $\rho = \ket{+}\bra{+} \otimes \ket{\boldsymbol{\phi}_0} \bra{\boldsymbol{\phi}_0}$ so that the overlap, \textit{e.g.}, $\Re \braket{\boldsymbol{\phi}_0|e^{-iHt}|\boldsymbol{\phi}_0}$ for $t>0$, under noisy time evolution becomes
\begin{align}
   {\rm Tr}\left[(X \otimes I) \Lambda_{t}[\rho]\right]
   = e^{-\zeta t} \Re \braket{\boldsymbol{\phi}_0|e^{-iHt}|\boldsymbol{\phi}_0}.
\end{align}
The ODMD observables under depolarizing noise exhibit exponential damping, which is reflected in the real part $\Re \log(\Tilde{\lambda}_{\ell})$ of the computed ODMD eigenphases. Because the real and imaginary parts of the ODMD eigenphases can be extracted separately, our energy estimate remains accurate despite such noise-induced spectral broadening. The generalization to arbitrary Markovian noise channels is explored in \cref{app:sec:noise}.


\section{Convergence}
\label{sec:convergence}

Alongside its algorithmic elegance, ODMD also converges rapidly. Here we present our main convergence analysis that demonstrates efficient estimation of the ground state energy via an algebraic argument. 

We first note that the overlap can always be expressed as a sum of complex sinusoids,
\begin{eqnarray}
    s_k = s(k \Delta t) = \sum_{n=0}^{N-1} p_n \lambda_n^{k},
\end{eqnarray}
where $N$ is the size of Hilbert space, $\lambda_n = e^{-iE_n \Delta t}$ are the eigenphases accumulated from a single time step, and $p_n = \lvert \langle \psi_n | \boldsymbol{\phi}_0  \rangle \rvert^2$ specify the eigenstate probabilities in the reference state
$|\boldsymbol{\phi}_0\rangle$. For the extreme case of sufficient sample data with $d = K+1 = N$, 
\begin{eqnarray}
   A = V \begin{bmatrix}
    \lambda_{0} & & & \\
    & \lambda_{1} & & \\
    & & \ddots & \\
    & & & \lambda_{N-1}
  \end{bmatrix} V^{-1},
\end{eqnarray}
where $(V_{ij})_{i,j=1}^{N} = \lambda_{i-1}^{j-1}$ is the Vandermonde matrix containing the eigenphases.  In this case, the eigenvalue problem of \cref{eq:dmd_eigval} resolves the entire Hamiltonian spectrum, echoing classic techniques of Prony~\cite{prony1795essai} and Pad\'e~\cite{Pade1892,Pade_overview}. Here we highlight the utility of our estimator as a powerful extension when working with limited data samples $d < K \ll N$. Under a mild assumption that we regulate our time step $\Delta t$ to keep the eigenphases in a $2\pi$-window satisfying \cref{eq:timestep_choice}, we can show a rapid convergence of the energy estimation,
\begin{align}
    \lvert \tilde{E}_0 - E_0 \rvert \leq \mathcal{O} \left(K e^{-\beta d} \right),
    \label{eq:convg_result}
\end{align}
where 
\begin{align}
    \beta > - \ln \frac{\sin((E_{N-1} - E_1)\Delta t/2 ) }{\sin((E_{N-1} - E_0)\Delta t/2 )},
\end{align}
denotes the leading-order exponent that depends on the normalized spectral gap $(E_1 - E_
0)/(E_{N-1} - E_0)$. 

A proof of this error bound exploits the minimal residual property of DMD solvers. We elaborate the proof in its entirety in \cref{app:sec:prony}, though sketch the main ideas here.

Our proof relies on the companion structure of the system matrix $A \in \mathbb{C}^{d \times d}$. The matrix is uniquely determined by its last row denoted as $\vec{a} = (a_0,a_1,\cdots,a_{d-1}) \in \mathbb{C}^d$. The eigenvalues $\tilde{\lambda}_{\ell}$ of $A$, corresponding to the roots of the following characteristic polynomial,
\begin{align}
    \mathcal{C}_{\rm DMD}(z) = \sum_{\ell=0}^{d} a_{\ell} z^\ell = \prod_{\ell=0}^{d-1} (z-\ritz_\ell),
    \label{eq:char_DMD}
\end{align}
exhibit a smooth dependence on $\vec{a}$. We note that $a_{d} \equiv 1$. The matrix elements are determined by the LS regression minimizing the total residual. In particular, we prove that the ODMD solution admits a residual at most exponentially small in the size $d$ of the system matrix. 

On the other hand, we can also construct a new companion matrix, $C$, that is of the same size as $A$ and has the two following key properties: (1) its maximal phase eigenvalue is precisely $\lambda_0 = e^{-i E_0 \Delta t}$, where $E_0$ is the true ground state energy; and (2) the difference between $A$ and $C$ can be effectively bounded from residual analysis.  The crux of the proof is that such a $C$ can indeed be constructed, which we show by extending Prony's idea~\cite{prony1795essai} from the full Hilbert space dimension $N$ to the much smaller dimension $d$.  

In particular, the eigenvalues of $C$, which has last row $\vec{c} = (c_0,c_1,\cdots,c_{d-1}) \in \mathbb{C}^d$, are determined by the characteristic polynomial, 
\begin{align}
    \mathcal{C}_{\Omega}(z) = \sum_{\ell=0}^{d} c_{\ell} z^\ell = (z-\lambda_0) \prod_{\ell=1}^{d-1} (z-\lambda_{n_\ell}),
\end{align}
where $\lambda_0$ is, again, the exact ground state phase and $(\lambda_{n_1}, \lambda_{n_2}, \cdots, \lambda_{n_{d-1}})$ are $d-1$ eigenphases different from $\lambda_0$.  We show in \cref{app:sec:prony} that there exists some choice of $\{ \lambda_{n_{\ell}}\}_{\ell=1}^{d-1}$ such that $C$ satisfies the above conditions (1) and (2).  Then based on an elementary smoothness argument, proximity in the characteristic polynomials $\Vec{a}$ and $\vec{c}$ implies proximity in their respective roots as the eigenvalues of the associated companion matrices. Specifically, we incur, using perturbation analysis, a nearly exponentially small error when approximating the exact ground state phase with the maximal phase ODMD eigenvalue $\Tilde{\lambda}_0 = e^{-i \Tilde{E}_0 \Dt}$.

As an aside, we observe that in the idealized case of a perfect initial state, where $\ket{\boldsymbol{\phi}_0} = \ket{\psi_0}$, ODMD requires only two time steps to recover the ground state energy, as the overlap signal contains a single frequency component. This reveals how a more informative reference state can further enhance the energy convergence. However, unlike alternative approaches that rely more critically on initial state preparation, ODMD does not require a dominantly high ground state overlap to perform effectively.


\section{Connections to classical techniques}
\label{sec:connections}

The ODMD approach is closely related to prominent eigenvalue methods from the classical literature.  In this section, we explore the connections to Krylov and matrix pencil methods to gain insight into the stable convergence of our algorithm and its ability to consistently determine the eigenenergies.

\subsection{Oblique two-sided Krylov method}

The ODMD algorithm can be interpreted as a hybrid quantum-classical oblique projection method onto two-sided Krylov subspaces~\cite{saadbook}. Krylov subspace approaches \cite{horn_johnson_1985} variationally target the ground state via searching in a systematically constructed low-energy subspace, and are thus especially powerful for large-scale eigenvalue problems found in quantum physics.

Defining $\cA := e^{-iH\Dt}$, we first observe that the shifted Hankel matrices in \cref{eq:hankel} can be combined linearly to form what is known in the numerical linear algebra community as a ``matrix pencil,'' $\bX' - \lambda\bX$. The matrix pencil can be decomposed as follows:
\begin{equation}
\mathbf{X}' - \lambda \mathbf{X} = \mathbf{W}_d^{\dagger} \, (\cA - \lambda I) \, \mathbf{V}_{K+1},
\label{eq:spectraltrans}
\end{equation}
where
\begin{equation}
\begin{aligned}
    \mathbf{W}_d & :=
    \begin{bmatrix}
        \ket{\boldsymbol{\phi}_0} & \cA^{\dagger} \ket{\boldsymbol{\phi}_0} & \cdots & (\cA^{d-1})^{\dagger} \ket{\boldsymbol{\phi}_0}
    \end{bmatrix}, \\
    \mathbf{V}_{K+1} & :=
    \begin{bmatrix}
        \ket{\boldsymbol{\phi}_0} & \cA \ket{\boldsymbol{\phi}_0} & \cdots & \cA^{K} \ket{\boldsymbol{\phi}_0}
    \end{bmatrix}.
\end{aligned}
\label{eq:spectraltrans2}
\end{equation}
When $d = K+1 = N$, \cref{eq:spectraltrans} represents a linear transformation that preserves the operator spectrum provided that $\mathbf{W}_N, \mathbf{V}_N$ possess full rank. In this case, we have that the spectrum of the matrix pencil $\mathbf{X}' - \lambda \mathbf{X}$ corresponds exactly to the eigenfrequencies of $e^{-iH\Dt}$. However, we typically do not run our algorithm to the full dimension $N$ and instead extract eigenvalue approximations for $d, K \ll N$. From~\cref{eq:spectraltrans,eq:spectraltrans2}, we observe that in this setting, the rectangular Hankel matrix pencil, $\mathbf{X}' - \lambda \mathbf{X}$, corresponds to a two-sided projection of $\cA - \lambda I$ onto the left and right Krylov subspaces,
\begin{equation}
\begin{aligned}
    \mathcal{L}_d &= \text{span}\left\{\ket{\boldsymbol{\phi}_0},  \cA^{\dagger} \ket{\boldsymbol{\phi}_0}, \hdots, (\cA^{d-1})^{\dagger} \ket{\boldsymbol{\phi}_0}\right\},\\
    \mathcal{R}_{K+1} &= \text{span}\left\{\ket{\boldsymbol{\phi}_0},  \cA\ket{\boldsymbol{\phi}_0}, \hdots, \cA^{K} \ket{\boldsymbol{\phi}_0}\right\},
\end{aligned}
\end{equation}
respectively, for which $\bW_d$ and $\bV_{K+1}$ form a pair of non-orthogonal bases. Specifically, $\mathbf{X}'$ is the oblique projection of $\cA$, the unitary evolution operator, onto $\cR_{K+1}$ along $\mathcal{L}_d$, while $\bX = \bW_d^{\dagger} \bV_{K+1}$ gives the two-sided overlap matrix. In the case of $\bX = I$, \textit{i.e.}, the bases for $\mathcal{L}_d$ and $\mathcal{R}_{K+1}$ are mutually orthogonal, this corresponds to a biorthogonal Krylov process~\cite{saadbook}. The procedure to compute the system matrix $A$ from \cref{eq:LS_ODMD} using the LS regularization gives a robust recipe to convert the generalized eigenvalue problem of \cref{eq:spectraltrans} to a well-conditioned standard eigenvalue problem of \cref{eq:dmd_eigval}.

Moreover, ODMD exploits the freedom of splitting the subspace dimensions $d$ and $K+1$ for the left and right Krylov spaces, respectively, in order to choose a combination such that each subspace individually is in general better-conditioned compared to other subspace methods exploiting real-time evolution. The most natural comparison is with the unitary variational quantum phase estimation algorithm (UVQPE)~\cite{klymko2022real}. In UVQPE, $\cA$ is projected \emph{one-sided} on $\mathbf{V}_{m}$, i.e., UVQPE solves the reduced generalized eigenvalue problem $\mathbf{V}_{m}^{\dagger} (\cA - \lambda I) \mathbf{V}_{m}$ which is constructed from measuring $m$ observables. ODMD uses the same $m$ observables and instead generates a two-sided projection using $\mathbf{W}_{m/3}$ and $\mathbf{V}_{2m/3}$. As the conditioning of $D$-dimensional non-orthogonal Krylov bases \eqref{eq:spectraltrans2} typically grows as $\mathcal{O}(\gamma^D)$ for some $\gamma > 1$~\cite{Beckermann2000}, we see that ODMD effectively reduces the problem of ill-conditioning compared to UVQPE by reducing the subspace dimensions ($D=m/3$ or $2m/3$ instead of $m$) while simultaneously preserving the information and distributing it asymmetrically over a pair of subspaces. The connection with projection methods provides an intuitive explanation for the validity of ODMD as Krylov subspace methods are known to extract accurate eigenvalue approximations from low-dimensional subspaces.

\subsection{Function Krylov method}

Here we highlight some additional theoretical intuition about the efficacy of data Hankelization for ODMD, further elucidating its connection to the Krylov subspace approaches. We establish this connection through the Koopman operator analysis~\cite{Koopman1931HamiltonianSA,Koopman1932,BruntonKutz2015,BruntonKutz2016,Arabi2017}. We consider a physical Hilbert space $\mathcal{H} = \C[N]$. The Koopman (super)operator, $\Ko$, probes the underlying dynamics of the system by mapping scalar functions, $f:\mathcal{H} \rightarrow \mathbb{C}$, or observables in our language, to scalar functions. For a function $f$, $\Ko$ is defined by
\begin{align}
    \Ko[f](\ket{\boldsymbol{\phi}}) = f ( e^{- i H \Delta t} \ket{\boldsymbol{\phi}} ),
\end{align}
which gives a push-forward of the dynamics via the time evolution $e^{-iH \Dt}$. For an eigenstate $\ket{\psi_n}$ of the quantum system, we define an associated function $f_{n}$ by
\begin{align}
    f_{n}(\ket{\boldsymbol{\phi}}) = \braket{\psi_n|\boldsymbol{\phi}},
\end{align}
which is an eigenfunction of $\Ko$ with eigenvalue $\lambda_n = e^{-i E_n \Delta t}$, 
\begin{align}
    \Ko [f_{n}]  = \lambda_n f_n .
\end{align}
Similarly, we define a function $f_{\boldsymbol{\psi}}$ from a state $\ket{\boldsymbol{\psi}}$ by
\begin{align}
f_{\boldsymbol{\psi}}(\ket{\boldsymbol{\phi}}) = \braket{\boldsymbol{\psi}|\boldsymbol{\phi}} = \sum_{n=0}^{N-1} \braket{\boldsymbol{\psi}| \psi_n} \braket{\psi_n|\boldsymbol{\phi}}. 
\end{align}
While the Koopman operator itself is infinite-dimensional, the functions $f_{\boldsymbol{\psi}}$ reside in an invariant subspace of $\Ko$ spanned by $\{ f_n \}_{n=0}^{N-1}$. If we use a state $\ket{\boldsymbol{\psi}}$ that maintains nonzero overlap with our target energy eigenstates, then we can translate the estimation of the Hamiltonian spectrum to a dual eigenvalue problem defined on the function space. For a delay embedding dimension $d$, we consider the collection of $d$ functions,
\begin{align}
\boldsymbol{g} = \begin{bmatrix}f_{\boldsymbol{\psi}} & \Ko[f_{\boldsymbol{\psi}}] & \cdots & (\Ko)^{d-1}[f_{\boldsymbol{\psi}}]\end{bmatrix}^{\top},
\end{align}
within the invariant subspace. For any such function $f_{\boldsymbol{\psi}}$, we can approximate $(\Ko)^d[f_{\boldsymbol{\psi}}]$ as a linear combination of the functions in $\boldsymbol{g}$, and this becomes exact as $d$ grows to the full Hilbert space dimension $N$.  When $d<N$, we can look for a linear operator $\bK_{\Dt} \in \mathbb{C}^{d \times d}$ that gives an optimal approximation on the invariant subspace.  To find $\bK_{\Dt}$, we formulate the dual problem,
\begin{widetext}
\begin{align}
    \underbrace{\begin{bmatrix}
        \vertbar & & \vertbar\\
        \boldsymbol{g}(e^{-iH\Dt} \ket{\boldsymbol{\phi}_0}) & \cdots & \boldsymbol{g}(e^{-iH\Dt}\ket{\boldsymbol{\phi}_K}) \\
        \vertbar & & \vertbar\\
    \end{bmatrix} }_{\bG' \in \mathbb{C}^{d \times (K+1)} } \lseq{} \bK_{\Dt} \underbrace{\begin{bmatrix}
        \vertbar & & \vertbar\\
        \boldsymbol{g}(\ket{\boldsymbol{\phi}_0}) & \cdots & \boldsymbol{g}(\ket{\boldsymbol{\phi}_K}) \\
        \vertbar & & \vertbar\\
    \end{bmatrix}}_{\bG \in \mathbb{C}^{d \times (K+1)} },
\end{align}
\end{widetext}
relaxing the pointwise equality on $\mathcal{H}$ to a LS equality on the states $\{ \ket{\boldsymbol{\phi}_k} \}_{k=0}^{K}$. This matrix approximation yields a factorization,
$\bK_{\Dt} = \bG' (\bG)^{+} = \bZ' (\bZ)^{+}$, with
\begin{align}
    \bZ' &= \frac{1}{K+1} \sum_{k=0}^{K} \boldsymbol{g}(e^{- iH \Dt}\ket{\boldsymbol{\phi}_k}) \boldsymbol{g}^{\dagger}(\ket{\boldsymbol{\phi}_k}),
    \label{eq:W_finite} \\
    \bZ &= \frac{1}{K+1} \sum_{k=0}^{K} \boldsymbol{g}(\ket{\boldsymbol{\phi}_k}) \boldsymbol{g}^{\dagger}(\ket{\boldsymbol{\phi}_k}), \label{eq:Z_finite}
\end{align}
from the definition $\bG^{+} = (\bG^{\dagger} \bG)^{-1} \bG^{\dagger}$. This factorization, as we shall show, has an intimate connection to the Krylov subspace approaches.~\cite{saadbook} For the  ODMD setting, we use $\ket{\boldsymbol{\psi}} = \ket{\boldsymbol{\phi}_0}$ and $\ket{\boldsymbol{\phi}_k} = e^{-i H k \Dt} \ket{\boldsymbol{\phi}_0}$ where $\ket{\boldsymbol{\phi}_0}$ is our reference state. To proceed, we define the functional inner product $\braket{f, f'}_{\mu}$ for functions $f$ and $f'$ as, 
\begin{align}
    \braket{f, f'}_{\mu} = \int_{\ket{\boldsymbol{\phi}} \in \mathcal{H}} d\mu\hspace{0.03cm} f^{\ast}(\ket{\boldsymbol{\phi}}) f'(\ket{\boldsymbol{\phi}}),
\end{align}
where $\mu$ is the empirical measure, 
\begin{align}
    \mu = \frac{1}{K+1} \sum_{k=0}^{K} \delta(\ket{\boldsymbol{\phi}} - \ket{\boldsymbol{\phi}_k}),
\end{align}
on the state trajectory $\{ \ket{\boldsymbol{\phi}_k} \}_{k=0}^{K}$ composed of $(K+1)$ Dirac $\delta$-functions $\delta(\ket{\boldsymbol{\phi}} - \ket{\boldsymbol{\phi}_k})$. \Cref{eq:Z_finite,eq:W_finite} can then be expressed using the $L^2(\mu)$ inner product defined above,
\begin{align}
    \bZ'_{ij} &= \braket{g_j, \Ko[ g_i]}_{\mu}, \\
    \bZ_{ij} &= \braket{g_j, g_i }_{\mu},
\end{align}
where we interpret $\bZ'$ and $\bZ$ as matrix representations of the functional ``target" and ``overlap" operators, respectively, projected onto a Krylov subspace spanned by the functions $\{\Ko^{i}[f_{\boldsymbol{\psi}}] \}_{i=0}^{d-1}$. An eigenpair $(\lambda, \vec{v})$ of $\bK_{\Dt}$ thus solves the generalized eigenvalue equation, $\bZ' \vec{v} = \lambda \bZ \vec{v}$. \Cref{eq:Z_finite,eq:W_finite} already point to the enhanced noise-resilience of our approach, since statistical errors on the observables $g_i(\ket{\boldsymbol{\phi}_k})$ become tighter when passed onto the functional matrix elements $\bZ'_{ij}$ and $\bZ_{ij}$. In other words, the variance in the matrix elements $\bZ'_{ij}$ and $\bZ_{ij}$ is reduced by a factor of $\mathcal{O}(K^{-1})$ relative to the variance in the measured observables due to central-limit scaling. For example, independent Gaussian errors $\mathcal{N}(0,\epsilon^2)$ on each observable combine to make an error of leading order $\mathcal{N}(\epsilon^2, K^{-1}\epsilon^2)$ and $\mathcal{N}(0, K^{-1} \epsilon^2)$ on $\bZ_{ij}$ for the respective cases $i=j$ and $i \neq j$.

We remark that for a general reference state $\ket{\boldsymbol{\phi}_0}$ and any finite $K$, the Koopman operator is non-normal and $\braket{f_n, f_m}_{\mu} \neq 0$. The canonical Krylov bounds on spectral convergence do not strictly apply here as they would require properties of orthogonal polynomials; nevertheless, we showed a rapid convergence of the ODMD estimation using pure algebraic tools in \cref{sec:convergence}.

\subsection{Matrix pencil methods}

We delve deeper into a common framework connecting our data-driven strategy to various well-known methods in signal processing such as the generalized matrix pencil method~\cite{mpm1989}, the Eigensystem Realization Algorithm (ERA)~\cite{era1984,era1985} and Estimation of Signal Parameters via Rotational Invariance Techniques (ESPRIT)~\cite{Roy1989ESPRITestimationOS,Sarkar1995}. We do so by explicitly showcasing an isomorphism between the Hankel matrix pencil of~\cref{eq:spectraltrans} used in ODMD and other matrix pencil approaches that are extensively used across scientific fields.

To elucidate the direct correspondence between DMD and matrix pencil methods used for solving \cref{eq:spectraltrans}, we consider the singular value decomposition (SVD) of $\mathbf{X} = U\Sigma V^{\dagger}$ from~\cref{eq:svd_x}. In terms of the SVD, we obtain a new matrix isospectral with the system matrix $A$,
\begin{equation}
    A = \mathbf{X}' \mathbf{X}^{+}
    \cong \Sigma^{-1/2} U^{{\dagger}} \mathbf{X}' V \Sigma^{-1/2},
    \label{eq:ERA}
\end{equation}
where $\cong$ denotes a spectral congruence that preserves the eigenvalues as any similarity transform leaves the spectrum invariant. The formulation of an eigenvalue problem with respect to the RHS of \cref{eq:ERA} is known as the ERA method~\cite{era1984,era1985,Gilbert1963ControllabilityAO,Klmn1963MathematicalDO,SanchezGasca1997ComputationOP} for system identification, which shows an equivalence to the DMD approach on the LHS. 

Next, we consider a simultaneous factorization of the data and shifted data matrix $\mathbf{X}$ and $\mathbf{X}'$ using the SVD of the base Hankel matrix,
\begin{equation}
\mathbf{X}_\text{b} =
    \begin{bmatrix}
        \mathbf{X}_{0} & \mathbf{X}_{1:K} & \mathbf{X}_{K+1} 
    \end{bmatrix}
    = U_{\text{b}} \Sigma_{\text{b}} \begin{bmatrix}
    V_1 \\
    V_2 \\
    V_3 \\
    \end{bmatrix}^{{\dagger}},
\end{equation}
for which $\mathbf{X} = [\mathbf{X}_{0} ~ \mathbf{X}_{1:K}]$, $\mathbf{X}' = [\mathbf{X}_{1:K} ~ \mathbf{X}_{K+1}]$, and $\Sigma_{\text{b}}$ constitutes three singular value blocks consistent with the partition. Since $\mathbf{X} = U_{\text{b}} \Sigma_{\text{b}} V_{\mathbf{X}}^{{\dagger}} = U_{\text{b}} \Sigma_{\text{b}}[V_1^{\dagger}~V_2^{\dagger}]$ and $\mathbf{X}' = U_{\text{b}} \Sigma_{\text{b}}  V_{\mathbf{X}'}^{\dagger} = U_{\text{b}} \Sigma_{\text{b}}[V_2^{\dagger}~V_3^{\dagger}]$, we have
\begin{align}
    A &= U_{\text{b}} \Sigma_{\text{b}} V_{\mathbf{X}'}^{{\dagger}} (V_{\mathbf{X}}^{{\dagger}})^{+} \Sigma_{\rm b}^{-1} U_{\rm b}^{{\dagger}},\\
    &\cong \big[ V_{\mathbf{X}}^{{\dagger}} (V_{\mathbf{X}'}^{{\dagger}})^{+} \big]^{\top} = \big[ ( V_{\mathbf{X}'}^{{\dagger}})^{\top} \big]^{+} ( V_{\mathbf{X}}^{{\dagger}})^{\top},
    \label{eq:ESPRIT}
\end{align}
where the congruence follows from invariance of the spectrum under matrix transpose. This SVD-based formulation of the eigenvalue problem associated with the RHS of \cref{eq:ESPRIT} is widely recognized as the ESPRIT method~\cite{Roy1989ESPRITestimationOS,Sarkar1995} for signal recovery.


\section{Numerical demonstration}
\label{sec:numerics}

We now empirically demonstrate ODMD's suitability for extracting spectral information from noisy measurements obtained on quantum hardware. Specifically, we demonstrate that our method achieves systematic convergence to the target many-body ground states and corresponding energies even with a large degree of perturbative noise. Our numerical experiments follow the prescription outlined in \cref{subsec:gs_from_odmd,subsec:eigenvector} above.

\input figure_Heisenberg_numerics.tex

Our test problems belong to two classes of many-body systems. First, we consider the Heisenberg model of interacting quantum spins arranged on a one-dimensional chain. This is a prototypical model describing the effects of strong interactions between electronic spins, and the model is commonly used in condensed matter physics for benchmarking new computational methods. Second, we estimate the ground state energies of strongly-correlated molecular Hamiltonians, a task of substantial pragmatic importance in chemical and materials sciences. We consider molecules of varying complexity, ranging from the relatively simple ${\rm LiH}$ to the more challenging ${\rm Cr}_2$ (displayed in \cref{app:sec:ground_state_estimation} and \cref{subsec:mol_sys}, respectively). 

In addition to investigating the performance of ODMD, we also examine recently-developed methods that use similar types of measurements and resources, including subspace expansion techniques exemplified by the Variational Quantum Phase Estimation (VQPE) and its unitary modification (UVQPE)~\cite{klymko2022real}, as well as optimization-based techniques represented by the Quantum Complex Exponential Least Squares (QCELS)~\cite{ding2022even,ding2023simultaneous}. 
A numerical comparison of the different approaches is a crucial initial step in quantifying the relative performance of different real-time algorithm classes, each developed for unique purposes and aimed at different strategies to leverage the capabilities of current quantum hardware. Out of the four methods considered, we illustrate that ODMD consistently reaches the target accuracy with both the fewest time steps and the fewest measurements, especially when the initial state has a small overlap with the target eigenstate.

\begin{table}[htbp!]
\resizebox{\columnwidth}{!}{
\begin{tabular}{|c|c|c|}
\hline
Algorithms & Real-time data & Tunable hyperparameters \\
\hline
ODMD & $d \times (T - d)$ Hankel
& SVD threshold $\tilde{\delta}$ \\
\hline QCELS~\cite{ding2022even} & $T \times 1$ time series
& Number of target modes \\
\hline
(U)VQPE~\cite{klymko2022real} & $T \times T$ Toeplitz
& SVD threshold $\tilde{\delta}$ \\
\hline
\end{tabular}
}
\centering
\caption{Summary of post-processing data structures and tunable hyperparameters for ODMD, QCELS, VQPE, and UVQPE. Each method processes the real-time data, assumed to be overlap matrix elements collected over $T$ time steps, in different formats and involves different hyperparameters that influence accuracy and stability. For ODMD, we set $d = \lfloor \frac{T}{3} \rfloor$.}
\label{tab:methods_tabulated}
\end{table}

For numerical efficiency, in each case we select the time step $\Delta t$ (to be specified within the captions of the following figures) and/or scale our target Hamiltonian $H$ such that
$\Vert H\Vert_2 \Dt \leq 3\pi/4$. We recall that the ODMD data matrices constructed from a total of $T$ time steps are of dimension $d \times (T-d)$ with $d = \lfloor \frac{T}{3} \rfloor$, a particular choice considered in \cref{subsec:gs_from_odmd}. For comparison, VQPE and UVQPE process the same real-time data as a pair of $T \times T$ Toeplitz matrices, while QCELS treats it as a length-$T$ time series. To simulate the noise effect, we first compute the exact overlap of \cref{eq:overlap}, then we introduce Gaussian errors $\mathcal{N}(0,\epsilon^2)$ on the overlap measurements. According to the central limit theorem, the Gaussian noise model correctly captures effects of the shot noise, \textit{i.e.}, the statistical error due to taking a finite number of measurements that scale as $\mathcal{O}(\epsilon^{-2})$ in the Hadamard test. Some types of hardware noise may also exhibit approximately Gaussian behavior, though we do not investigate specific hardware noise models in this section. To mitigate the impact of the added noise, we apply singular-value thresholding as described at the end of \cref{subsec:gs_from_odmd} for ODMD; an SVD-based thresholding strategy is also applicable to VQPE and UVQPE~\cite{klymko2022real}, which we adopt to promote optimized numerical stability across all real-time methods.

We summarize these structural and parametric aspects in \cref{tab:methods_tabulated}.  We remark that the precise noise variance $\epsilon^2$ is not directly accessible in practice, as this would require an infinite number of measurement shots. However, if $\epsilon$ is known up to its order of magnitude, we find empirically that choosing $\tilde{\delta}/\epsilon = 10$ for $\epsilon \lesssim 10^{-2}$ typically leads to an optimal performance for ODMD, VQPE, and UVQPE. Moreover, throughout this section, we primarily consider reference states that do not overlap dominantly with the low energy eigenstates, where the dominance condition is given by $\min_{n \in \mathcal{S}} p_n > 1-\sum_{n \in \mathcal{S}} p_n$ for target eigenindices $\mathcal{S} \subset \{0,1,\ldots,N-1\}$. This condition is satisfied only if $\mathcal{S}$ approaches the full set of eigenindices, which corresponds to an intractable classical cost because of the exponential scaling of the nonlinear optimization with the number of target modes. In the few-mode regime, QCELS lacks a convergence guarantee. Hence, we restrict our attention to the single-mode QCELS for ground state estimation without loss of generality.

In \cref{subsec:spin_sys,subsec:mol_sys,subsec:excited_state}, we present numerical results on spin and molecular systems for benchmarking the performance of ODMD. All dataset and codes are available through the repository: \faGithub\href{https://github.com/QuantumComputingLab/odmd}{ODMD}.


\subsection{Eigenstates and eigenenergies of spin systems}
\label{subsec:spin_sys}

To demonstrate the usefulness of ODMD in condensed matter physics, we consider the Heisenberg spin model in 1D,
\begin{equation}
    H = J\sum_{i=0}^{L-1} \mathbf{S}_i^{ } \cdot \mathbf{S}_{i+1}^{ },
\end{equation}
where $J>0$ is the antiferromagnetic coupling strength and $\mathbf{S}_i$ is the spin-1/2 operator $\frac{\hbar}{2}\left(\sigma^x_{ },\sigma^y_{ },\sigma^z_{ }\right)$ on site $i$, with the standard Pauli operators $\boldsymbol{\sigma}_{ }$.  We take periodic boundary conditions and examine systems of size $L=8$ and $L=12$ spins.

The spectrum of the 1D Heisenberg Hamiltonian is bounded by $-3L J < \braket{\boldsymbol{\phi}|H|\boldsymbol{\phi}} \leq L J$ for any state $\ket{\boldsymbol{\phi}}$~\footnote{The lower bound comes from the sum of singlet energies on every bond, which is lower than the true ground state energy. The upper bound comes from a fully ferromagnetic state with all spins up.}.  Accordingly, we compute real-time overlaps in the two cases with time step sizes of $\Delta t = 0.15$ and $0.1$, respectively, satisfying the bound from \cref{eq:timestep_choice}. For numerical convenience we have set $J=4$ and $\hbar=1$, so that $J\mathbf{S}\cdot\mathbf{S}$ becomes simply $\boldsymbol{\sigma}\cdot\boldsymbol{\sigma}$.  We use two different types of initial states: an antiferromagnetic product state $|\boldsymbol{\phi}_0\rangle = |01\rangle^{\otimes L/2}$ and an equal superposition of two such product states, $|\boldsymbol{\phi}_0\rangle = (|01\rangle^{\otimes L/2} + |10\rangle^{\otimes L/2})/\sqrt{2}$.  Both have reasonable overlap with the true ground state; the overlap for the superposition state is exactly twice that of the product state.  We compute the expectation values $s(k\Dt) = \langle \boldsymbol{\phi}_0|e^{-iH k\Delta t}|\boldsymbol{\phi}_0 \rangle$ for $0 \leq k \leq 250$, and after adding Gaussian errors, we feed the noisy data into ODMD as well as several of the advanced hybrid methods discussed in \cref{sec:background}.  For SVD thresholding, we adopt $\tilde{\delta}/\epsilon = 10$ for a given noise level $\epsilon$.

We show the convergence of the approximate ground state energies $\tilde{E}_0$ in \cref{fig:Heisenberg_convergence}, left and center panels, where the reference energies come from exact Hamiltonian diagonalization.  First, we explain the horizontal axis.  All four algorithms we test, namely ODMD, QCELS, VQPE, and UVQPE, take as input a list of measured observables $\langle\phi_0|e^{-iHk\Delta t}|\phi_0\rangle$ for $k=1, 2, \ldots$; VQPE additionally uses the Hamiltonian matrix elements $\langle\phi_0|e^{-iHk\Delta t}H|\phi_0\rangle$.  
One way to measure the cost of running one of these algorithms would be to count the number of time steps used, \emph{i.e.}, the number of integers $k$ at which the expectation values are measured.  However, a more relevant estimate of the cost is the total number of circuits that need to be run on a quantum computer.  

If we use the Hadamard test to measure the expectation values $\braket{U_k}$ of $U_k = e^{-iHk\Delta t}$, we need to run repeated shots of two distinct circuits, one for the real part and the other for the imaginary part.  Hence ODMD, which needs only the real part, actually requires half as many circuits as QCELS or UVQPE.  We show this in the figure by plotting against the number of observables, separating the real and imaginary parts.  We show the same number of time steps for each algorithm, which is why the ODMD results (blue stars) terminate only at 250 on the horizontal axis while QCELS (red circles) and UVQPE (purple squares) go to 500.  Thus our figure contains both metrics of convergence speed.  By comparing in a single vertical column, we see the relative speed of convergence of the algorithms as a function of the number of circuits to be run, \emph{i.e.}, the practical cost.  By comparing the end of the blue markers at 250 to the end of the purple or red markers at 500, we see the relative speed of convergence as a function of the number of time steps.

Because VQPE additionally calculates the expectation values of $U_k H$, we count it as requiring four observables per time step: the real and imaginary parts of $\braket{U_k}$ and of $\braket{ U_k H}$. In fact, this underestimates the true cost as the many-body Hamiltonians are generally non-unitary. Thus $\braket{U_k H}$ will require either multiple circuits for each of the real and imaginary parts or multiple ancillas for methods such as block encoding~\cite{qsvt2018}. Since QCELS and UVQPE double the number of observables, while VQPE increases it by at least a factor of four, the total number of measurement shots required to achieve $\epsilon$-precision across all observables scales accordingly.

ODMD results show rapid convergence despite a significant level of noise.  The convergence is slightly faster than with UVQPE, though the latter converges reasonably well.  In contrast, QCELS, which is more reliant on the overlap between the initial state and the true ground state, converges more slowly.  VQPE may not converge at all. The comparison with VQPE in particular highlights the noise-resilience of ODMD; with no noise (including no statistical noise from sampling the wavefunction, \emph{i.e.}, infinite shots), VQPE also converges to the correct energy.  However, VQPE involves an ill-conditioned generalized eigenvalue problem, so that even with thresholding strategies it fails to converge in the presence of moderate noise. Such instability is not observed for ODMD. We also notice that ODMD's performance is robust when the initial state overlap is relatively small. 

Now we investigate the convergence of the corresponding ODMD state approximation $\ket{\tilde{\psi}_0}$ to the ground state $\ket{\psi_0}$.  We measure convergence using the residual norm, $|\!|H|\tilde{\psi}_0 \rangle - \tilde{E}_0|\tilde{\psi}_0\rangle|\!|_2/\norm{H}_2$.  We note that although ODMD does not directly prepare the ground state, the residual (up to the normalization factor $\norm{H}_2$) can be accessed on the quantum computer.  Expanding the square of the residual norm gives 
\begin{equation}
    \langle \tilde{\psi}_0|H^2|\tilde{\psi}_0\rangle - 2\tilde{E}_0\langle \tilde{\psi}_0|H|\tilde{\psi}_0\rangle + \tilde{E}_0^2,
\end{equation}
and each term can be computed from \cref{eq:GS_properties} using real-time evolution.  We observe that the residual norm is smaller for initial state with a better ground state overlap; see, for example, the black crosses (15\% overlap) versus black squares (7\%). On the other hand, the system size does not seem to have a strong impact: the black crosses and green circles correspond to different system sizes, with approximately the same ground state overlap, and behavior similarly as a function of time step.

\input figure_molecule_comparison.tex

\subsection{Eigenenergies of molecular systems}
\label{subsec:mol_sys}

We focus on the ground state energy calculation of ${\rm Cr}_2$, a transition metal dimer known for exhibiting among the strongest electronic correlations. 
Here we work with a truncated sector of the molecular Hilbert space by employing the adaptive sampling configuration interaction algorithm (ASCI)~\cite{tubman2016deterministic,tubman2020modern} (see \cref{app:sec:ground_state_estimation} for further details). \Cref{fig:compare_methods} shows the convergence of the real-time algorithms and illuminates the efficacy and noise-resilience of the ODMD approach. For these calculations, we have rescaled the Hamiltonian so that its eigenenergies fall in the range $[-\pi/4, \pi/4]$, enabling us to choose $\Dt < 4$.  The three panels display the convergence of estimated ground state energy; note that in the figures we use the original (unscaled) energy scale in order to show where the energy estimate reaches chemical accuracy. We examine both small ($\Dt = 1$, left and center panels) and large ($\Dt = 3$, right panel) time step regimes, with the same number of time steps in each case,  favoring shorter time evolution and thus lower circuit depth on the one hand and longer time evolution and thus richer dynamical information on the other.

\input figure_molecule_convergence.tex

The left panel of \cref{fig:compare_methods} shows a lower noise level and a reference state with small ground state probability $p_0 = \lvert \langle \psi_0 | \boldsymbol{\phi}_0  \rangle \rvert^2 = 0.05 =  \max_{n} p_n $. The blue stars marking the ODMD estimation exhibit satisfying performance compared to the purple squares marking the UVQPE estimation that converges variationally to the true ground state energy in the noiseless limit. The slower convergence of QCELS estimation, fitted to a uniform - time step model, results from a rugged optimization landscape as the reference $\ket{\boldsymbol{\phi_0}}$ is almost orthogonal to the target state $\ket{\psi_0}$.  VQPE fails to converge to high precision even with this small level of noise due to the ill-conditioning of the generalized eigenvalue problem when the initial state overlap is small.

On the other hand, the right panel shows numerical simulations for a Hartree-Fock reference state with large ground state overlap $p_0 > 0.7$.  In this case, even though the noise level is also higher, all the algorithms converge.  However, ODMD still has a significant advantage coming from the fact that it needs only half as many observables/circuits per time step.

Finally, the center panel shows an intermediate case with higher noise and small ground state probability. The ODMD estimation is the only one converging past chemical accuracy.
Due to a relatively low measurement cost and robust continuous convergence, the ODMD approach emerges as a promising hybrid eigensolver on near-term platforms.

\Cref{fig:convergence} further shows the energy convergence of ODMD for varying noise levels, singular value truncation thresholds, and ground state probabilities. The left panel plots the estimation for different reference states with increasing ground state probability. We observe that for any initial overlap size, the ODMD result progressively improves with additional observables. The right panels show the noise-resilience of ODMD for specific ground state probabilities $p_0 = 0.2$ and $p_0 > 0.7$. The two panels present the number of time steps needed to achieve chemical accuracy, highlighting the compactness of our measurement-driven approach.

\subsection{Eigenenergies beyond the ground state}
\label{subsec:excited_state}
So far we have numerically demonstrated the utility of ODMD for ground state calculations. In particular, we observe that our ODMD ground state estimate converges much more rapidly than VQPE when noise on the overlap matrix elements becomes large, and than QCELS when the initial state lacks a dominant overlap with the ground state. In both regimes, ODMD's significant performance advantage is robust to fine-tuning of the post-processing hyperparameters of the real-time algorithms.

To further substantiate the performance of ODMD relative to UVQPE, we proceed to examine their extended applicability to excited state problems. We first note that although ODMD and UVQPE are constructed distinctly, both can be regarded as a subspace-based approach and thereby face a common challenge: convergence to excited states tends to be more gradual than to the ground state -- a limitation typical of many subspace methods~\cite{saadbook}. For VQPE and UVQPE to access excited state information, we require multiple reference states $ \{\ket{\boldsymbol{\phi}_0^{(i)} }: 1 \leq i \leq D \}$, which readily extends the single-reference method. This leads to the generalized eigenvalue problem $ \textbf{H}\textbf{$\Psi$} = \tilde{E} \textbf{S} \textbf{$\Psi$}$ with a block structure, \textit{i.e.}, 
\begin{align}
    \textbf{H} = \begin{bmatrix}
  \textbf{H}^{(11)} & \cdots &  \textbf{H}^{(1D)} \\
  \vdots & \ddots & \vdots \\
  \textbf{H}^{(D1)} & \cdots & \textbf{H}^{(DD)}
\end{bmatrix},
\end{align}
and 
\begin{align}
    \textbf{S} =\begin{bmatrix}
  \textbf{S}^{(11)} & \cdots &  \textbf{S}^{(1D)} \\
  \vdots & \ddots & \vdots \\
  \textbf{S}^{(D1)} & \cdots & \textbf{S}^{(DD)}
\end{bmatrix},
\end{align}
where the individual matrix blocks arising from the pairing of different reference states are defined as,
\begin{align}
    \textbf{H}_{jk}^{(ii')} &=  \braket{\boldsymbol{\phi}_0^{(i)}| H e^{iH(t_j-t_k)} |\boldsymbol{\phi}_0^{(i')}}, \\
    \textbf{S}_{jk}^{(ii')} &= \braket{\boldsymbol{\phi}_0^{(i)}| e^{iH(t_j-t_k)} |\boldsymbol{\phi}_0^{(i')}},
    \label{eq:multi-VQPE_S}
\end{align}
for $1 \leq i,i' \leq D$. The multi-reference (U)VQPE thus accelerates excited state convergence by initializing with a $D$-dimensional subspace spanned by the reference states, indicating a runtime \textit{quadratic} in the number of reference states.

Similarly, ODMD admits a novel extension to a multi-observable setting, mirroring a multi-reference subspace approach. To introduce the multi-observable framework, we solve the LS problem $\mathbf{X}' \lseq{} A \mathbf{X}$ with a vectorial signal,
\begin{align}
    \mathbf{X} = \begin{bmatrix}
  \textbf{s}_0 & \cdots &  \textbf{s}_{K} \\
  \vdots & \ddots & \vdots \\
  \textbf{s}_{d-1} & \cdots & \textbf{s}_{K+d}
\end{bmatrix},
\end{align}
and 
\begin{align}
    \mathbf{X}' = \begin{bmatrix}
  \textbf{s}_{1} & \cdots &  \textbf{s}_{K+1} \\
  \vdots & \ddots & \vdots \\
  \textbf{s}_{d} & \cdots & \textbf{s}_{d+K+1}
\end{bmatrix},
\end{align}
where the vector $\mathbf{s}_k$ at the $k$-th timestep has components,
\begin{align}
    \mathbf{s}^{(i)}_k = \braket{\boldsymbol{\phi}_0|O_i e^{-iHt_k} |\boldsymbol{\phi}_0},
    \label{eq:multi-ODMD_s}
\end{align}
for a set of Hermitian operators $ \{ O_i = O_i^{\dagger} : 1 \leq i \leq D \}$, each determining a sequence of time-delayed observables. These operators define a $D$-dimensional ``observable subspace''. Under our generalized framework, the observable space formulation implies a runtime \textit{linear} in the number of operators $O_i$. This key difference in runtime is directly reflected in the number of time evolution circuits required for generating the real-time data.

\input figure_TFIM_numerics.tex

To evaluate the excited state performance of the multi-observable ODMD and multi-reference UVQPE methods, we consider the transverse-field Ising spin model in 1D,
\begin{equation}
    H =  - J\sum_{i=0}^{L-1} \sigma_i^{z} \sigma_{i+1}^{z} - \sum_{i=0}^{L-1} \sigma_i^{x},
\end{equation}
with ferromagnetic coupling of $J > 0$ and system size of $L=12$. For numerical convenience, we have set $J=1$ and rescaled the Hamiltonian spectrum such that $\Dt = 1$. We start with the single reference state as a superposition of two product states, $|\boldsymbol{\phi}_0\rangle = (3|+\rangle^{\otimes L} + 2|-\rangle^{\otimes L})/\sqrt{13}$ where $\ket{\pm}$ are the $\pm 1$ eigenstates of the Pauli $X$ operator, $\sigma^x_{ }$. For ODMD, we consider the operators
\begin{align}
     \left \{ O_1 = I,~ O_2 = \sigma^{z}_{i},~ O_3 =  \sigma^{y}_{j}: i,j~{\rm random}  \right \},
\end{align}
which compose a $3$-dimensional observable subspace. For UVQPE, we initialize with the associated reference states $\{ \ket{\boldsymbol{\phi}_0^{(i)} } = O_i |\boldsymbol{\phi}_0\rangle \}_{i=1}^{3}$. We compute the expectation values using \cref{eq:multi-ODMD_s,eq:multi-VQPE_S} for time steps $0 \leq k \leq 200$, and after adding Gaussian errors, we feed the noisy real-time data into ODMD and UVQPE. For SVD thresholding, we adopt $\tilde{\delta}/\epsilon = 10$ for a given noise level $\epsilon$.

We illustrate the convergence of the approximate first excited state energy $\tilde{E}_1$ in \cref{fig:excited_state}, where we compare it against a reference energy obtained by exact Hamiltonian diagonalization.  To assess the quantum resource cost, we follow the same metric used in \cref{subsec:spin_sys,subsec:mol_sys} by counting the total number of circuits required on a quantum computer. Specifically, the multi-observable ODMD measures $D=3$ observables, the real part of $\langle O_i U_k \rangle$, per time step, whereas the multi-reference UVQPE measures $2D^2=18$ observables, the real and imaginary parts of $\langle O_i U_k O_{i'} \rangle$, per time step. As visualized in \cref{fig:excited_state}, ODMD converges with fewer total observables -- and hence fewer quantum circuits -- highlighting its efficiency for excited state estimation.

Finally, we comment that the runtime of multi-observable ODMD may be further reduced to \textit{logarithmic} in the number of observables $D$. Such favorable reduction is achievable with randomization techniques, as carefully developed in our recent work~\cite{shen2024efficientmeasurementdriveneigenenergyestimation}. In contrast, applying similar randomization techniques to UVQPE still results in a runtime \textit{superlinear} in $D$. This is because each initial state $\ket{\boldsymbol{\phi}_0^{(i)} }$ requires its own state preparation. Consequently, the ODMD framework lays the foundation for a novel subspace approach capable of simultaneously resolving many-body ground and excited state properties at a significantly lower cost.


\section{Discussion}
\label{sec:diss}

In this work, we developed a hybrid quantum-classical algorithm, ODMD, that leverages real-time evolution to extract ground state information of quantum many-body systems. We proved rigorously that the ODMD estimate converges rapidly to the ground state energy, and we numerically demonstrated robust convergence against noise using Hamiltonians drawn from both condensed matter physics and quantum chemistry. We also related ODMD to a variety of established methods in the numerical linear algebra community.

ODMD presents several distinct advantages over previous hybrid eigensolvers. Most notably, ODMD is more versatile and resilient to perturbative noise. While our framework shares salient perspectives with state-of-the-art subspace and signal processing approaches, it remains effective in regimes where those methods converge slowly or even fail to converge. By formulating a least-squares problem in the space of time-delayed observables, ODMD improves the numerical conditioning and admits a powerful multi-observable generalization best tailored for excited state computation. Moreover, it does not presume a dominant overlap between the initial state and ground state, alleviating the resource overhead associated with state preparation. Consequently, the algorithm converges more reliably as we increase the number of time steps, and also requires fewer shots to sample each measured observable. Additionally, ODMD allows us to recover the eigenstate information and desired eigenstate properties along side the corresponding eigenenergies. Finally, our algorithm only samples the real or imaginary part of the observables, making it even more measurement-efficient for near-term implementation. This last advantage may be adaptable to other real-time based algorithms. 

We have further established a unified theoretical foundation for ODMD. We explicitly showed that it can be understood as a Krylov subspace method, which provides an intuitive variational picture to explain convergence to the ground state energy. Also, we demonstrated the connection between ODMD and matrix factorization methods, highlighting the signal processing nature of our algorithm and suggesting a clear interpretation of its noise-resilience.  

The ODMD framework hence marks a significant step towards realizing viable real-time evolution methods on quantum hardware, and can be extended to applications beyond ground state problems.

\begin{acknowledgments}
This work was funded by the U.S. Department of Energy (DOE) under Contract No.~DE-AC0205CH11231, through the Office of Science, Office of Advanced Scientific Computing Research (ASCR) Exploratory Research for Extreme-Scale Science (YS, DC, KK, SD, RVB) and Accelerated Research for Quantum Computing Programs (AS, DBWY).
This research used resources of the National Energy Research Scientific Computing Center (NERSC), a U.S. Department of Energy Office of Science User Facility located at Lawrence Berkeley National Laboratory, operated under Contract No. DE-AC02-05CH11231. 
NMT is grateful for support from NASA Ames Research Center and from the NASA-DOE interagency agreement SAA2-403862. 

\end{acknowledgments}

\bibliographystyle{quantum}
\bibliography{references}

\onecolumngrid


\appendix
\input appendix.tex

\end{document}

%% file: figures_commands.tex
\newcommand{\plotconv}[6][]{%
\begin{tikzpicture}
\begin{semilogyaxis}[
  width=\columnwidth,%
  xmin=#3,xmax=#4,%
  ymin=#5,ymax=#6,%
  xlabel={\# Observables (Real)},%
  ylabel={Absolute Error},%
  legend style={draw=none,row sep=-2pt},%
  legend pos=north east,%
  #1,%
]
#2
\end{semilogyaxis}
\end{tikzpicture}%
}
\newcommand{\plotconvN}[3][1e-10]{%
\plotconv[%
  xtick={0,150,...,750},%
  ytick={1e-10,1e-8,1e-6,1e-4,1e-2,1e0},%
  title={#2},%
]{%
\addplot[myColOne,thick,only marks,mark=asterisk]%
   table[x index=0,y index=1] {\datfile{#3}};
\addplot[myColTwo,thick,only marks,mark=o,mark size=1.5]%
   table[x index=0,y index=2] {\datfile{#3}};
\addplot[myColThr,thick,only marks,mark=diamond]%
   table[x index=0,y index=3] {\datfile{#3}};
\legend{$\Tilde{\delta}_{\rm DMD} = 10^{-1}$,$\Tilde{\delta}_{\rm DMD} = 10^{-2}$,$\Tilde{\delta}_{\rm DMD} = 10^{-3}$};%
}{0}{750}{#1}{1e0}%
}

\newcommand{\plotheatmap}[3][]{%
\begin{tikzpicture}
\begin{axis}[%
  axis equal image,%
  width=0.75\columnwidth,%
  view={0}{90},%
  xmin=-0.5,xmax=9.5,%
  ymin=-0.5,ymax=9.5,%
  xtick={0,3,...,9},%
  xticklabels={$10^{-1}$,$10^{-2}$,$10^{-3}$,$10^{-4}$},%
  ytick={0,3,...,9},%
  yticklabels={$10^{-1}$,$10^{-2}$,$10^{-3}$,$10^{-4}$,$10^{-5}$},%
  xlabel={Noise Level $\epsilon$},%
  ylabel={$\Tilde{\delta}_{\rm DMD}$},%
  title={#2},%
  point meta min=0,%
  point meta max=600,%
  #1,%
]
#3
\end{axis}
\end{tikzpicture}%
}

%% file: figure_hadamard.tex

\begin{figure}[hbtp]
\centering
\pgfdeclarelayer{circbox}
\pgfsetlayers{circbox,main}

\tikzstyle{operator} = [thick,draw,fill=white,minimum size=1.5em] 
\tikzstyle{measurement} = [thick,draw,rounded rectangle, rounded rectangle west arc=0pt,minimum size=1.5em] 
\tikzstyle{control} = [draw,fill,shape=circle,minimum size=5pt,inner sep=0pt]
    
\newcommand{\xxos}{0}

\begin{tikzpicture}[thick]
%
\matrix[row sep=0.3cm, column sep=0.2cm] (circuit) at (0,0) {
\node (q1) {$\ket{0}$}; & 
\node[operator] (H11) {${\rm H}$}; &
\node[control] (C) {}; &
\node[measurement,fill=white] (P13) {$X$}; &[-0.2cm]
\coordinate (end1); \\
\node (q2) {$\ket{\boldsymbol{\phi}}$}; &
&
\node[operator] (U) {$U$}; &
&
\coordinate (end2);\\
};

\begin{pgfonlayer}{circbox}
\node[fill=white,thick,draw=black,rounded corners=2mm] (background) [fit = (q1) (U) (end2)] {};
\draw[thick] (q1) -- (end1)  (q2) -- (end2) (C) -- (U);
\end{pgfonlayer}
%
\node[above = 0.15 of circuit](overlap) {$\Re \bra{\boldsymbol{\phi}} U\ket{\boldsymbol{\phi}}$};

\renewcommand{\xxos}{3.5}
%
\matrix[row sep=0.3cm, column sep=0.2cm] (circuit) at (0+\xxos,0) {
\node (q1) {$\ket{0}$}; & 
\node[operator] (H11) {${\rm H}$}; &
\node[control] (C) {}; &
\node[measurement,fill=white] (P13) {$Y$}; &[-0.2cm]
\coordinate (end1); \\
\node (q2) {$\ket{\boldsymbol{\phi}}$}; &
&
\node[operator] (U) {$U$}; &
&
\coordinate (end2);\\
};

\begin{pgfonlayer}{circbox}
\node[fill=white,thick,draw=black,rounded corners=2mm] (background) [fit = (q1) (U) (end2)] {};
\draw[thick] (q1) -- (end1)  (q2) -- (end2) (C) -- (U);
\end{pgfonlayer}
%
\node[above = 0.15 of circuit](overlap) {$\Im \bra{\boldsymbol{\phi}} U\ket{\boldsymbol{\phi}}$};
\end{tikzpicture}
\caption{Hadamard test circuits for measuring the real part (\textbf{left}) and imaginary part (\textbf{right}) of the state overlap $\braket{\boldsymbol{\phi}|U|\boldsymbol{\phi}} \in \mathbb{C}$. A Hadamard gate is applied to an ancilla qubit initialized in the $\ket{0}$ state. The ancilla then acts as the control qubit of the unitary $U$ operating on the register qubits. Ancilla measurements are performed in the Pauli basis to evaluate the expectation.}
\label{fig:hadamard_test}
\end{figure}

%% file: figure_quantum_vs_classical.tex
\begin{tikzpicture}
\draw (-.25,-.6) rectangle (8.25,4.6);
\draw[gray,thick,densely dotted] (4,-.6) -- (4,4.6);
\draw (1.875,4.5) node[below] {\textbf{Quantum states}};
\draw (6.125,4.5) node[below] {\textbf{Delayed observables}};
%
\draw (2,3.4) node (psi2) {\footnotesize$\ket{\boldsymbol{\phi}(t_2)}$};
\draw[rotate around={40:(1.9,1.1)}] (1.9,1.1) ellipse (1.7 and 1.3);
\draw (2.75,0.75) node {\footnotesize$\C[N]$};
\coordinate (p0) at (0.75,0.25);
\coordinate (p1) at (1.25,1);
\coordinate (p2) at (1.60,1.5);
\coordinate (p3) at (2.25,2);
\draw[->,thick,myblue]   (p0.center) to[out=85,in=210] (p1.center);
\draw[->,thick,myred]    (p1.center) to[out=25,in=260] (p2.center);
\draw[->,thick,myorange] (p2.center) to[out=80,in=180] (p3.center);
\draw (p0) node {$\bullet$};
\draw (p1) node {\textcolor{myblue}{$\bullet$}};
\draw (p2) node {\textcolor{myred}{$\bullet$}};
\draw (p3) node {\textcolor{myorange}{$\bullet$}};
\draw (p0) node[right] {\footnotesize$\ket{\boldsymbol{\phi}_0}$};
\draw (p1) node[below right,xshift=-1pt,yshift=2pt] {\footnotesize$\ket{\boldsymbol{\phi}_1}$};
\draw (p2) node[right] {\footnotesize$\ket{\boldsymbol{\phi}_2}$};
\draw (p3) node[right] {\footnotesize$\ket{\boldsymbol{\phi}_3}$};
%
\draw (4,3.4) node[fill=white] {$\Qcircuit @C=1em {& \meter & \cw}$};
%
\draw (6,3.4) node {\footnotesize$o(t_2) \inC$};
\draw[->] (4.35,2.5) -- (4.35,-0.2) node[right] {\footnotesize$t$};
\draw[->] (4.35,2.5) -- (5.75,2.5) node[below] {};
\pgfmathsetseed{1}
\draw[thick,smooth,domain=0.1:2.5,samples=6,name path=obs] plot (rand+5.2,\x);
\path [name path=line0] (4,2.499) -- (7,2.499);
\path [name path=line1] (4,2.200) -- (7,2.200);
\path [name path=line2] (4,1.900) -- (7,1.900);
\path [name path=line3] (4,1.600) -- (7,1.600);
\path [name path=line4] (4,1.300) -- (7,1.300);
\path [name path=line5] (4,1.000) -- (7,1.000);
\path [name path=line6] (4,0.700) -- (7,0.700);
\path [name path=line7] (4,0.400) -- (7,0.400);
\path [name intersections={of=obs and line0,by=P0}];
\path [name intersections={of=obs and line1,by=P1}];
\path [name intersections={of=obs and line2,by=P2}];
\path [name intersections={of=obs and line3,by=P3}];
\path [name intersections={of=obs and line4,by=P4}];
\path [name intersections={of=obs and line5,by=P5}];
\path [name intersections={of=obs and line6,by=P6}];
\path [name intersections={of=obs and line7,by=P7}];
\draw (P0) node {$\star$};
\draw (P1) node {\textcolor{myblue}{$\star$}};
\draw (P2) node {\textcolor{myred}{$\star$}};
\draw (P3) node {\textcolor{myorange}{$\star$}};
\draw (P4) node {\textcolor{mypurple}{$\star$}};
\draw (P5) node {\textcolor{mygreen}{$\star$}};
\draw (P6) node {\textcolor{mylightblue}{$\star$}};
\draw (P7) node {\textcolor{mydarkred}{$\star$}};
\draw (6.5,2.5) node {$\star$};
\draw (6.5,2.2) node {\textcolor{myblue}{$\star$}};
\draw (6.5,1.9) node {\textcolor{myred}{$\star$}};
\draw (6.8,2.2) node {\textcolor{myblue}{$\star$}};
\draw (6.8,1.9) node {\textcolor{myred}{$\star$}};
\draw (6.8,1.6) node {\textcolor{myorange}{$\star$}};
\draw (7.1,1.9) node {\textcolor{myred}{$\star$}};
\draw (7.1,1.6) node {\textcolor{myorange}{$\star$}};
\draw (7.1,1.3) node {\textcolor{mypurple}{$\star$}};
\draw (7.4,1.6) node {\textcolor{myorange}{$\star$}};
\draw (7.4,1.3) node {\textcolor{mypurple}{$\star$}};
\draw (7.4,1.0) node {\textcolor{mygreen}{$\star$}};
\draw (7.7,1.3) node {\textcolor{mypurple}{$\star$}};
\draw (7.7,1.0) node {\textcolor{mygreen}{$\star$}};
\draw (7.7,0.7) node {\textcolor{mylightblue}{$\star$}};
\draw[thin,gray] (6.4,2.6) rectangle +(0.2,-0.8);
\draw[thin,gray] (6.7,2.3) rectangle +(0.2,-0.8);
\draw[thin,gray] (7.0,2.0) rectangle +(0.2,-0.8);
\draw[thin,gray] (7.3,1.7) rectangle +(0.2,-0.8);
\draw[thin,gray] (7.6,1.4) rectangle +(0.2,-0.8);
\draw[densely dotted]             (P0) -- (6.35,2.5);
\draw[densely dotted,myblue]      (P1) -- (6.35,2.2);
\draw[densely dotted,myred]       (P2) -- (6.35,1.9);
\draw[densely dotted,myorange]    (P3) -- (6.65,1.6);
\draw[densely dotted,mypurple]    (P4) -- (6.95,1.3);
\draw[densely dotted,mygreen]     (P5) -- (7.25,1.0);
\draw[densely dotted,mylightblue] (P6) -- (7.55,0.7);
\draw[<->] (6.25,1.8) -- (6.25,2.6);
\draw[<->] (6.4,0.45) -- (7.8,0.45);
\draw (6.25,2.2) node[left,fill=white] {\footnotesize$d$};
\draw (7.1,0.45) node[below] {\footnotesize$K$};
\draw (8.1,2.65) node[left] {\footnotesize$\bfo_{t_2,d}$};
\draw (8.1,-0.25) node[left] {\footnotesize Takens' embedding};
\draw[myred,densely dashed] (p2) to[out=120,in=-90] (psi2);
\draw[myred,densely dashed] (P2) to[out=  0,in=-120] (5.70,3.15);
\draw[gray,densely dashed] (7.1,2.05) to[out=90,in=-90] (7.425,2.55);
\end{tikzpicture}

%% file: figure_overview.tex

\begin{figure*}[t!]
\centering
\pgfdeclarelayer{bg}
\pgfdeclarelayer{circbox}
\pgfsetlayers{bg,circbox,main}
\begin{adjustbox}{width=\textwidth}
\begin{tikzpicture}[thick]
\tikzstyle{olOne} = [fill=blue!70!white]
\tikzstyle{olTwo} = [fill=blue!55!white]
\tikzstyle{olThr} = [fill=blue!40!white]
\tikzstyle{olFou} = [fill=blue!20!white]	
\tikzstyle{olFiv} = [fill=blue!10!white]
\tikzstyle{olSix} = [fill=blue!5!white]
\tikzstyle{olSev} = [fill=blue!2!white]
\tikzstyle{olOne} = [fill=myblue]
\tikzstyle{olTwo} = [fill=myred]
\tikzstyle{olThr} = [fill=myorange]
\tikzstyle{olFou} = [fill=mypurple]	
\tikzstyle{olFiv} = [fill=mygreen]
\tikzstyle{olSix} = [fill=mylightblue]
\tikzstyle{olSev} = [fill=mydarkred]

\tikzstyle{operator} = [thick,draw,fill=white,minimum size=1.5em] 
\tikzstyle{measurement} = [thick,draw,rounded rectangle, rounded rectangle west arc=0pt, fill=white,minimum size=1.5em] 
\tikzstyle{control} = [draw,fill,shape=circle,minimum size=5pt,inner sep=0pt]
    
\newcommand{\xos}{0}
\newcommand{\yos}{0}
	
\matrix[row sep=0.3cm, column sep=0.2cm] (circuit) at (0,0) {
\node (q1) {$\ket{0}$}; & 
\node[operator] (H11) {${\rm H}$}; &
\node[control] (C) {}; &
\node[measurement] (P13) {$X$}; &[-0.2cm]
\coordinate (end1); \\
\node (q2) {$\ket{\boldsymbol{\phi}_0}$}; &
&
\node[operator,olOne] (U) {\color{white}$e^{-i H \Delta t}$}; &
&
\coordinate (end2);\\
};
\begin{pgfonlayer}{circbox}
\node[fill=white,thick,draw=black,rounded corners=2mm] (background) [fit = (q1) (U) (end2)] {};
\draw[thick] (q1) -- (end1)  (q2) -- (end2) (C) -- (U);
\end{pgfonlayer}
        
\node[above = 0.15 of circuit](overlap) {$\Re s(\Delta t) = \Re \bra{\boldsymbol{\phi}_0} e^{-i H \Delta t}\ket{\boldsymbol{\phi}_0}$};

\renewcommand{\yos}{-3}
    
\matrix[row sep=0.3cm, column sep=0.2cm] (circuit2) at (0,0+\yos) {
\node (q1) {$\ket{0}$}; &
\node[thick,operator] (H11) {${\rm H}$}; &
\node[control] (C) {}; &
\node[thick,measurement] (P13) {$X$};
&[-0.2cm]
\coordinate (end1); \\
\node (q2) {$\ket{\boldsymbol{\phi}_0}$}; &
&
\node[thick,operator,olSev] (U) {\color{white}$e^{-i H 7\Delta t}$}; &
&
\coordinate (end2);\\
};
\begin{pgfonlayer}{circbox}
\node[fill=white,thick,draw=black,rounded corners=2mm] (background) [fit = (q1) (U) (end2)] {};
\draw[thick] (q1) -- (end1)  (q2) -- (end2) (C) -- (U);
\end{pgfonlayer}
    
\node[above = 0.15 of circuit2](overlap2) {$\Re s(7\Delta t) = \Re \bra{\boldsymbol{\phi}_0} e^{-i H 7\Delta t}\ket{\boldsymbol{\phi}_0}$};
    
\node[below = -0.15 of circuit] (vd1) {\vdots};
\begin{pgfonlayer}{bg}
\node[fill=blue!5!white,thin,draw=black,rounded corners=2mm,fit=(overlap)(circuit)(vd1)(circuit2)(overlap2),draw] (qbox) {};
\end{pgfonlayer}
\node[above = 0.3 of qbox] {\textbf{Quantum Measurements}};
    
\renewcommand{\xos}{3} 
\renewcommand{\yos}{-2.0} 
\filldraw [olOne,draw=white] (\xos+0,\yos+3) rectangle (\xos+0.5,\yos+3.5);
\filldraw [olTwo,draw=white] (\xos+0,\yos+2.5) rectangle (\xos+0.5,\yos+3.0);
\filldraw [olThr,draw=white] (\xos+0,\yos+2.0) rectangle (\xos+0.5,\yos+2.5);
\filldraw [olFou,draw=white] (\xos+0,\yos+1.5) rectangle (\xos+0.5,\yos+2.0);
\filldraw [olFiv,draw=white] (\xos+0,\yos+1.0) rectangle (\xos+0.5,\yos+1.5);
\filldraw [olSix,draw=white] (\xos+0,\yos+0.5) rectangle (\xos+0.5,\yos+1.0);
\filldraw [olSev,draw=white] (\xos+0,\yos+0.0) rectangle (\xos+0.5,\yos+0.5);
\draw (\xos+0.5,\yos+3.25) node[right] {\footnotesize$o(t_1)$};
\draw (\xos+0.5,\yos+2.75) node[right] {\footnotesize$o(t_2)$};
\draw (\xos+0.5,\yos+2.25) node[right] {\footnotesize$o(t_3)$};
\draw (\xos+0.5,\yos+1.75) node[right] {\footnotesize$o(t_4)$};
\draw (\xos+0.5,\yos+1.25) node[right] {\footnotesize$o(t_5)$};
\draw (\xos+0.5,\yos+0.75) node[right] {\footnotesize$o(t_6)$};
\draw (\xos+0.5,\yos+0.25) node[right] {\footnotesize$o(t_7)$};
\draw [thick] (\xos,\yos) rectangle (\xos+0.5,\yos+3.5) node[] (dvec){};

\draw[very thick,color=cyan!70,latex-] (\xos+2.00,\yos+1.75) -- (\xos+1.35,\yos+1.75);
\draw[very thick, color=cyan!70, latex-] (\xos-0.05,\yos+3.25) -- (overlap.east);
\draw[very thick, color=cyan!70, latex-] (\xos-0.05,\yos+0.25) -- (overlap2.east);

\newcommand{\xosh}{\xos+2.0}
\newcommand{\yosh}{\yos+1.0} 
	
\filldraw [olOne,draw=white] (\xosh+0,\yosh+2) rectangle (\xosh+0.5,\yosh+2.5);
\filldraw [olTwo,draw=white] (\xosh+0.5,\yosh+2) rectangle (\xosh+1,\yosh+2.5);
\filldraw [olTwo,draw=white] (\xosh,\yosh+1.5) rectangle (\xosh+0.5,\yosh+2);
\filldraw [olThr,draw=white] (\xosh,\yosh+1) rectangle (\xosh+0.5,\yosh+1.5);
\filldraw [olThr,draw=white] (\xosh+0.5,\yosh+1.5) rectangle (\xosh+1,\yosh+2);
\filldraw [olThr,draw=white] (\xosh+1,\yosh+2) rectangle (\xosh+1.5,\yosh+2.5);
\filldraw [olFou,draw=white] (\xosh+0.5,\yosh+1) rectangle (\xosh+1,\yosh+1.5);
\filldraw [olFou,draw=white] (\xosh+1,\yosh+1.5) rectangle (\xosh+1.5,\yosh+2);
\filldraw [olFou,draw=white] (\xosh+1.5,\yosh+2) rectangle (\xosh+2,\yosh+2.5);
\filldraw [olFiv,draw=white] (\xosh+1,\yosh+1) rectangle (\xosh+1.5,\yosh+1.5);
\filldraw [olFiv,draw=white] (\xosh+1.5,\yosh+1.5) rectangle (\xosh+2,\yosh+2);
\filldraw [olSix,draw=white] (\xosh+1.5,\yosh+1) rectangle (\xosh+2,\yosh+1.5);
	
\draw [thick] (\xosh,\yosh+1) rectangle (\xosh+2,\yosh+2.5) node[] (h1) {};
    
\node at (\xosh+1,\yosh+2.75) (x) {\footnotesize\color{black}$\mathbf{X}\in\mathbb{R}^{d\times (K+1)}$};    
    
\newcommand{\xoshh}{\xosh+0.5}
\newcommand{\yoshh}{\yosh-1.55} 
	
\filldraw [olTwo,draw=white] (\xoshh,\yoshh+1.5) rectangle (\xoshh+0.5,\yoshh+2);
\filldraw [olThr,draw=white] (\xoshh,\yoshh+1) rectangle (\xoshh+0.5,\yoshh+1.5);
\filldraw [olThr,draw=white] (\xoshh+0.5,\yoshh+1.5) rectangle (\xoshh+1,\yoshh+2);
\filldraw [olFou,draw=white] (\xoshh,\yoshh+0.5) rectangle (\xoshh+0.5,\yoshh+1);
\filldraw [olFou,draw=white] (\xoshh+0.5,\yoshh+1) rectangle (\xoshh+1,\yoshh+1.5);
\filldraw [olFou,draw=white] (\xoshh+1,\yoshh+1.5) rectangle (\xoshh+1.5,\yoshh+2);
\filldraw [olFiv,draw=white] (\xoshh+0.5,\yoshh+0.5) rectangle (\xoshh+1,\yoshh+1);
\filldraw [olFiv,draw=white] (\xoshh+1,\yoshh+1) rectangle (\xoshh+1.5,\yoshh+1.5);
\filldraw [olFiv,draw=white] (\xoshh+1.5,\yoshh+1.5) rectangle (\xoshh+2,\yoshh+2);
\filldraw [olSix,draw=white] (\xoshh+1,\yoshh+0.5) rectangle (\xoshh+1.5,\yoshh+1);
\filldraw [olSix,draw=white] (\xoshh+1.5,\yoshh+1) rectangle (\xoshh+2,\yoshh+1.5);
\filldraw [olSev,draw=white] (\xoshh+1.5,\yoshh+0.5) rectangle (\xoshh+2,\yoshh+1);	
	
\draw [thick] (\xoshh,\yoshh+0.5) rectangle (\xoshh+2,\yoshh+2) node[] (h2){};
    
\node at (\xoshh+1,\yoshh+2.25) {\footnotesize\color{black}$\mathbf{X}^{\prime}\in\mathbb{R}^{d\times (K+1)}$};      
    
\renewcommand{\xos}{8.25}
\renewcommand{\yos}{0}
\filldraw [fill=blue!90!white,draw=white] (\xos+0.5,\yos+1.0) rectangle (\xos+1.0,\yos+1.5);
\filldraw [fill=blue!90!white,draw=white] (\xos+1.0,\yos+0.5) rectangle (\xos+1.5,\yos+1.0);
\filldraw [fill=blue!30!white,draw=white] (\xos+0.0,\yos+0.0) rectangle (\xos+0.5,\yos+0.5);
\filldraw [fill=blue!15!white,draw=white] (\xos+0.5,\yos+0.0) rectangle (\xos+1.0,\yos+0.5);
\filldraw [fill=blue!60!white,draw=white] (\xos+1.0,\yos+0.0) rectangle (\xos+1.5,\yos+0.5);

\draw [thick] (\xos,\yos) rectangle (\xos+1.5,\yos+1.5) node[pos=.5] (dmd) {};
\node at (\xos+0.75,\yos+1.75) {\footnotesize\color{black}$A\in\mathbb{R}^{d\times d}$};
\node[right = 0.625 of dmd] (eq) {$=$};
\draw [thick,fill=blue!5!white] (\xos+2,\yos) rectangle (\xos+3.5,\yos+1.5) node[pos=.5] (v) {$V$};
\draw [thick] (\xos+3.65,\yos) rectangle (\xos+5.25,\yos+1.5);
\node[below right] at (\xos+3.60,\yos+1.55) {\footnotesize$\tilde{\lambda}_0$};
\node              at (\xos+4.30,\yos+0.85) {$\ddots$};
\node[above left]  at (\xos+5.35,\yos-0.05) {\footnotesize$\tilde{\lambda}_{d-1}$};
\draw [thick,fill=blue!5!white] (\xos+5.4,\yos) rectangle (\xos+6.9,\yos+1.5) node[pos=.5] (v) {$V^{-1}$};

\draw[very thick,color=cyan!70,latex-] (8.025,+0.75) -- (7.275,+0.75);
\draw[very thick,color=cyan!70,latex-] (9,-0.2) |- (9,-1.32) -- (7.75,-1.32);
\draw[very thick,color=cyan!70,latex-] (12.7,-0.75) -- (12.7,-0.2);

\renewcommand{\xos}{12.7}
\renewcommand{\yos}{-2.5}
\newcommand{\scale}{1.25} 
	
\filldraw[fill=gray!10!white] (0+\xos,0+\yos) -- ({\scale*cos(135)+\xos},{\scale*sin(135)+\yos}) arc (135:225:\scale*1) -- cycle;
\draw[very thick,domain=135:225,gray] plot ({\scale*cos(\x)+\xos}, {\scale*sin(\x)+\yos});
\draw[very thick,domain=-135:135]      plot ({\scale*cos(\x)+\xos}, {\scale*sin(\x)+\yos});
\draw[->] (-1.6+\xos,0+\yos) -- (1.6+\xos,0+\yos);
\draw[->] (0+\xos,-1.6+\yos) -- (0+\xos,1.6+\yos);
\draw[blue!55!white,thick] (0+\xos,0+\yos) -- ({\scale*1.25*cos( 101)+\xos},{\scale*1.25*sin( 101)+\yos});
\draw[blue!55!white,thick] (0+\xos,0+\yos) -- ({\scale*1.25*cos(125)+\xos},{\scale*1.25*sin(125)+\yos});
\draw[blue!55!white,thick,densely dotted] ({\scale*cos(113)+\xos},{\scale*sin(113)+\yos}) circle ({\scale*sin(12)});
\node at ({\scale*cos(113)+\xos},{\scale*sin(113)+\yos}) {\textbullet};
\node[above right] at ({\scale*cos(113)+\xos-0.3},{\scale*sin(113)+\yos+0.05}) {$\lambda_0$};
\node at ({\scale*cos(93)+\xos},{\scale*sin(93)+\yos}) {\textbullet};
\foreach \x in {-116,-102,-93,-84,-70,-60,-45,-30,-20,-5,3,17,25,35,45,60,71}
\node at ({\scale*cos(\x)+\xos},{\scale*sin(\x)+\yos}) {\textbullet};
\node[below right] at ({\scale*cos(-116)+\xos-0.5},{\scale*sin(-116)+\yos}) {$\lambda_{N-1}$};
\node at ({\scale*1.1*cos(12)*cos(125)+\xos},{\scale*1.1*cos(12)*sin(125)+\yos}) {\textcolor{blue!55!white}{\scriptsize\textbullet}};
\node at ({\scale*1.3*cos(125)+\xos-0.1},{\scale*1.3*sin(125)+\yos-0.05}) {\textcolor{blue!55!white}{$\ritz_0$}};

\renewcommand{\xos}{2.5}    
\node at (\xos+7, 2.24) {\textbf{ODMD for Eigenenergy Estimation}};

\end{tikzpicture}
\end{adjustbox}
\caption{ODMD applied to eigenenergy estimation.  ODMD uses the real (or imaginary) part of the expectation value, $s(k\Delta t) = \bra{\boldsymbol{\phi}_0} e^{-i H k\Delta t}\ket{\boldsymbol{\phi}_0}$, of time evolution with respect to a reference state $\ket{\boldsymbol{\phi}_0}$. This data can be measured efficiently on a quantum processor, for example by means of the Hadamard test. ODMD constructs a pair of Hankel matrices $\mathbf{X},\mathbf{X}^{\prime}$ using Takens' embedding (matrix elements of the same color are equal) and generates the DMD system matrix $A$, which has a companion structure. Under mild assumptions on $\Delta t$, the eigenvalue $\ritz_0$ of $A$ with $\text{arg}(\ritz_0) = \max_{\ell} \text{arg}(\ritz_\ell) $ converges to the true ground state phase $\lambda_0$ as the size of $\mathbf{X},\mathbf{X}^{\prime}$ increases. The ground state energy is estimated as $\Tilde{E}_0 = -\text{arg}(\ritz_0)/\Delta t$.\label{fig:overview}}
\end{figure*}

%% file: figure_aliasing.tex

\pgfdeclarelayer{bg}
\pgfdeclarelayer{circbox}
\pgfsetlayers{bg,circbox,main}
\begin{tikzpicture}[thick]
\tikzstyle{olOne} = [fill=blue!70!white]
\tikzstyle{olTwo} = [fill=blue!55!white]
\tikzstyle{olThr} = [fill=blue!40!white]
\tikzstyle{olFou} = [fill=blue!20!white]	
\tikzstyle{olFiv} = [fill=blue!10!white]
\tikzstyle{olSix} = [fill=blue!5!white]
\tikzstyle{olSev} = [fill=blue!2!white]
\tikzstyle{olOne} = [fill=myblue]
\tikzstyle{olTwo} = [fill=myred]
\tikzstyle{olThr} = [fill=myorange]
\tikzstyle{olFou} = [fill=mypurple]	
\tikzstyle{olFiv} = [fill=mygreen]
\tikzstyle{olSix} = [fill=mylightblue]
\tikzstyle{olSev} = [fill=mydarkred]
    
\newcommand{\xos}{0}
\newcommand{\yos}{0}
\newcommand{\lambdaphase}{0}
\newcommand{\scaledE}{0}
\newcommand{\scale}{1.25} 

 
\draw[->] (0+\xos,-1.6+\yos) -- (0+\xos,1.6+\yos);
\draw[thick,-] (-0.2+\xos, 0 + \yos) -- (0.2+\xos, 0 + \yos);
\node[above] at (\xos, 1.6+\yos) {$E$};
\foreach \E in {-167, -147, -134, -121, -101, -86, -65, -43, -29, -7, 4, 24, 36, 50, 65, 86, 102, 134, 163}
    \draw[-] (0+\xos, \yos + \E*1.6/180) -- (0.2+\xos, \yos + \E*1.6/180);

\renewcommand{\xos}{\scale + 0.7}
	
\filldraw[fill=gray!10!white] (0+\xos,0+\yos) -- ({\scale*cos(170)+\xos},{\scale*sin(170)+\yos}) arc (170:190:\scale*1) -- cycle;
\draw[very thick,domain=170:190,gray] plot ({\scale*cos(\x)+\xos}, {\scale*sin(\x)+\yos});
\draw[very thick,domain=-170:170]      plot ({\scale*cos(\x)+\xos}, {\scale*sin(\x)+\yos});

\draw[->] (-1.6+\xos,0+\yos) -- (1.6+\xos,0+\yos);
\draw[->] (0+\xos,-1.6+\yos) -- (0+\xos,1.6+\yos);

\foreach \x in {-167, -147, -134, -121, -101, -86, -65, -43, -29, -7, 4, 24, 36, 50, 65, 86, 102, 134, 163}
    \node at ({\scale*cos(-\x)+\xos},{\scale*sin(-\x)+\yos}) {\textbullet};
\node[right] at ({\scale*cos(-163)+\xos},{\scale*sin(-163)+\yos-0.05}) {$\lambda_{N-1}$};
\node[right] at ({\scale*cos(167)+\xos},{\scale*sin(167)+ \yos+0.12}) {$\lambda_0$};


\renewcommand{\xos}{3*\scale + 0.7}
 
\draw[->] (0+\xos,-1.6+\yos) -- (0+\xos,1.6+\yos);
\draw[thick,-] (-0.2+\xos, 0 + \yos) -- (0.2+\xos, 0 + \yos);
\node[above] at (\xos, 1.6+\yos) {$E$};
\foreach \E in {6, 17, 23, 30, 40, 47, 58, 68, 76, 86, 92, 102, 108, 115, 122, 133, 141, 157, 171}
    \def\scaledE{(\E+180)/2}
    \draw[-] (0+\xos, \yos + \E*1.6/180) -- (0.2+\xos, \yos + \E*1.6/180);

\renewcommand{\xos}{4*\scale + 1.4}
	
\filldraw[fill=gray!10!white] (0+\xos,0+\yos) -- ({\scale*cos(0)+\xos},{\scale*sin(0)+\yos}) arc (0:20:\scale*1) -- cycle;
\draw[very thick,domain=0:20,gray] plot ({\scale*cos(\x)+\xos}, {\scale*sin(\x)+\yos});
\draw[very thick,domain=0:-340]      plot ({\scale*cos(\x)+\xos}, {\scale*sin(\x)+\yos});

\draw[->] (-1.6+\xos,0+\yos) -- (1.6+\xos,0+\yos);
\draw[->] (0+\xos,-1.6+\yos) -- (0+\xos,1.6+\yos);

\foreach \x in {-167, -147, -134, -121, -101, -86, -65, -43, -29, -7, 4, 24, 36, 50, 65, 86, 102, 134, 163}
    \renewcommand{\scaledE}{(-\x-180)/1.05}
    \node at ({\scale*cos(\scaledE)+\xos},{\scale*sin(\scaledE)+\yos}) {\textbullet};
\node[right] at ({\scale*cos(-5)+\xos},{\scale*sin(-5)+\yos-0.15}) {$\lambda_0$};
\node[above right] at ({\scale*cos(20)+\xos-0.01},{\scale*sin((20)+\yos+0.45}) {$\lambda_{N-1}$};

\end{tikzpicture}

%% file: figure_Heisenberg_numerics.tex

\begin{figure*}[hbtp]
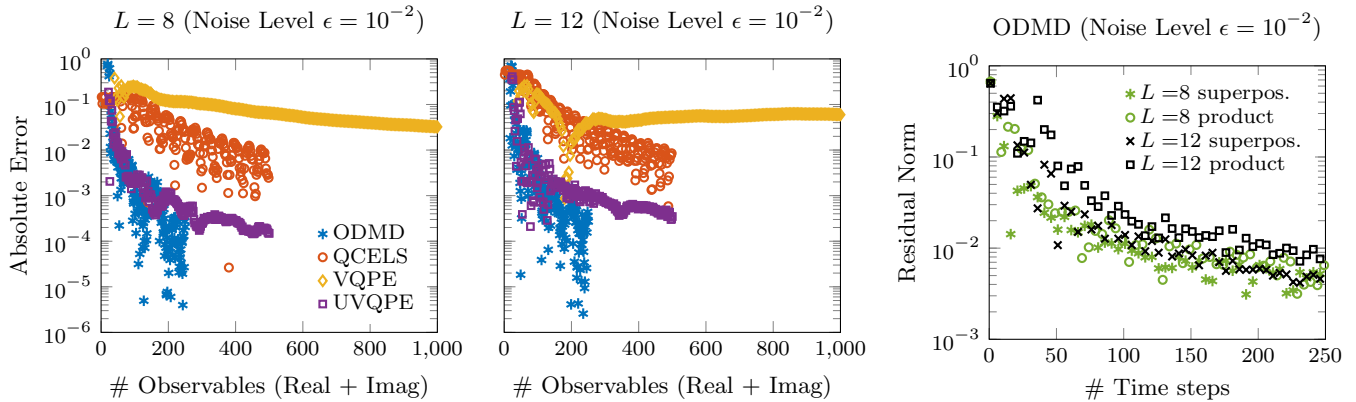

\plotconv[%
width=0.70\columnwidth,%
xlabel={\# Observables (Real + Imag)},%
xmax=1000,%
ytick={1e-6,1e-5,1e-4,1e-3,1e-2,1e-1,1e0},%
title={$L=8$ (Noise Level $\epsilon = 10^{-2}$)},%
  legend pos=south east,%
]{\addplot[myColOne,thick,only marks,mark=asterisk]%
table[x index=0,y index=1] {\datfile{Heis_8_lo-n0.01-CS-dmd}};
\addplot[myColTwo,thick,only marks,mark=o,mark size=1.5]%
  table[x index=0,y index=1] {\datfile{qcels_Heis_8_lo_eps0.01}};
\addplot[myColThr,thick,only marks,mark=diamond]%
table[x index=0,y index=1] {\datfile{Heis_8_lo-n0.01-CS-vqpe}};
\addplot[myColFou,thick,only marks,mark=square,mark size=1.25]%
table[x index=0,y index=1] {\datfile{Heis_8_lo-n0.01-CS-uvqpe}};
\legend{ODMD,QCELS,VQPE,UVQPE};
}{0}{1000}{1e-6}{1e0}%
\hfill%
\plotconv[%
  width=0.70\columnwidth,%
  xlabel={\# Observables (Real + Imag)},%
  ylabel={},%
  xmax=1000,%
  ytick={1e-6,1e-5,1e-4,1e-3,1e-2,1e-1,1e0},%
  yticklabels={},%
  title={$L=12$ (Noise Level $\epsilon = 10^{-2}$)},%
]{
\addplot[myColOne,thick,only marks,mark=asterisk]%
  table[x index=0,y index=1] {\datfile{Heis_12-n0.01-CS-dmd}};
\addplot[myColTwo,thick,only marks,mark=o,mark size=1.5]%
  table[x index=0,y index=1] {\datfile{qcels_Heis_12_eps0.01}};
\addplot[myColThr,thick,only marks,mark=diamond]%
  table[x index=0,y index=1] {\datfile{Heis_12-n0.01-CS-vqpe}};
\addplot[myColFou,thick,only marks,mark=square,mark size=1.25]%
  table[x index=0,y index=1] {\datfile{Heis_12-n0.01-CS-uvqpe}};
}{0}{1000}{1e-6}{1e0}%
\hfill%
\plotconv[%
  width=0.70\columnwidth,%
  xlabel={\# Time steps},%
  ylabel={Residual Norm},%
  xmin=0,xmax=250,%
  xtick={0,50,...,250},%
  ytick={1e-3,1e-2,1e-1,1e0},%
  title={ODMD (Noise Level $\epsilon = 10^{-2}$)},%
]{
\addplot[myColFiv,thick,only marks,mark repeat=5,mark=asterisk]%
  table[x index=0,y index=1] {\datfile{Heisenberg_residual_norm}};
\addplot[myColFiv,thick,only marks,mark repeat=5,mark=o,mark size=1.5]%
  table[x index=0,y index=2] {\datfile{Heisenberg_residual_norm}};
\addplot[black,thick,only marks,mark repeat=5,mark=x]%
  table[x index=0,y index=3] {\datfile{Heisenberg_residual_norm}};
\addplot[black,thick,only marks,mark repeat=5,mark=square,mark size=1.25]%
  table[x index=0,y index=4] {\datfile{Heisenberg_residual_norm}};
\legend{$L=$8 superpos., $L=$8 product, $L=$12 superpos., $L=$12 product};%
}{0}{250}{1e-3}{1e0}%
\caption{Convergence of ODMD compared with state-of-the-art hybrid algorithms based on real-time evolution (ODMD, QCELS, VQPE, UVQPE) for extracting ground state information of Heisenberg spin chain from overlap observables, $\langle \boldsymbol{\phi}_0|e^{-iHk\Delta t}O|\boldsymbol{\phi}_0\rangle$ with $O = I$ or $H$. The maximum number of time steps shown is the same for each algorithm; however, the maximum number of observables can differ, for example because we count real and imaginary parts separately and ODMD measures only the real part.  \textbf{Left}: Convergence of the ground state energy for $L=8$. For the reference state $\ket{\boldsymbol{\phi}_0}$ used for generating the overlaps, we use the product state $|01010101\rangle$. The noise level $\epsilon$, singular value threshold $\tilde{\delta}$ (for ODMD, VQPE, UVQPE), reference-ground overlap probability $p_0 = \lvert \braket{\psi_0|\boldsymbol{\phi}_0} \rvert^2$, and timestep $\Delta t$ are $(10^{-2},10^{-1},0.16,0.15)$.  \textbf{Center}: Convergence of the ground state energy for $L=12$ with $(\epsilon, \tilde{\delta}, p_0, \Dt) = (10^{-2},10^{-1},0.15,0.1)$. For the initial reference state, we use the superposition of two product states specified in \cref{subsec:spin_sys}. \textbf{Right}: Convergence of ground state eigenvector found with ODMD, quantified by the residual norm.  We show the convergence for four cases: $L=8$ with product and superposition initial states (16\% and 32\% overlaps, respectively), and $L=12$ with product and superposition initial states (7\% and 15\%). }
\label{fig:Heisenberg_convergence}
\end{figure*}

%% file: figure_molecule_comparison.tex

\begin{figure*}[t!]
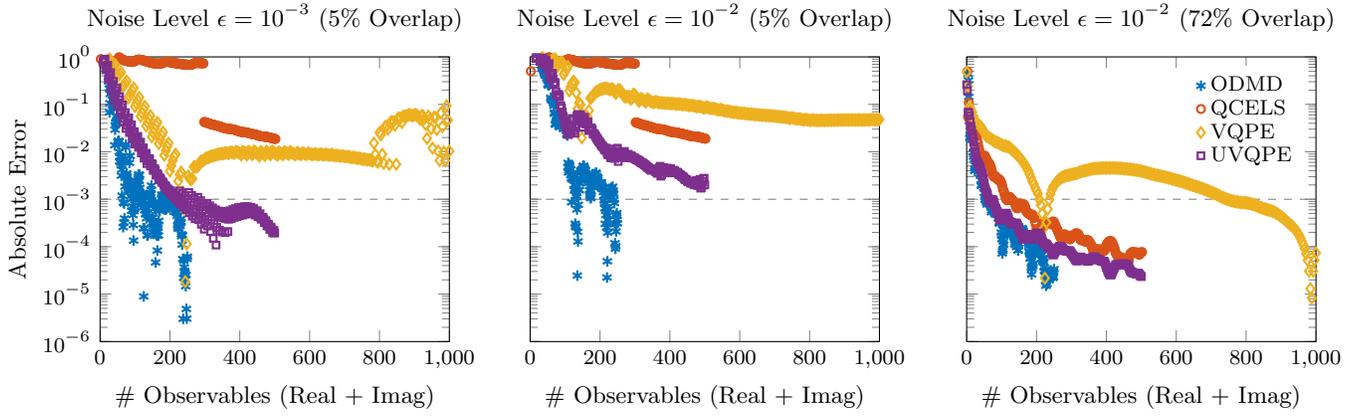

\plotconv[%
  width=0.7\columnwidth,%
  xlabel={\# Observables (Real + Imag)},%
  xmax=1000,%
  ytick={1e-6,1e-5,1e-4,1e-3,1e-2,1e-1,1e0},%
  title={Noise Level $\epsilon = 10^{-3}$ (5\% Overlap)},%
]{%
\addplot[gray,dashed] coordinates { (0,1e-3) (1000,1e-3) };
\addplot[myColOne,thick,only marks,mark=asterisk]%
  table[x index=0,y index=1] {\datfile{Cr2-n0.001-o5-dmd}};
\addplot[myColTwo,thick,only marks,mark=o,mark size=1.5]%
  table[x index=0,y index=1] {\datfile{qcelsCr2_o0.05_eps0.001}};
\addplot[myColThr,thick,only marks,mark=diamond]%
  table[x index=0,y index=1] {\datfile{vqpeCr2_o0.05_eps0.001}};
\addplot[myColFou,thick,only marks,mark=square,mark size=1.25]%
  table[x index=0,y index=1] {\datfile{uvqpeCr2_o0.05_eps0.001}};
}{0}{1000}{1e-6}{1e0}%
%
%
\plotconv[%
  width=0.7\columnwidth,%
  xlabel={\# Observables (Real + Imag)},%
  ylabel={},%
  xmax=1000,%
  ytick={1e-6,1e-5,1e-4,1e-3,1e-2,1e-1,1e0},%
  yticklabels={},%
  title={Noise Level $\epsilon = 10^{-2}$ (5\% Overlap)},%
]{%
\addplot[gray,dashed] coordinates { (0,1e-3) (1000,1e-3) };
\addplot[myColOne,thick,only marks,mark=asterisk]%
  table[x index=0,y index=1] {\datfile{Cr2-n0.01-o5-dmd}};
\addplot[myColTwo,thick,only marks,mark=o,mark size=1.5]%
  table[x index=0,y index=1] {\datfile{qcelsCr2_o0.05_eps0.01}};
\addplot[myColThr,thick,only marks,mark=diamond]%
  table[x index=0,y index=1] {\datfile{vqpeCr2_o0.05_eps0.01}};
\addplot[myColFou,thick,only marks,mark=square,mark size=1.25]%
  table[x index=0,y index=1] {\datfile{uvqpeCr2_o0.05_eps0.01}};
}{0}{1000}{1e-6}{1e0}%
%
%
\plotconv[%
  width=0.7\columnwidth,%
  xlabel={\# Observables (Real + Imag)},%
  ylabel={},%
  xmax=1000,%
  ytick={1e-6,1e-5,1e-4,1e-3,1e-2,1e-1,1e0},%
  yticklabels={},%
  title={Noise Level $\epsilon = 10^{-2}$ (72\% Overlap)},%
]{%
\addplot[gray,dashed] coordinates { (0,1e-3) (1000,1e-3) };
\addplot[myColOne,thick,only marks,mark=asterisk]%
  table[x index=0,y index=1] {\datfile{Cr2-n0.01-HF-dmd}};
\addplot[myColTwo,thick,only marks,mark=o,mark size=1.5]%
  table[x index=0,y index=1] {\datfile{qcelsCr2_HF_eps0.01}};
\addplot[myColThr,thick,only marks,mark=diamond]%
  table[x index=0,y index=1] {\datfile{vqpeCr2_HF_eps0.01}};
\addplot[myColFou,thick,only marks,mark=square,mark size=1.25]%
  table[x index=0,y index=1] {\datfile{uvqpeCr2_HF_eps0.01}};
\legend{,ODMD,QCELS,VQPE,UVQPE};%
}{0}{1000}{1e-6}{1e0}%

\caption{Absolute error in the Cr$_2$ ground state energy computed from four state-of-the-art hybrid algorithms based on real-time evolution (ODMD, QCELS, VQPE, UVQPE) as a function of the number of observables, $\langle \boldsymbol{\phi}_0|e^{-iHk\Delta t}O|\boldsymbol{\phi}_0\rangle$ with $O = I$ or $H$,  measured.  The maximum number of timesteps shown is the same for each algorithm; the maximum number of observables can differ, for example because we count real and imaginary parts separately and ODMD uses only the real part. The three panels exhibit convergence for different values $(\epsilon, \tilde{\delta}, p_0, \Delta t)$ of the noise level $\epsilon$, singular value threshold $\tilde{\delta}$ (for ODMD, VQPE, UVQPE), reference-ground overlap probabilities $p_0 = \lvert \braket{\psi_0|\boldsymbol{\phi}_0} \rvert^2$, and timestep $\Delta t$. \textbf{Left}: $(10^{-3}, 10^{-2}, 5\%, 1)$; \textbf{Center}: $(10^{-2}, 10^{-1}, 5\%, 1)$; and \textbf{Right}: $(10^{-2}, 10^{-1}, 72\%, 3)$ (for a Hartree-Fock reference state). The horizontal dashed line indicates chemical accuracy.}
\label{fig:compare_methods}
\end{figure*}

%% file: figure_molecule_convergence.tex

\begin{figure*}[hbtp]
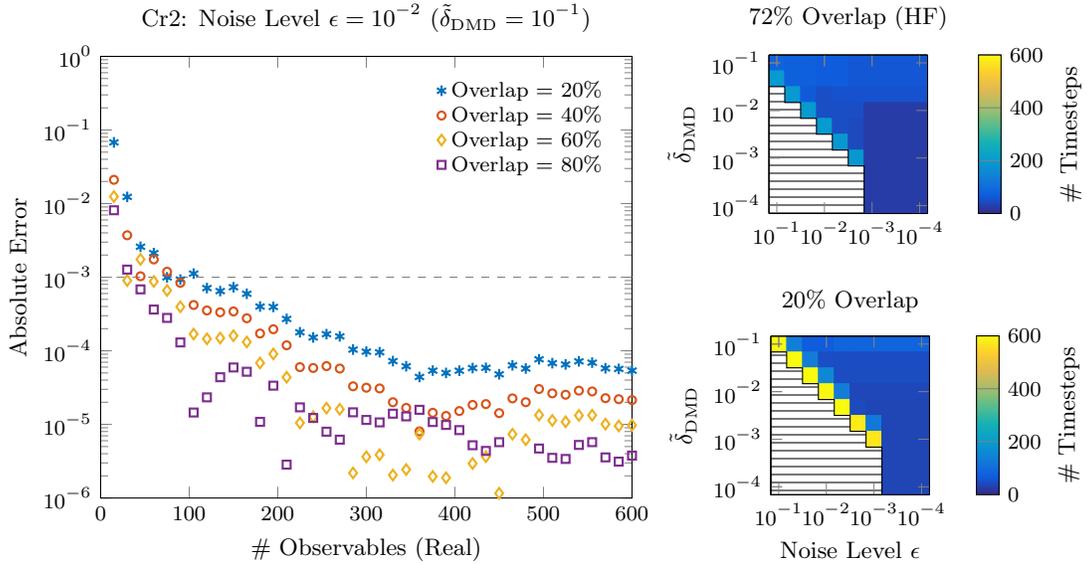

\centering
\begin{minipage}{1\columnwidth}
\plotconv[%
  xtick={0,100,...,600},%
  ytick={1e-6,1e-5,1e-4,1e-3,1e-2,1e-1,1e0},%
  title={Cr2: Noise Level $\epsilon = 10^{-2}$ ($\tilde{\delta}_{\rm DMD} = 10^{-1}$)},%
]{%
\addplot[gray,dashed] coordinates { (0,1e-3) (1000,1e-3) };
\addplot[myColOne,thick,only marks,mark=asterisk]%
   table[x index=0,y index=1] {\datfile{Cr2-n0.01-o20-dmd}};
\addplot[myColTwo,thick,only marks,mark=o,mark size=1.5]%
   table[x index=0,y index=1] {\datfile{Cr2-n0.01-o40-dmd}};
\addplot[myColThr,thick,only marks,mark=diamond]%
   table[x index=0,y index=1] {\datfile{Cr2-n0.01-o60-dmd}};
\addplot[myColFou,thick,only marks,mark=square,mark size=1.5]%
   table[x index=0,y index=1] {\datfile{Cr2-n0.01-o80-dmd}};
\legend{,Overlap = 20\%,Overlap = 40\%,Overlap = 60\%,Overlap = 80\%};%
}{0}{600}{1e-6}{1e0}%
\end{minipage}
\begin{minipage}{0.66 \columnwidth}
\plotheatmap[colorbar,colorbar style={ylabel={\# Timesteps}},xlabel={}, colorbar style={width=0.3 cm}]{72\% Overlap (HF)}{%
\addplot3[matrix plot] coordinates {
 (0,0, 81) (0,1,195) (0,2,  0) (0,3,  0) (0,4,  0) (0,5,  0) (0,6,  0) (0,7,  0) (0,8,  0) (0,9,  0)

 (1,0, 75) (1,1, 75) (1,2,195) (1,3,  0) (1,4,  0) (1,5,  0) (1,6,  0) (1,7,  0) (1,8,  0) (1,9,  0)

 (2,0, 75) (2,1, 75) (2,2, 64) (2,3,195) (2,4,  0) (2,5,  0) (2,6,  0) (2,7,  0) (2,8,  0) (2,9,  0)

 (3,0, 69) (3,1, 69) (3,2, 48) (3,3, 48) (3,4,198) (3,5,  0) (3,6,  0) (3,7,  0) (3,8,  0) (3,9,  0)

 (4,0, 69) (4,1, 69) (4,2, 54) (4,3, 48) (4,4, 48) (4,5,189) (4,6,  0) (4,7,  0) (4,8,  0) (4,9,  0)

 (5,0, 60) (5,1, 60) (5,2, 54) (5,3, 48) (5,4, 42) (5,5, 42) (5,6, 192) (5,7, 0) (5,8, 0) (5,9, 0)

 (6,0, 60) (6,1, 60) (6,2, 54) (6,3, 24) (6,4, 24) (6,5, 24) (6,6, 24) (6,7, 24) (6,8, 24) (6,9, 24)

 (7,0, 60) (7,1, 60) (7,2, 54) (7,3, 24) (7,4, 24) (7,5, 24) (7,6, 24) (7,7, 24) (7,8, 24) (7,9, 24)

 (8,0, 60) (8,1, 60) (8,2, 54) (8,3, 24) (8,4, 24) (8,5, 24) (8,6, 24) (8,7, 24) (8,8, 24) (8,9, 24)

 (9,0, 60) (9,1, 60) (9,2, 54) (9,3, 24) (9,4, 24) (9,5, 24) (9,6, 24) (9,7, 24) (9,8, 24) (9,9, 24)
};
\fill[white] (-0.5,1.5) -- (0.5,1.5) -- (0.5,2.5) -- (1.5,2.5) -- (1.5,3.5) -- (2.5,3.5) -- (2.5,4.5) -- (3.5,4.5) -- (3.5,5.5) -- (4.5,5.5) -- (4.5,6.5) -- (5.5,6.5) -- (5.5,9.5) -- (-0.5,9.5) -- cycle;
\draw[pattern={Lines[angle=45]}] (-0.5,1.5) -- (0.5,1.5) -- (0.5,2.5) -- (1.5,2.5) -- (1.5,3.5) -- (2.5,3.5) -- (2.5,4.5) -- (3.5,4.5) -- (3.5,5.5) -- (4.5,5.5) -- (4.5,6.5) -- (5.5,6.5) -- (5.5,9.5) -- (-0.5,9.5) -- cycle;%
}

\vspace{10pt}

\plotheatmap[colorbar,colorbar style={ylabel={\# Timesteps}},colorbar style={width=0.3 cm}]{20\% Overlap}{%
\addplot3[matrix plot] coordinates {
 (0,0, 630) (0,1,  0) (0,2,  0) (0,3,  0) (0,4,  0) (0,5,  0) (0,6,  0) (0,7,  0) (0,8,  0) (0,9,  0)

 (1,0, 135) (1,1, 654) (1,2,  0) (1,3,  0) (1,4,  0) (1,5,  0) (1,6,  0) (1,7,  0) (1,8,  0) (1,9,  0)

 (2,0, 69) (2,1, 132) (2,2, 606) (2,3, 0) (2,4,  0) (2,5,  0) (2,6,  0) (2,7, 0) (2,8,  0) (2,9,  0)

 (3,0, 69) (3,1, 48) (3,2, 156) (3,3, 594) (3,4,  0) (3,5,  0) (3,6,  0) (3,7,  0) (3,8,  0) (3,9,  0)

 (4,0, 84) (4,1, 45) (4,2, 45) (4,3, 132) (4,4,612) (4,5,  0) (4,6,  0) (4,7,  0) (4,8,  0) (4,9,  0)

 (5,0, 84) (5,1, 45) (5,2, 45) (5,3, 39) (5,4, 42) (5,5, 576) (5,6,  0) (5,7,  0) (5,8,  0) (5,9,  0)

 (6,0, 84) (6,1, 45) (6,2, 45) (6,3, 39) (6,4, 36) (6,5, 123) (6,6, 570) (6,7,  0) (6,8,  0) (6,9,  0)

 (7,0, 84) (7,1, 45) (7,2, 45) (7,3, 39) (7,4, 36) (7,5, 36) (7,6, 36) (7,7, 36) (7,8, 36) (7,9, 36)

 (8,0, 84) (8,1, 45) (8,2, 45) (8,3, 39) (8,4, 36) (8,5, 36) (8,6, 36) (8,7, 36) (8,8, 36) (8,9, 36)

 (9,0, 84) (9,1, 45) (9,2, 45) (9,3, 39) (9,4, 36) (9,5, 36) (9,6, 36) (9,7, 36) (9,8, 36) (9,9, 36)
};
\fill[white] (-0.5,0.5) -- (0.5,0.5) -- (0.5,1.5) -- (1.5,1.5) -- (1.5,2.5) -- (2.5,2.5) -- (2.5,3.5) -- (3.5,3.5) -- (3.5,4.5) -- (4.5,4.5) -- (4.5,5.5) -- (5.5,5.5) -- (5.5,6.5) -- (6.5,6.5) -- (6.5,9.5) -- (-0.5,9.5) -- cycle;
\draw[pattern={Lines[angle=45]}] (-0.5,0.5) -- (0.5,0.5) -- (0.5,1.5) -- (1.5,1.5) -- (1.5,2.5) -- (2.5,2.5) -- (2.5,3.5) -- (3.5,3.5) -- (3.5,4.5) -- (4.5,4.5) -- (4.5,5.5) -- (5.5,5.5) -- (5.5,6.5) -- (6.5,6.5) -- (6.5,9.5) -- (-0.5,9.5) -- cycle;%
}
\end{minipage}
\caption{\textbf{Left}: Convergence of the ODMD estimate of the ground state energy for Cr$_2$, with overlap measurements given by the exact result $s(t)$ plus Gaussian noise $\mathcal{N}(0,\epsilon^2)$ with standard deviation $\epsilon=10^{-2}$. A timestep of $\Delta t =3$ is used for generating the results. The energy converges rapidly even when the reference state $|\boldsymbol{\phi}_0\rangle$ has only 20\% overlap probability with the true ground state. In solving the least-squares problem, we filter singular values below the threshold $\tilde{\delta}_\text{DMD}=10^{-1}$. \textbf{Right}: The number of DMD timesteps needed to reach chemical accuracy (horizontal dashed line) as a function of the noise level (horizontal axis) and DMD truncation (vertical axis).  We consider ground state probabilities of 72\% (Hartree-Fock reference state) and 20\%. In the black-dotted regions, absolute error remains above chemical accuracy after 750 timesteps.
\label{fig:convergence}}
\end{figure*}

%% file: figure_TFIM_numerics.tex
\begin{figure*}[htp!]
\centering
\plotconv[%
  xtick={0,600,...,36000},%
  ytick={1e-6,1e-5,1e-4,1e-3,1e-2,1e-1,1e0},%
  xlabel={\# Total Observables},%
  title={TFIM: Noise Level $\epsilon = 10^{-6}$ ($\tilde{\delta}_{\rm DMD} = 10^{-5}$)},%
]{%
\addplot[myColOne,thick,only marks,mark=asterisk]%
   table[x index=0,y index=1] {\datfile{TFIM_multi_dmd}};
\addplot[myColTwo,thick,only marks,mark=o,mark size=1.5]%
   table[x index=0,y index=1] {\datfile{TFIM_multi_uvqpe}};
\legend{ODMD (Multi-observable), UVQPE (Multi-reference)};%
}{0}{3600}{1e-7}{1e0}%
\caption{ Absolute error in the TFIM first excited state energy computed using two real-time algorithms (ODMD and UVQPE) as a function of the total number of observables, $\langle \boldsymbol{\phi}_0|O_i e^{-iHk\Delta t}|\boldsymbol{\phi}_0\rangle$ (with $O_1 = I$ for ODMD) and $\langle \boldsymbol{\phi}_0^{(i)}|e^{-iHk\Delta t}|\boldsymbol{\phi}_0^{(i')}\rangle$ (with $\ket{\boldsymbol{\phi}_0^{(1)}} = \ket{\boldsymbol{\phi}_0}$ for UVQPE), measured. The maximum number of timesteps shown is the same for each algorithm; the maximum number of observables can differ, due to ODMD's observable-based formulation versus UVPQE's state-based formulation. The noise level $\epsilon$, singular value threshold $\tilde{\delta}$, and timestep $\Delta t$ are $(\epsilon, \tilde{\delta}, \Delta t) = (10^{-6}, 10^{-5}, 1)$. We start with the reference state $\ket{\boldsymbol{\phi}_0} = (3 \ket{+}^{\otimes L} + 2 \ket{-}^{\otimes L})/\sqrt{13}$. For ODMD we use the additional observables $O_2 = \sigma^{z}_i$ and $O_3 = \sigma^{y}_j$ with random sites $i$ and $j$ while for multi-reference UVQPE we use the corresponding additional reference states $\ket{\boldsymbol{\phi}_0^{(2)}} = O_2 \ket{\boldsymbol{\phi}_0}$ and $\ket{\boldsymbol{\phi}_0^{(3)}} = O_3 \ket{\boldsymbol{\phi}_0}$.}
\label{fig:excited_state}
\end{figure*}

%% file: appendix.tex
\newpage
\section{ODMD under general noise model}
\label{app:sec:noise}

The depolarizing noise model discussed within \cref{subsec:practical_performance} can be extended to more general error models, including Pauli noise, where the channel acts as $\Lambda_t[\rho] = e^{-\mathcal{L}t}[\rho]$ for some Lindbladian,
\begin{align}
    \mathcal{L}[\rho;\{\zeta_i\}] &= \mathcal{L}_{\rm U}[\rho;H] + \mathcal{L}_{\rm D}[\rho;\{\zeta_i\}], \\
    &= \frac{i}{2} [(I_{2} - Z) \otimes H,\rho] - \sum_{i} \zeta_i \left( M_i \rho M_i^{\dagger} - \frac{1}{2} \{M_i^{\dagger} M_i,  \rho \}   \right),
\end{align}
defined by a set of jump operators $\{ M_i \}_{i \geq 1}$ with jumping rates $\{ \zeta_i \}_{i \geq 1}$. The generators $\mathcal{L}_{\rm U}$ and $\mathcal{L}_{\rm D}$ govern the unitary and dissipative components of the dynamics, respectively. For simplicity, we examine the special case of a single Pauli jump operator $M_1 = I_2 \otimes M_{1,{\rm sys}}$ where $M_{1, {\rm sys}} =  M_{1, {\rm sys}}^{\dagger}$ commutes with $H$. The Lindbladian determines the evolution of the input state as, 
\begin{align}
    \Lambda_t[\rho] = \frac{1+e^{-2\zeta_1 t}}{2} \mathcal{U}_t \rho \mathcal{U}_t^{\dagger} + \frac{1-e^{-2\zeta_1 t}}{2} \mathcal{U}_t \tilde{\rho} \mathcal{U}_t^{\dagger},
\end{align}
where $\tilde{\rho}= \ket{+}\bra{+} \otimes \ket{\tilde{\boldsymbol{\phi}}_0} \bra{\tilde{\boldsymbol{\phi}}_0}$ represents the noise-corrupted reference state with $\ket{\tilde{\boldsymbol{\phi}}_0} = M_{1, {\rm sys}} \ket{\boldsymbol{\phi}_0}$. In this case, the overlap becomes $\frac{1+e^{-2\zeta_1 t}}{2} \Re \braket{\boldsymbol{\phi}_0|e^{-iHt}|\boldsymbol{\phi}_0} + \frac{1-e^{-2\zeta_1 t}}{2} \Re \braket{\tilde{\boldsymbol{\phi}}_0|e^{-iHt}|\tilde{\boldsymbol{\phi}}_0}$, indicating that the ODMD eigenphases still provide an approximation to the exact eigenenergies. For general jump operator(s), we may instead consider measuring extended observables of the form,
\begin{align}
     o_j(t) = \frac{1}{2} {\rm Tr}\left[(X \otimes O_j) \Lambda_{t}[\rho]\right],
\end{align}
for some pool of operators $\{O_j\}_{j\geq 1}$ chosen to efficiently probe the eigenstructure of the Lindbladian $\mathcal{L}$. Suitable choices of $\{O_j\}_{j \geq 1}$ can be guided by the invariant subspaces of the Pauli transfer matrix associated with the Lindbladian~\cite{greenbaum2015introductionquantumgateset}, which in itself presents a theoretically inspiring direction for future investigation.

\newpage
\section{Generalized Extension to Prony's Method}
\label{app:sec:prony}

In this appendix, we include a comprehensive proof of the convergence bound from the main text. 
Prony's method~\cite{prony1795essai} is a pioneering classic approach that has laid the foundation for many modern signal processing and recovery techniques. To understand its underlying idea, let us consider the task of extracting principal frequency components from temporally uniform samples,
\begin{equation}
    s_k = \sum_{n=0}^{N-1} p_n e^{-iE_n k \Dt} = \sum_{n=0}^{N-1} p_n \lambda_n^k = \left( \sum_{\Omega} + \sum_{ \Omega^{c} } \right) p_n \lambda_n^k,
    \label{eq:sk}
\end{equation}
where $p_n = |\langle \psi_n | \phi_0 \rangle|^2$ so that $\sum_{n} p_n = 1$, $\Omega \subseteq [N] = \{0,1,\ldots, N-1\}$ denotes a set of $|\Omega| \ll N$ eigenindices satisfying $\sum_{n \in \Omega} p_n \gg \sum_{n \in \Omega^c} p_n$, and $\Omega^{c} = [N] \setminus \Omega$ denotes the complement of $\Omega$. For simplicity, let us assume $\Omega = [d]$. Recall that the eigenvalues are the zeros of the characteristic polynomial
\begin{equation}
    \cC_{\Omega}(z) = \prod_{\ell \in \Omega} (z - \lambda_\ell)
                    = \sum_{\ell=0}^d c_\ell z^\ell,
    \label{eq:charac_poly}
\end{equation}
where $\{ c_\ell \}_{\ell=0}^d$ are a set of  monomial coefficients that depend on $\{\lambda_{\ell} \}_{\ell \in \Omega}$ with the convention $c_d \equiv 1$.
Next, observe that evaluating \cref{eq:charac_poly} at the samples \eqref{eq:sk}, \textit{i.e.}, $\cC_\Omega(z = s_k)$ for $k \geq 0$, yields
\begin{align}
    \sum_{\ell=0}^d c_{\ell} s_{\ell + k}
    &= \sum_{\ell=0}^d c_{\ell} \sum_{n=0}^{N-1} p_n \lambda_n^{\ell + k}, \\
    &= \sum_{n=0}^{N-1} p_n \lambda_n^k \sum_{\ell=0}^d c_{\ell} \lambda_n^\ell
     = \sum_{n=0}^{N-1} p_n \lambda_n^k \cC_\Omega(\lambda_n), \\
    &= \sum_{n=d}^{N-1} p_n \lambda_n^k \cC_\Omega(\lambda_n)
    := \res_{k,\Omega},
\end{align}
where the sum $\sum_{n < d}$ vanishes since $\cC_{\Omega}(\lambda_n) \equiv 0$ for $n < d$.
By using the triangle inequality, the residual term $\res_{k,\Omega}$ can be bounded uniformly from above
\begin{equation}
    \abs{\res_{k,\Omega}} \leq \sum_{n \in \Omega^c} p_n \abs{\cC_\Omega(\lambda_n)} \leq \max_{n \in \Omega^c} \abs{\cC_\Omega(\lambda_n)} \sum_{n \in \Omega^c} p_n  \leq \normone{\Vec{c}\,} \sum_{n \in \Omega^c} p_n,
    \label{eq:r_kOmega}
\end{equation}
so
\begin{equation}
\sum_{\ell=0}^{d-1} c_\ell s_{\ell+k} = - s_{d+k} + \res_{k,\Omega} \approx -s_{d+k},
\end{equation}
when $\normone{\Vec{c}\,} \sum_{n \in \Omega^c} p_n \ll 1$. In matrix form, the evolution of $\{s_k \}_{d \leq k \leq K+d}$ for some $K \inN$ therefore follows
\begin{equation}
    \begin{bmatrix}
        s_0 & s_1 & & \cdots & & s_{d-1} \\
        s_1 & s_2 & & \cdots & & s_{d}  \\
        \vdots & \vdots & & \ddots & & \vdots \\
        s_K & s_{K+1} & & \cdots & & s_{K+d-1} \\
    \end{bmatrix} 
    \begin{bmatrix}
        c_0 \\
        c_1 \\
        \vdots \\
        c_{d-1} \\
    \end{bmatrix} = -
    \begin{bmatrix}
        s_{d} \\
        s_{d+1}\\
        \vdots \\
        s_{K+d} \\
    \end{bmatrix} +
    \begin{bmatrix}
        \res_{0,\Omega} \\
        \res_{1,\Omega}\\
        \vdots \\
        \res_{K,\Omega} \\
    \end{bmatrix},
    \label{eq:Prony_DMD}
\end{equation}
which describes the linear time-invariant dynamics,
\begin{equation}
     \underbrace{\begin{bmatrix}
        s_1 & s_2 & \cdots & s_{K+1} \\
        s_2 & s_3 & \cdots & s_{K+2}  \\
        \vdots & \vdots & \ddots & \vdots \\
        s_{d} & s_{d+1} & \cdots & s_{K+d} \\
    \end{bmatrix}}_{\mathbf{X}'} \approx
    \underbrace{\left[
    \begin{array}{c|ccc}
    \begin{matrix}
        0 \\ \vdots \\ 0
    \end{matrix} & \eye[d-1] \\
    \hline
    -c_0 & \begin{matrix}
        -c_1 & \cdots & -c_{d-1}
    \end{matrix}
    \end{array}
    \right]  }_{A}
    \underbrace{\begin{bmatrix}
        s_0 & s_1 & \cdots & s_K \\
        s_1 & s_2 & \cdots & s_{K+1}  \\
        \vdots & \vdots & \ddots & \vdots \\
        s_{d-1} & s_{d} & \cdots & s_{K+d-1} \\
    \end{bmatrix}}_{\mathbf{X}},
    \label{eq:prony_recur}
\end{equation}
where $\mathbf{X},\mathbf{X}' \in \mathbb{C}^{d \times (K+1)}$ are time-delayed Hankel matrices, $A \in \mathbb{C}^{d \times d}$ is a system matrix of companion structure, and $\eye[d-1] \in \mathbb{C}^{(d-1) \times (d-1)}$ denotes the identity matrix. For interested readers, the connection between the characteristic polynomial $\cC_\Omega$ and the associated linear recurrence relation above are explicitly discussed in Lemma 2.2 from Ref~\cite{POTTS20131024}. Prony's method essentially solves for the monomial coefficients $\Vec{c}$ and thus the eigenfrequencies $\{ \lambda_n \}_{n \in \Omega}$ as the roots of $\cC_\Omega$ when \cref{eq:Prony_DMD,eq:prony_recur} hold exact for $N=d$ (so that $r_{k,\Omega} \equiv 0$).

Here, we consider the dynamic mode decomposition (DMD) approach formulated in \cref{eq:dmd_lls} as a powerful extension of Prony's method when $d \ll N$, which solves the linear least-squares (LLSQ) problems of the same type, \textit{i.e.}, $\mathbf{X}' = A\mathbf{X}$, where $A = \argmin_{\Tilde{A} \in \mathbb{C}^{d \times d}} \norm{\mathbf{X}' - \Tilde{A} \mathbf{X}}_{F}^2$. In other words, the DMD post-processing identifies the set of monomial coefficients $\Vec{a} \approx \Vec{c}$ that minimize the total residual.
Let
\begin{equation}
    \cC_{\rm DMD}(z) = \prod_{\ell=0}^{d-1} (z-\ritz_\ell) = \sum_{\ell=0}^{d} a_{\ell} z^\ell,
\end{equation}
represent the characteristic polynomial corresponding to the system matrix $A$, with $a_d \equiv 1$.
Then, we define the DMD residual as follows
\begin{equation}
    \resritz_{k, \rm DMD} = \sum_{n=0}^{N-1} p_n \lambda_n^k \cC_{\rm DMD}(\lambda_n),
    \label{eq:DMD_residual}
\end{equation}
$p_n$ and $\lambda_n$ are given in \cref{eq:sk}.
Note that in case $\ritz_n \approx \lambda_n$, for $n = 0,1,\ldots,d-1$, the contribution of the $d$ first terms in \cref{eq:DMD_residual} will remain minimal. In fact, we may now show that the eigenvalues $\{ \ritz_\ell \}_{\ell=0}^{d-1}$ of the DMD system matrix $A$ well approximate the target eigenvalues $\{ \lambda_n \}_{n \in \Omega}$ for some $\Omega$. 

Let $\cP_d$ be the set of polynomials of at most degree $d$ and $\cP'_d \subset \cP_d$ the set of monic polynomials, \textit{i.e.}, $\cP'_d = \{ \cC \in \cP_d: \cC- z^d \in \cP_{d-1} \}$. For any polynomial $\cC \in \cP_d$, the triangle inequality implies
\begin{equation}
    \left|\sum_{n=0}^{N-1} p_n \lambda_n^k \cC(\lambda_n)\right| \leq \sum_{n=0}^{N-1} p_n \abs{\cC(\lambda_n)},
\end{equation}
which allows us to establish a uniform upper bound on the DMD residual
\begin{align}
    \sum_{k=0}^{K} \abs{\resritz_{k, \rm DMD}}^2
    &= \min_{\cC \in \cP'_d} \sum_{k=0}^{K} \left|\sum_{n=0}^{N-1} p_n \lambda_n^k \cC(\lambda_n)\right|^2, \\
    &\leq (K+1) \min_{\cC \in \cP'_d} \left( \sum_{n=0}^{N-1} p_n \abs{\cC(\lambda_n)} \right)^2, \\
    &\leq (K+1) \min_{\cC \in \cP'_d} \max_{0 \leq n \leq N-1} \abs{\cC(\lambda_n)}^2.
    \label{eq:DMD_polyapprox}
\end{align}

We now proceed by saturating a tight bound on the DMD residual through the min-max inequality above. Given our freedom to pick the time increment $\Dt$ such that $(E_{N-1} - E_0) \Dt = 2\Theta \leq 2\pi$, we further bound the RHS of \cref{eq:DMD_polyapprox} via polynomials over the circular arc $\cA_{\Theta} = \{ z \inC: \abs{z}=1, {\rm -\Theta \leq arg}(z) \leq \Theta  \} $. An example of such circular arc with $\Theta = \pi/2$ is sketched in \cref{fig:circle} below.

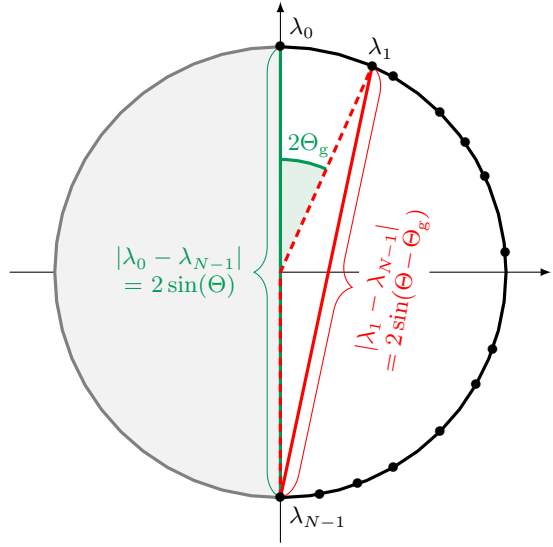
\begin{figure}[hbtp]
\centering
\begin{tikzpicture}[scale=3]
\draw[->] (-1.2,0) -- (1.2,0);
\draw[->] (0,-1.2) -- (0,1.2);
\fill[gray!10!white] (0,1) arc (90:270:1);
\draw[very thick,domain=90:270,gray] plot ({cos(\x)}, {sin(\x)});
\draw[very thick,domain=-90:90]      plot ({cos(\x)}, {sin(\x)});
\fill[ForestGreen!10!white] (0,0) -- (0,0.5) arc (90:66:0.5);
\draw[ForestGreen,very thick] (0,-1) -- (0,1);
\draw[ForestGreen,very thick] (0,0.5) arc (90:66:0.5) node[above,midway,xshift=0.5ex,yshift=-0.5ex] {\textcolor{ForestGreen}{$2\Theta_{\rm g}$}};
\draw[ForestGreen,decorate,decoration={brace,amplitude=10pt}] (0,-1) -- (0,1) node[left,midway,xshift=-1em,text width=1.75cm] {\textcolor{ForestGreen}{$\abs{\lambda_0 - \lambda_{N-1}}$ $~= 2\sin(\Theta)$}};
\draw[red,very thick,densely dashed] (0,-1) -- (0,0) -- ({cos(66)},{sin(66)});
\draw[red,very thick] (0,-1) -- ({cos(66)},{sin(66)});
\draw[red,decorate,decoration={brace,amplitude=10pt}] ({cos(66)},{sin(66)}) -- (0,-1) node[below,midway,xshift=1.25em,text width=2.2cm,fill=white,rotate=80] {\textcolor{red}{$~~\abs{\lambda_1 - \lambda_{N-1}}$ $= 2\sin(\Theta - \Theta_{\rm g})$}};
\node at (0,1) {\textbullet};
\node[above right] at (0,1) {$\lambda_0$};
\node at ({cos(66)},{sin(66)}) {\textbullet};
\node[above] at ({1.01*cos(64)},{1.01*sin(64)}) {$\lambda_1$};
\foreach \x in {-80,-70,-60,-45,-30,-20,5,25,35,45,60}
  \node at ({cos(\x)},{sin(\x)}) {\textbullet};
\node at (0,-1) {\textbullet};
\node[below right] at (0,-1) {$\lambda_{N-1}$};
\end{tikzpicture}
\caption{ Characteristic polynomial $\cC_{\Omega}$ with roots along the arc $\mathcal{A}_{\Theta = \pi/2}$ and $0 \in \Omega$. The eigenphases $\{ \lambda_n \}_{n=0}^{N-1}$ are represented schematically by black dots on the unit circle. Geometrically, $\Vert \Tilde{\cC}_{\Omega \setminus \{0\} } \Vert_{\Theta - \Theta_{\rm g}}$ is bounded by the solid red chord of length $|\lambda_1 - \lambda_{N-1}| = 2\sin{(\Theta-\Theta_{\rm g})}$, whereas $\Vert \cC_{\Omega} \Vert_{\Theta}$ is bounded by the green solid chord of length $|\lambda_0 - \lambda_{N-1}| = 2\sin{(\Theta)}$. Our strategy for selecting appropriate reference eigenindex set $\Omega$ is aimed at minimizing $\Vert \Tilde{\cC}_{\Omega \setminus \{0\} } \Vert_{\Theta - \Theta_{\rm g}}/ \abs{\partial_z \cC_{\Omega}(\lambda_0)}$, where $\abs{\partial_z \cC_{\Omega}(\lambda_0)} \approx \Vert \cC_{\Omega} \Vert_{\Theta}^{d-1}$ for $\{\lambda_{n_{\ell}} \}_{\ell = 1}^{d-1}$ far away from $\lambda_0$. Note that $\Tilde{\cC}_{\Omega \setminus \{0\}}$ undergoes a global phase shift of $e^{-i\Theta_{\rm g}}$ relative to $\cC_{\Omega \setminus \{0\}}$, which center the arc $\mathcal{A}_{\Theta-\Theta_{\rm g}}$ around $z=1$.}
\label{fig:circle}
\end{figure}

In particular, let us consider the choice of complex-valued Chebyshev polynomials,
\begin{align}
    \hat{\cC}(z)
    = \argmin_{\cC \in \cP'_d} \sup_{z \in \cA_{\Theta}} \abs{\cC(z)},
\end{align}
whose parametric representations can be explicitly constructed from tools such as Jacobi’s elliptic and theta functions~\cite{akhiezer1928functionen}. For notational convenience, we define $\norm{\cC(z)}_{\Theta} = \sup_{z \in \cA_{\Theta}} \abs{\cC(z)}$. Similar to real-valued Chebyshev polynomials over the unit interval, $\hat{\cC}(z)$ retains the minimal norm property on $\mathcal{A}_{\Theta}$ with
\begin{align}
    \norm{\hat{\cC}}_{\Theta} \sim 2 \sin^{d} \left( \Theta/2 \right) \cos^2 \left( \Theta/4 \right),
\end{align}
asymptotically for large $d$~\cite{schiefermayr2019chebyshev}. Consequently, the difference in residuals associated with the Prony-like and DMD linear systems of the type \cref{eq:Prony_DMD} can be bounded as,
\begin{align}
    \hspace{-1 cm} \Vert \Vec{\res}_\Omega - \Vec{\resritz}_{\rm DMD} \Vert_2^2
    &= \sum_{k=0}^{K} \left| \res_{k,\Omega} - \resritz_{k,\rm DMD} \right|^2, \\
    &\leq 2 \sum_{k=0}^{K} \abs{\res_{k,\Omega}}^2
        + 2 \sum_{k=0}^{K} \abs{\resritz_{k,\rm DMD}}^2, \label{eq:20} \\
    &\leq 2 \sum_{k=0}^{K} \max_{1 \leq n \leq N-1} \abs{\cC_{\Omega}(\lambda_n)}^2
        + 2 \sum_{k=0}^{K} \abs{\resritz_{k, \rm DMD}}^2,
        \label{eq:21}
\end{align}
where we now work with a generic set $\Omega = \{ n_0=0, n_1, \cdots, n_{d-1} \}$ of $d$ eigenindices. \cref{eq:20} follows from the application of triangle and Cauchy–Schwarz inequalities $|z-z'|^2 \leq \abs{z}^2 + 2\abs{z}\abs{z'} + \abs{z'}^2 \leq 2\abs{z}^2 + 2\abs{z'}^2$ while \cref{eq:21} exploits the upper bound derived in \cref{eq:r_kOmega}. Using the minimal norm property of the DMD residual, we have
\begin{align}
    \Vert \Vec{\res}_\Omega - \Vec{\resritz}_{\rm DMD} \Vert_2^2 &\leq 2(K+1) \max_{1 \leq n \leq N-1} \abs{\cC_{\Omega}(\lambda_n)}^2 + 2(K+1) \min_{\cC \in \cP'_d} \max_{0\leq n\leq N-1} \abs{\cC(\lambda_n)}^2, \\
    &\leq 2 (K+1) \left[ \max_{1 \leq n \leq N-1} \abs{\lambda_n - \lambda_0}^2
                         \max_{1 \leq n \leq N-1} \left| \prod_{\ell = 1}^{d-1} (\lambda_n - \lambda_{n_{\ell}}) \right|^2
                       + \norm{\hat{\cC}}_{\Theta}^2 \right], \\
    &\leq 2 (K+1) \left[ \left( 2 \max_{1 \leq n \leq N-1} \left| \prod_{\ell = 1}^{d-1} (\lambda_n - \lambda_{n_{\ell}}) \right| \right)^2
                       + \norm{\hat{\cC}}_{\Theta}^2 \right], \\
    &\leq 2(K+1) \left[ 2 \norm{\Tilde{\cC}_{\Omega\setminus\{0\}}}_{\Theta - \Theta_{\rm g}}
                       + \norm{\hat{\cC}}_{\Theta}
                       \right]^2,
\end{align}
where $\Theta_{\rm g} = (E_1 - E_0)\Dt/2$ represents the gap between the dimensionless ground state and first excited state angle. It is worth noting that we can always assume $\Theta_{\rm g} > 0$ since otherwise $p_0 \lambda_0^k + p_1 \lambda_1^k = (p_0+p_1) \lambda_0^k \implies N \mapsto N-1$ and $\{ E_{n} \}_{n = 1}^{N-1} \mapsto \{ E_{n-1} \}_{n = 1}^{N-1}$. For notational consistency, we let $\Tilde{\cC}_{\Omega \setminus \{0\}}(z) = \prod_{\ell =1}^{d} (e^{-i \Theta_g} z - \lambda_{n_\ell}) \in \cP_{d-1}$ denote the polynomial of degree $(d-1)$ with a  proper phase shift included. Since $\Vert \hat{\cC} \Vert_{\Theta}$ is optimally bounded, next we show that $\Vert \Tilde{\cC}_{\Omega \setminus \{0\}} \Vert_{\Theta-\Theta_{\rm g}}$ can also be bounded in a controlled way. Specifically, we claim that there exists $(\Theta_{0}, \Theta_1) \subseteq [0, \pi]$ so that $\forall \Theta \in [\Theta_0,  \Theta_1],$ $\Vert \Tilde{\cC}_{\Omega \setminus \{0\}} \Vert_{\Theta-\Theta_{\rm g}} \ll 2^{d-1}$ asymptotically for a collection of index sets: for concreteness, we examine the example $\Theta = \pi/2$ and $\Omega \setminus \{0\} = \{ n_{\ell} \}_{\ell=1}^{d-1}$ which clusters around the antipodal point of $\lambda_0$, \textit{i.e.}, $\lambda_{n_\ell} \approx -\lambda_0$, $\forall \ell \neq 0$. A direct calculation shows that 
\begin{align}
    \abs{z - \lambda_{n_\ell}} =  2 \sin{ \left| \frac{{\rm arg}(z) - {\rm arg}(\lambda_{n_\ell})}{2} \right| } \implies \Vert \Tilde{\cC}_{\Omega \setminus \{0\}}\Vert_{\pi/2 -\Theta_{\rm g}} \sim 2^{d-1} \sin^{d-1}{(\pi/2 - \Theta_{\rm g})},
\end{align}
Graphically, \cref{fig:circle} helps elucidate the bounded behavior of $\cC_{\Omega}$ and thereby $\Tilde{\cC}_{\Omega \setminus \{0\}}$ in the presence of the spectral gap $\Theta_{\rm g}$.

Going forward, for the sake of illustration, let us restrict our attention to any index set $\Omega$ that admits near maximal product of distances $\prod_{\ell = 1}^{d-1} |\lambda_0 - \lambda_{n_{\ell}}|$. Recall from \cref{eq:Prony_DMD} that $\Vec{a} - \Vec{c} = \mathbf{X}^{\top} (\Vec{r}_{\Omega} - \Vec{r}_{\rm DMD})$. Thus combining our previous results, we see that the DMD coefficients $\Vec{a}$ and the target coefficients $\Vec{c}$ converge towards each other at a rate,
\begin{align}
    \Vert \Vec{a} - \Vec{c} \Vert_2 &\lesssim \Vert (\mathbf{X}^{\top})^{+} \Vert_2 \Vert \Vec{r}_{\Omega} - \Vec{r}_{\rm DMD} \Vert_2, \\
    &\leq \frac{ [2(K+1)]^{1/2} \Big\{2 \Vert \Tilde{\cC}_{\Omega \setminus \{0\}} \Vert_{\Theta-\Theta_{\rm g}} + \Vert \hat{\cC} \Vert_{\Theta} \Big\} }{\delta},
\end{align}
where recall that $\mathbf{X} \in \mathbb{C}^{d \times (K+1)}$ is the data matrix and $\delta > 0$ denotes a lower bound on its singular values. To monitor the scaling of our DMD approximation, we write $d =\lfloor \alpha (K+1) \rfloor$ for which $0 \leq \alpha \leq 1$ is kept constant throughout (here it suffices to discuss $0 \leq \alpha \leq 1$ due to the Hankel structure of our data matrix). Moreover, previous literature supports the common practice of working with $\alpha \lesssim 1$ as near-optimal choice~\cite{mpm1989,mpm1990}. The error in the ground state energy can then be estimated and analyzed through first-order perturbation theory,
\begin{align}
    \partial_{c_{\ell}} \cC_{\Omega}(\lambda_0) = 0,~ \forall \ell &\iff \lambda_0^{\ell} + \sum_{k=1}^{d}  c_{k} k \left( \lambda_0 \right)^{k-1} \partial_{c_\ell} \lambda_0 = 0,\\
    & \iff \partial_{c_\ell} \lambda_0 = - \frac{\lambda_0^{\ell}}{\partial_z \cC_{\Omega} (\lambda_0)},
\end{align}
so we have $\ritz_{0} - \lambda_{0} = \sum_{\ell=0}^{d-1} (a_{\ell} - c_{\ell}) \partial_{c_\ell} \lambda_0$ up to leading order and,

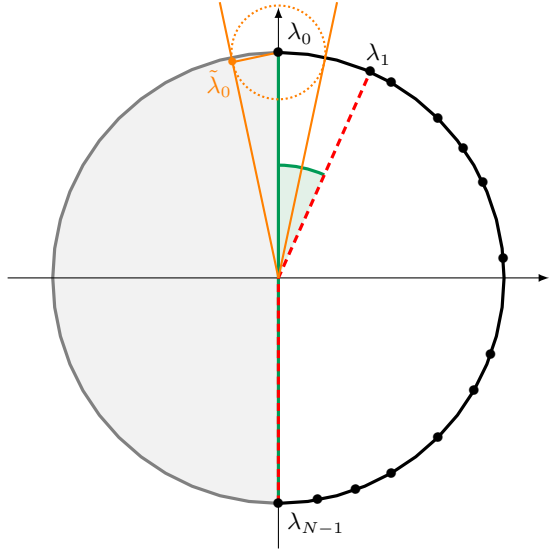
\begin{figure}[tp]
\centering
\begin{tikzpicture}[scale=3]
\fill[gray!10!white] (0,1) arc (90:270:1);
\draw[very thick,domain=90:270,gray] plot ({cos(\x)}, {sin(\x)});
\draw[very thick,domain=-90:90]      plot ({cos(\x)}, {sin(\x)});
\draw[->] (-1.2,0) -- (1.2,0);
\draw[->] (0,-1.2) -- (0,1.2);
\fill[ForestGreen!10!white] (0,0) -- (0,0.5) arc (90:66:0.5);
\draw[ForestGreen,very thick] (0,-1) -- (0,1);
\draw[ForestGreen,very thick] (0,0.5) arc (90:66:0.5);
\draw[red,very thick,densely dashed] (0,-1) -- (0,0) -- ({cos(66)},{sin(66)});
\draw[orange,thick] (0,0) -- ({1.25*cos( 78)},{1.25*sin( 78)});
\draw[orange,thick] (0,0) -- ({1.25*cos(102)},{1.25*sin(102)});
\draw[orange,thick,densely dotted] (0,1) circle ({sin(12)});
\draw[orange,thick] (0,1) -- ({cos(12)*cos(102)},{cos(12)*sin(102)});
\node at (0,1) {\textbullet};
\node[above right] at (0,1) {$\lambda_0$};
\node at ({cos(66)},{sin(66)}) {\textbullet};
\node[above] at ({1.01*cos(64)},{1.01*sin(64)}) {$\lambda_1$};
\foreach \x in {-80,-70,-60,-45,-30,-20,5,25,35,45,60}
  \node at ({cos(\x)},{sin(\x)}) {\textbullet};
\node at (0,-1) {\textbullet};
\node[below right] at (0,-1) {$\lambda_{N-1}$};
\node at ({cos(12)*cos(102)},{cos(12)*sin(102)}) {\textcolor{orange}{\scriptsize\textbullet}};
\node at ({0.9*cos(107)},{0.9*sin(107)}) {\textcolor{orange}{$\ritz_0$}};
\end{tikzpicture}
\caption{Error in the ground state energy inferred from error in the ground state eigenphase. The eigenphases $\{ \lambda_n \}_{n=0}^{N-1}$ are shown schematically as black dots on the unit circle. The dotted orange circle indicates an upper bound on the eigenphase difference $\abs{\ritz_0 - \lambda_0}$. Deviation in the angles $|\Tilde{E}_0 - E_0| \Dt$ can be determined through the construction of tangent lines to the orange circle centered at $\lambda_0$. }
\label{fig:circle_2}
\end{figure}

\begin{align}
    \abs{\ritz_{0} - \lambda_{0}} = \left| \frac{ \sum_{\ell=0}^{d-1} (a_{\ell} - c_{\ell}) \lambda_0^{\ell}  }{ \partial_z \cC_{\Omega} (\lambda_0) } \right| &\leq \frac{\Vert \Vec{a} - \Vec{c} \Vert_1}{\displaystyle \prod_{\ell \in \Omega \setminus \{0\}} \abs{\lambda_0 - \lambda_{\ell}}}, \\
    &\leq \frac{ (2\alpha)^{1/2} (K+1) \Big\{2 \Vert \Tilde{\cC}_{\Omega \setminus \{0\}} \Vert_{\Theta-\Theta_{\rm g}} + \Vert \hat{\cC} \Vert_{\Theta} \Big\} }{\displaystyle \delta \prod_{\ell \in \Omega \setminus \{0\} } 2 \abs{\sin \left( \theta_{0l} \right) } },
    \label{eq:circle_bd}
\end{align}
where we use the norm inequality $\Vert \Vec{a} - \Vec{c} \Vert_1 \leq d^{1/2} \Vert \Vec{a} - \Vec{c} \Vert_2$ and define the eigenangle variables $\theta_{j \ell} = (E_\ell - E_{j})\Dt /2$. Here we order the DMD eigenvalues $\{ \ritz_0, \ldots, \ritz_{d-1} \}$ based on their arguments in a descending order, \textit{i.e.}, ${\rm arg}{\ritz_0} \geq {\rm arg}{\ritz_1} \geq \cdots \geq {\rm arg}{\ritz_{d-1}}$, so that $\ritz_0 \approx \lambda_0$ contains the ground state information. For the case $\Theta_0 \leq \Theta \leq \pi/2$, we note that picking $\theta_{0l} \sim 2\Theta$ gives $\abs{\sin( \theta_{0l})} \sim \sin{(\Theta)}$. Therefore \cref{eq:circle_bd} can be bounded asymptotically as,
\begin{align}
    \abs{\ritz_{0} - \lambda_{0}} &\lesssim  \frac{ (2\alpha)^{1/2}(K+1) }{\displaystyle \delta 2^{d-1} \sin^{d-1}(\Theta) } \Big\{ 2 \Vert \Tilde{\cC}_{\Omega \setminus \{0\}} \Vert_{\Theta-\Theta_{\rm g}} + \Vert \hat{\cC} \Vert_{\Theta} \Big\}, \\
    &\leq \frac{2^{3/2} \alpha^{1/2} (K+1) }{\displaystyle \delta 2^{d-1} \sin^{d-1}(\Theta) } \Big\{ 2^{d-1} \sin^{d-1}{(\Theta - \Theta_{\rm g})} + \sin^{d} ( \Theta/2 ) \cos^2 ( \Theta/4 ) \Big\}, \\
    &\approx
        \frac{2^{3/2} \alpha^{1/2} K}{\delta} \left[ \frac{\sin^{\alpha}{(\Theta - \Theta_{\rm g})}}{\sin^{\alpha}(\Theta)} \right]^K 
     + \frac{2^{3/2} \alpha^{1/2} K \sin(\Theta_1/2) \cos^2{(\Theta_0/4)} }{ \delta (2^{\alpha})^K} \left[ \frac{\sin^{\alpha} ( \Theta/2 )}{\sin^{\alpha}(\Theta)} \right]^{K},
    \label{eq:exp_bd}
\end{align}
with $K \gg 1$. Since typically $\Theta > 2\Theta_{\rm g}$ for practical Hamiltonians, it follows that the leading term in \cref{eq:exp_bd} decays proportional to $K e^{-\beta_1 K}$ where $\beta_1 (\Theta, \Theta_{\rm g}) = - \alpha \ln[\sin(\Theta-\Theta_{\rm g}) /\sin(\Theta)] > 0$ and the subleading term decays proportional to $K e^{-\beta_2 K}$ where $\beta_2 (\Theta) = - \alpha \ln[\sin(\Theta/2) /2\sin(\Theta)] > \beta_1 > 0$. Finally using a simple tangent line argument as sketched in \cref{fig:circle_2}, we have for sufficiently small $\abs{\ritz_0 - \lambda_0}$,
\begin{equation}
    \abs{\rm{arg}(\ritz_0) - \rm{arg}(\lambda_0)} \leq \sin^{-1} \abs{\ritz_0 - \lambda_0} \leq \abs{\ritz_0 - \lambda_0} + \frac{\abs{\ritz_0 - \lambda_0}^3}{2},
\end{equation}
leading to an asymptotic error bound on the ground state energy,
\begin{align}
    \delta E_0 &= \frac{\abs{\rm{arg}(\ritz_0) - \rm{arg}(\lambda_0)}}{\Dt}, \\
    &\leq \frac{2^{3/2} \alpha^{1/2}}{\delta \Dt} \left[ K e^{-\beta_1 K} + \sin(\Theta_1/2) \cos^2{(\Theta_0/4)} K e^{-\beta_2 K} \right] + {\rm h.o.t},
\end{align}
which holds for an arbitrary initial state $|\boldsymbol{\phi}_0\rangle$. We remark that the error in the excited state energies can be derived in a similar manner. However, the proof requires careful selection of different index sets $\Omega$ (since we want to control $|\partial_{z}\cC_{\Omega}(z_{\ell})|$ from below in the perturbative analysis).

Our theoretical result above suggests that the DMD approach has the potential of adeptly approximating the ground state energy, provided that we have access to the overlap $\{ s_k \}_k$ whose real and imaginary parts are measured separately using a quantum computer. In practice, we can trivially replicate the preceding proof by considering only the real (or imaginary) parts of the overlaps, \textit{e.g.}, $\Re s_k = \sum_{n=0}^{N-1} p_n \cos{(k E_n \Dt)}$. By Euler's formula $\Re s_k = \sum_{n=0}^{N-1} p_n(\lambda_n^{k} + \lambda_n^{-k})/2$, we immediately see that,
\begin{equation}
    \sum_{\ell = 0}^{d} c_{\ell} s_{\ell + k} = 0 \implies \sum_{\ell = 0}^{2d} \underline{c}_{\ell} \underline{s}_{l + k} = 0,
\end{equation}
where
\begin{equation}
    \sum_{\ell=0}^{2d} \underline{c}_{\ell} z^{\ell} = \prod_{\ell=1}^{d} (z-\lambda_{n_{\ell}})(z-\lambda_{n_{\ell}}^{\ast}), 
\end{equation}
is an extended characteristic polynomial containing conjugate pairs eigenphases $(\lambda_n, \lambda_n^{\ast})$ and $\underline{S}_{k} = \sum_{n=0}^{2N-1} (p_n/2) \lambda_n^k$ is the extended overlap matrix element with $(p_n, E_n) = (p_{n-N}, - E_{n-N})$ for $N \leq n \leq 2N-1$. Hence given the data matrix $\{ \Re s_k \}_{k}$, taking $d \mapsto 2d$ or $\alpha \mapsto 2\alpha$ also effectively recovers the ground state energy in the large $K$ limit.


\newpage
\section{Connection to Pad\'e's rational approximation}
\label{app:sec:pade}

In this appendix, we present how the DMD approach can be heuristically conceptualized under the framework of Pad\'e approximation~\cite{Pade1892,Pade_overview}. 

We return to our simple assumption $\Omega = [d]$ and work with the efficient approximating sequence $\{ \Tilde{s}_k \}_{k}$ with $\Tilde{s}_{k} = \sum_{n \in \Omega} p_n \lambda_n^k \approx s_k$ (recall that $\abs{\sum_{n \in \Omega^c} p_n \lambda_n^k} \leq \sum_{n \in \Omega^c} p_n \ll 1$). Let us now consider the $Z$-transform of $\{ \Tilde{s}_k \}_{k \in \mathbb{N}}$,
\begin{align}
    \mathcal{Z}\big[ \{ \Tilde{s}_k \}\big](z) &= \sum_{k = 0}^{\infty} \Tilde{s}_k z^{-k}, \\
    &= \sum_{n=0}^{d-1} p_n \sum_{k = 0}^{\infty} \left( \frac{\lambda_n}{z} \right)^k = \sum_{n=0}^{d-1} p_n \frac{z}{z- \lambda_n},
    \label{eq:Z_trans}
\end{align}
which converges in the region $\{ z \in \mathbb{C}: \abs{z} > \limsup_{k} |\Tilde{s}_k|^{1/k} = 0 \}$. The RHS of \cref{eq:Z_trans} consists of a rational function,
\begin{align}
   \sum_{n=0}^{d-1} p_n \frac{z}{z- \lambda_n} = \sum_{n=0}^{d-1} p_n  \frac{\displaystyle z \prod\nolimits_{\ell \neq n \in \Omega}(z-\lambda_\ell)}{\displaystyle \prod\nolimits_{\ell \in \Omega} (z- \lambda_\ell)} = \frac{\cB_\Omega(z)}{\cC_\Omega(z)},
\end{align}
where we recognize that $\cC_\Omega(z) = \prod_{\ell \in \Omega}(z - \lambda_{\ell}) = z^d + \sum_{\ell=0}^{d-1} c_{\ell} z^{\ell}$ is the characteristic polynomial in \cref{eq:charac_poly} and $\cB_\Omega(z) = \sum_{n \in \Omega} p_n z \prod_{\ell \neq n \in \Omega}(z-\lambda_\ell)  \approx z^{d} + \sum_{\ell=1}^{d-1} b_{\ell} z^{\ell}$. Associated with these two polynomials is a rational approximation of the power series,
\begin{align}
    \sum_{k=0}^{\infty} \Tilde{s}_k z^{k} = \frac{\cB_\Omega(z^{-1})}{\cC_\Omega(z^{-1})} \equiv \frac{\cB_{\Omega,P}(z)}{\cC_{\Omega,P}(z)},
    \label{eq:pade_approx}
\end{align}
where $\cB_{\Omega,P} = z^d \cB_\Omega(z^{-1}) \approx 1 + \sum_{\ell=1}^{d-1} b_{d-\ell} z^{\ell}$ and $\cC_{\Omega,P} = z^d \cC_\Omega(z^{-1}) = 1 + \sum_{\ell=1}^{d} c_{d-\ell} z^{\ell}$ constitute the Pad\'e approximant. In our context, Pad\'e approximation finds the best rational approximation $(\cB_\Omega, \cC_\Omega)$ from the time-accumulated overlap data,
\begin{align}
     \int_0^{\infty} \frac{dt}{\Dt} s(t) e^{i E t} \approx \sum_{k=0}^{\infty} \Tilde{s}_k \left( e^{i E \Dt} \right)^k,
\end{align}
whose poles contain the target eigenfrequencies $\{ \lambda_n^{\ast} \}_{n \in \Omega}$ (up to a complex conjugation). Rearranging \cref{eq:pade_approx} above, we have
\begin{align}
    \sum_{k=0}^{\infty} \Tilde{s}_k z^k \cC_{\Omega,P}(z) = \cB_{\Omega,P}(z) \implies \begin{cases}
    \displaystyle \sum_{\ell = d-k}^{d} c_\ell \Tilde{s}_{\ell + k - d} = b_{d-k},  &~~ k \in [d] \\ 
    \displaystyle \sum_{\ell=0}^{d} c_\ell \Tilde{s}_{\ell+k} = 0, &~~  k \inN \\   
    \end{cases},
\end{align}
upon matching $z^k$ terms on the LHS and the RHS order by order. The first set of derived relations reveals an inherent linkage of the monomial coefficients $\Vec{c}$ obtained from the characteristic polynomial $\cC_\Omega(z)$ to the set of coefficients $\Vec{b}$ used in Pad\'e's construction, whereas the second set of relations simply reiterate the linear dynamics modeled in the DMD approach.

\newpage
\section{Ground State Energy Estimation of Molecular Hamiltonians}
\label{app:sec:ground_state_estimation}

Here we provide additional data and details for the range of molecular Hamiltonians considered in this study. We recall that a general electronic Hamiltonian takes the form,
\begin{align}
    H = \sum_{ij} h_{ij} a_i^{\dagger} a_j + \frac{1}{2} \sum_{ijk\ell} a_i^{\dagger} a_j^{\dagger} a_k a_{\ell},
\end{align}
where $a_i$ and $a_i^{\dagger}$ denote the fermionic annihilation and creation operators respectively in the second quantized picture, while $i$ and $j$ index the sets of virtual and occupied electron spin-orbitals accordingly. The electronic Hamiltonian can be systematically mapped onto its qubit representation via techniques
such as the Jordan-Wigner transformation.~\cite{JW_transform} An advantageous reference state for molecular problems is the Hartree-Fock ground state, $\ket{\boldsymbol{\phi}_0} = \ket{\boldsymbol{\phi}_{\rm HF}}$, derived from a self-consistent mean-field calculation.~\cite{HF_Seaton1977} Under the Jordan-Wigner transformation,
\begin{align}
\ket{\boldsymbol{\phi}_{\rm HF}} = \ket{0}^{\otimes (N_{\rm orb} - N_{\rm el})} \ket{1}^{\otimes N_{\rm el}},
\end{align}
where $N_{\rm orb}$ and $N_{\rm el}$ denote the number of spin-orbitals and electrons. Thus $\ket{\boldsymbol{\phi}_{\rm HF}}$ takes the simple form of a product state with each qubit indicating
the occupancy of a spin-orbital.

For ${\rm Cr}_2$ dimer (both here and in the main text), we work with the def2-SVP basis set~\cite{weigend2005balanced} (bond length 1.5 Å, 30 orbitals, and 24 electrons).
We restrict our simulation to the widely studied 30 orbital active space and truncate the Hilbert space, considering only a subset of Slater determinants that we choose using ASCI~\cite{tubman2016deterministic, tubman2020modern}. 
All data shown in the main text and here are a truncation of size 4000, selected from a one million determinant ASCI calculation where the 4000 determinants with the largest coefficients are included.
In addition to ${\rm Cr}_2$, we include LiH (3-21g basis set~\cite{binkley1980self}, bond length 1.5 Å, 11 orbitals, and 4 electrons) and ${\rm H}_6$ chain (STO-6g basis~\cite{hehre1969self}, bond length 1.5 Å, 6 orbitals, and 6 electrons).

\begin{figure}[hbtp]
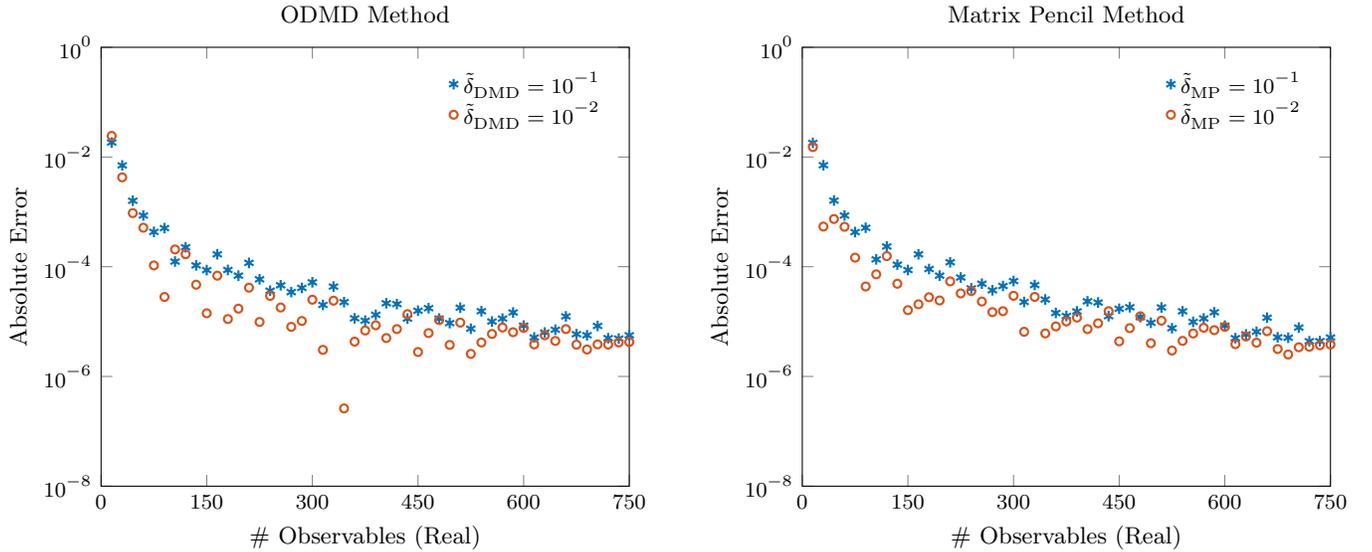

\centering
\begin{minipage}{0.48\columnwidth}
\plotconv[%
  xtick={0,150,...,750},%
  ytick={1e-10,1e-8,1e-6,1e-4,1e-2,1e0},%
  title={ODMD Method},%
]{%
\addplot[myColOne,thick,only marks,mark=asterisk]%
   table[x index=0,y index=1] {\datfile{Cr2-n0.001-HF-dmd}};
\addplot[myColTwo,thick,only marks,mark=o,mark size=1.5]%
   table[x index=0,y index=2] {\datfile{Cr2-n0.001-HF-dmd}};
\legend{$\Tilde{\delta}_{\rm DMD} = 10^{-1}$,$\Tilde{\delta}_{\rm DMD} = 10^{-2}$,$\Tilde{\delta}_{\rm DMD} = 10^{-3}$};%
}{0}{750}{1e-8}{1e0}
\end{minipage}%
\hfill%
\begin{minipage}{0.48\columnwidth}
\plotconv[%
  xtick={0,150,...,750},%
  ytick={1e-10,1e-8,1e-6,1e-4,1e-2,1e0},%
  title={Matrix Pencil Method},%
]{%
\addplot[myColOne,thick,only marks,mark=asterisk]%
   table[x index=0,y index=1] {\datfile{Cr2-n0.001-HF-mp}};
\addplot[myColTwo,thick,only marks,mark=o,mark size=1.5]%
   table[x index=0,y index=2] {\datfile{Cr2-n0.001-HF-mp}};
\legend{$\Tilde{\delta}_{\rm MP} = 10^{-1}$,$\Tilde{\delta}_{\rm MP} = 10^{-2}$,$\Tilde{\delta}_{\rm MP} = 10^{-3}$};%
}{0}{750}{1e-8}{1e0}
\end{minipage}%
\caption{Data comparing ODMD and the matrix pencil method with different singular value truncations ($\Tilde{\delta}_{\rm DMD}$ and $\Tilde{\delta}_{\rm MP}$) for ${\rm Cr}_2$ with noise level $\epsilon = 10^{-3}$. A timestep of $\Dt = 3$ is used for generating the results.}
\label{fig:odmd_mpm}
\end{figure}

\begin{figure}[hbtp]
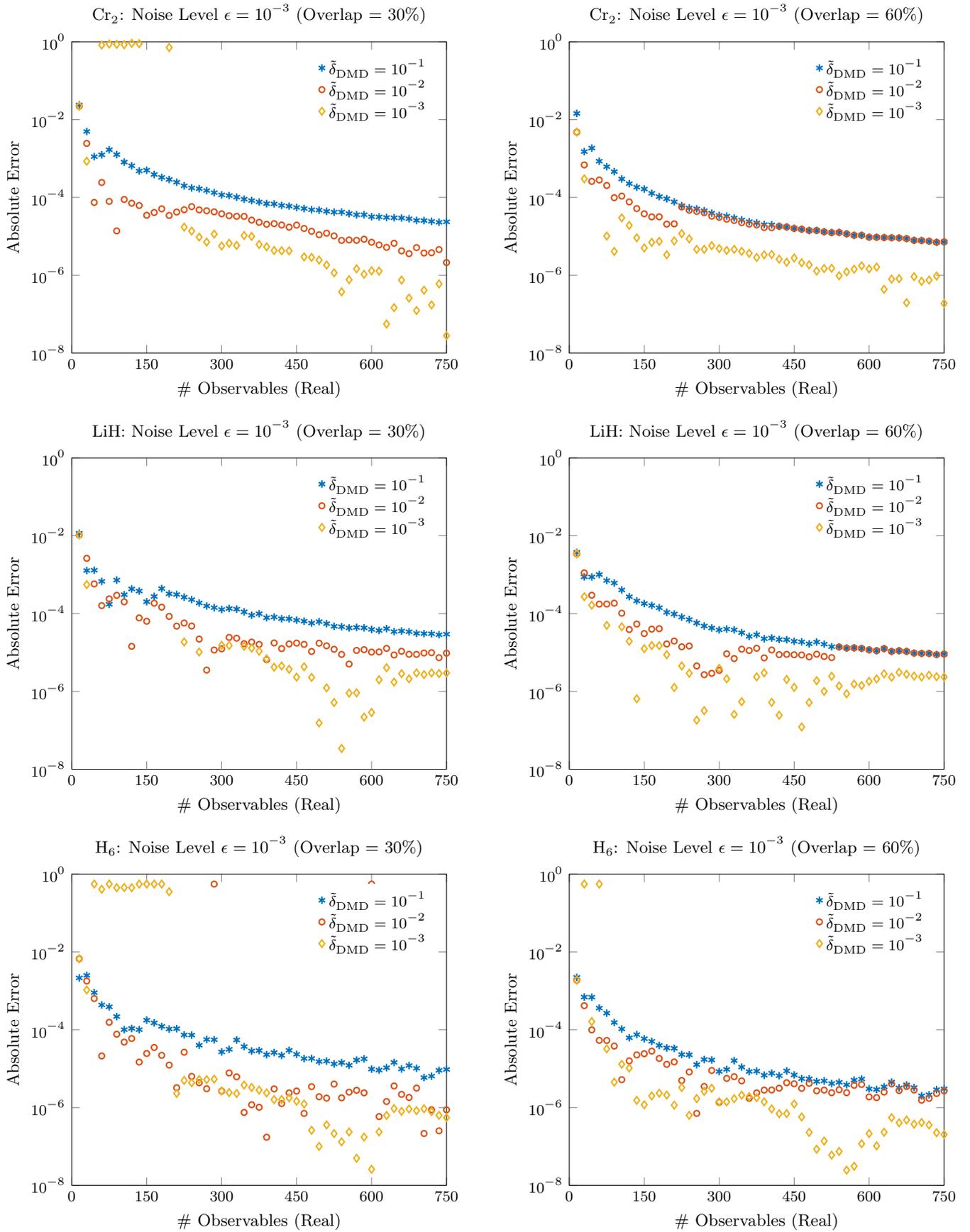

\centering
\begin{minipage}{0.48\columnwidth}
\plotconvN[1e-8]{Cr$_2$: Noise Level $\epsilon = 10^{-3}$ (Overlap = 30\%)}{Cr2-n0.001-o30}\\[5pt]
\plotconvN[1e-8]{LiH: Noise Level $\epsilon = 10^{-3}$ (Overlap = 30\%)}{LiH-n0.001-o30}\\[5pt]
\plotconvN[1e-8]{ H$_6$: Noise Level $\epsilon = 10^{-3}$ (Overlap = 30\%)}{H6-n0.001-o30}
\end{minipage}%
\hfill%
\begin{minipage}{0.48\columnwidth}
\plotconvN[1e-8]{Cr$_2$: Noise Level $\epsilon = 10^{-3}$ (Overlap = 60\%)}{Cr2-n0.001-o60}\\[5pt]
\plotconvN[1e-8]{LiH: Noise Level $\epsilon = 10^{-3}$ (Overlap = 60\%)}{LiH-n0.001-o60}\\[5pt]
\plotconvN[1e-8]{H$_6$: Noise Level $\epsilon = 10^{-3}$ (Overlap = 60\%)}{H6-n0.001-o60}
\end{minipage}%
\caption{Comparison of ODMD ground state energy convergence for a range of molecules with varying reference-ground state overlap probabilities $p_0$ and singular value truncations $\tilde{\delta}_{\rm DMD}$. A timestep of $\Dt = 3$ is used for generating the results.}
\end{figure}